\documentclass[final,3p,times]{elsarticle}

\usepackage{amssymb,amsmath,amsfonts,amsthm}
\usepackage{caption}
\usepackage{subcaption}
\journal{Journal of \LaTeX\ Templates}

\newtheorem{theorem}{Theorem}[section]
\newtheorem{corollary}{Corollary}[theorem]

\newtheorem{remark}[theorem]{Remark}
\newtheorem{example}[theorem]{Example}
\newtheorem{definition}[theorem]{Definition}

\newtheorem{property}[theorem]{Property}
\usepackage{color}
\usepackage{hyperref}

\usepackage{bigints}

\usepackage{overpic}
\usepackage{tikz}
\usepackage{graphicx}
\usepackage{array}
\usepackage{multirow}

\usepackage{amssymb}
\usepackage{tikz}

\usepackage{comment}

\usetikzlibrary{matrix,arrows.meta}
\newlength{\ml}
\renewcommand{\epsilon}{\varepsilon}
\renewcommand{\phi}{\varphi}

\newcommand{\sign}{\operatorname{sign}}

\begin{document}

\begin{frontmatter}

\title{Shape Reconstruction of Trapezoidal Surfaces}

\author{Arvin Rasoulzadeh \fnref{Arvin}}
\author{Martin Kilian \fnref{Martin}}
\author{Georg Nawratil \fnref{Georg}}
\address{Institute of Discrete Mathematics and Geometry \& Center for Geometry and Computational Design, TU Wien \\
Wiedner Hauptstra{\ss}e~8-10/104, Vienna, Austria}
\fntext[Arvin]{E-mail address: rasoulzadeh@geometrie.tuwien.ac.at.}
\fntext[Martin]{E-mail address: kilian@geometrie.tuwien.ac.at.}
\fntext[Georg]{E-mail address: nawratil@geometrie.tuwien.ac.at.}

\begin{abstract}
A smooth T-surface can be thought of as a generalization of a surface of revolution in such a way that the axis of rotation is not fixed at one point but rather traces a smooth path on the base plane. Furthermore, the action, by which the aforementioned surface is obtained does not need to be merely rotation but any ``suitable" planar equiform transformation applied to the points of a certain smooth profile curve.
In analogy to the smooth setting, if the axis footpoints sweep a polyline on the base plane and if the profile curve is discretely chosen then a T-hedra (discrete T-surface) with trapezoidal faces is obtained.\\ 
The goal of this article is to reconstruct a T-hedron from an already given point cloud of a T-surface. In doing so, a kinematic approach is taken into account, where the algorithm at first tries to find the aforementioned axis direction associated with the point cloud. Then the algorithm finds a polygonal path through which the axis footpoint moves. Finally, by properly cutting the point cloud with the planes passing through the axis and its footpoints, it reconstructs the surface. 
The presented method is demonstrated on base of  examples.\\
From an applied point of view, the straightforwardness of the generation of these surfaces predestines them for building and design processes. In fact, one can find many built objects belonging to the sub-classes of T-surfaces such as \emph{surfaces of revolution} and \emph{moulding surfaces}.  Furthermore, the planarity of the faces of  the discrete version paves the way for steel/glass construction in industry. Finally, these surfaces 
are also suitable for transformable designs as they allow an isometric deformation. 
\end{abstract}
\end{frontmatter}

\section{Introduction}

The reconstruction of geometric models from point clouds (e.g.\ 3D scan data) is also known under the term {\it reverse engineering} \cite{VARADY}.  One important phase in this process is the {\it segmentation and surface fitting} step (cf.\ \cite[Fig.\ 1]{VARADY}); i.e.\ the point cloud has to be partitioned into suitable subsets, which are then fitted by certain primitive surfaces \cite{ANDREWS}. 
We focus on the latter step, where we restrict on the class of surfaces swept by a {\it profile curve} $\mathbf{p}$ under a one-parametric motion $\mu$. 
Several papers are dealing with the fitting of these {\it kinematic surfaces}, which are reviewed next.

\subsection{Review on fitting with kinematic surfaces}

Most work has been done on the recognition of surfaces, which are invariant with respect to the generating one-parametric motion $\mu$. A necessary and sufficient condition for this invariance is that $\mu$ is a so-called {\it uniform motion} having the property that the associated velocity vector filed is constant; i.e.\ it does not depend on the time $t$.

\begin{tabular}[t]{ |p{0.7\columnwidth} p{0.2\textwidth}|}
 \hline
 \multicolumn{2}{|c|}{Nomenclature} \\
 \hline
 Real numbers                                                               & $\mathbb{R}$ \\
 Integers                                                                   & $\mathbb{Z}$ \\
 Time                                                                       & $t$          \\
 Scaling factor of equiform motion                                          & $\eta$       \\
 Smooth scaling factor                                                      & $\xi$       \\
 Directing surface                                                          & $\Delta$     \\
 Profile surface
              & $\Psi$     \\
 Profile planes in smooth/discrete scenarios                                & $\delta(t)$\,/\,$\delta_j$  \\
 Trajectorty planes in smooth/discrete scenarios                            & $\tau(s)$\,/\,$\tau_i$    \\
 Base plane                                                                 & $\tau_0$     \\
 Generating motion                                                          & $\mu$        \\
 Planar Frenet frame (tangent vectors, normal vector)                              & ($\mathbf{f}_t$,$\mathbf{f}_n$)\\
 Derivative of a single variable vector $\mathbf{c}(t)$                     & $\dot{\mathbf{c}}  = \frac{\mathrm{d}}{\mathrm{d} t}\mathbf{c}$\\
 $\beta$-evolute of planar curve $\mathbf{c}$                               & $\mathbf{c}_{*}(\beta)$ \\
 $\beta$-involute of $\mathbf{c}$ at certain distance $d$                   & $\mathbf{c}^{*}_d(\beta)$ \\    
 Rotation matrix with angle $\beta$                                         & $\mathbf{R}(\beta)$\\
 Unorganized point cloud                                                    & $\mathcal{X}$\\
 Unorganized trajectories of $\mathcal{X}$                                  & $\mathcal{X}_{\text{cut}}$\\
 Organized point cloud of a T-hedron/T-surface                & $\mathbf{x}(i,j)$\,/\,$\mathbf{x}(s,t)$\\
Initial guess for the global optimization
& $\mathbf{x_0}(i,j)$\\
 Organized trajectory                                                       & $\mathbf{t}$     \\
 Organized profile curve                                                    & $\mathbf{p}$     \\
 Directrix                                                                  & $\mathbf{d}$     \\
 Generator of the directing surface $\Delta$                                                                  & $\mathbf{g}$     \\
 \hline
\end{tabular}

\paragraph{Invariant surfaces}
In the case of Euclidean motions these are uniform rotations, uniform translations and uniform helical motions and the associated invariant surfaces are rotational, translational and helical, respectively.  
These surfaces can also be characterized by the fact that the surface normals belong to the set of path normal of the underlying uniform motion \cite[Section 4.2]{PW}. 
This  line-geometric characterisation can be exploited computationally  for solving the optimization problem of determining the best approximating invariant surface\footnote{It is also possible to detect spheres, planes and cylinders of revolution as they are invariant with respect to a higher-dimensional set of uniform motions. \label{fn:multi}} to a given point cloud (see also \cite{PPR_intro} and the references given therein), but it can also be used for the segmentation task \cite{gelfand}. A practical application is in detail discussed in \cite{gfrerrer}.

\paragraph{Profile surfaces}
A further interesting class of kinematic surfaces are {\it profile surfaces} $\Psi$, which can be generated by the rolling (without sliding) of the carrier plane $\delta$ of a planar profile curve $\mathbf{p}$ on a developable surface $\Delta$. Due to the rolling the instantaneous motion is at any time instant $t$ a pure rotation about the generator $\mathbf{g}(t)$ of $\Delta$ along which $\Delta$ is touched by $\delta(t)$. 
 As a result of this special generation, there exists for all $t$ a surface of revolution $\Phi(t)$ which osculates  $\Psi$ along the profile curve $\mathbf{p}(t)$. This latter property motivated the reconstruction of these surfaces by smoothly joined surfaces of revolutions\footnote{To the best of the authors' knowledge, this method was only applied to examples of moulding surfaces in the literature so far.} (trajectory curve is a $G^1$ curve composed or circular arcs) \cite{PCL}. 
Another characterisation of $\Psi$ implied by the family of osculating surfaces $\Phi(t)$ is that the principal curvature lines correspond to  $\mathbf{p}(t)$ and the trajectories of its points. This property was exploited in \cite{kovacs} for the surface reconstruction. 

By restricting $\Delta$ to a cylindrical surface one gets a special class of profile surfaces known as {\it moulding surfaces}. For this special type a further method for the fitting was presented in \cite{PHOW} based on more sophisticated line-geometric considerations using the theory of normal congruences. Note that this approach also allows the detection of further special surface classes;  e.g.\ canal surfaces and developable surfaces. 

\paragraph{Equiform invariant surfaces}
The concept of invariant surfaces can also applied to the group of equiform motions, which is the composition of Euclidean displacements and similarities with scale-factor $\eta$. This motion group contains two further invariant surfaces; namely conical surfaces and spiral surfaces which are implied by uniform central scalings and uniform spiral motions, respectively.  
Up to the knowledge of the authors' this idea was first published in \cite{lee}. By extending the geometry of lines to the geometry of line-elements (line plus a point on it) in \cite{boris} the invariant surfaces under equiform motions can be characterized by the 
fact that the surface normal-elements belong to the set of path normal-elements of the underlying uniform motion. This theory is applied in \cite{hofer} where it is also used for the segmentation task\footnote{
Beside the already mentioned invariant surfaces it is also possible to detect cones of revolution and spiral cylinders for the same reasons as given in footnote \ref{fn:multi}.}.

All the mentioned approaches for reconstructing 
(equiform) invariant surfaces rely on a constrained optimization, where the normalization condition can lead to problems\footnote{This also holds for the mentioned detection based on normal congruences \cite{PHOW}, as it involves the normalization of the instantaneous screw, which is problematic to the dimensional inhomogeneity.} in terms of robustness \cite{ANDREWS}. The authors of  \cite{ANDREWS}  suggested alternative normalization methods (see also \cite{lin}), which can also be applied to generalized velocity fields allowing the detection of further surface classes, such as affinely-scaled surfaces of revolution.

\paragraph{Equiform profile surfaces} 
These surfaces can be generated as the profile surface mentioned above but superposed by a scaling whose center is located on $\mathbf{g}(t)$. The reconstruction of such surfaces was not done but only mentioned as future research in \cite{kovacs}.

\paragraph{Kinematically generated freeform surfaces} In \cite{barton} the most general approach was presented for detecting parts of freeform surfaces that can be (approximately) generated by an Euclidean motion of a planar profile curve. The given algorithm is of high complexity and very time consuming, as it is based on an exhaustive sampling of planar cuts. 

Moreover, in \cite{snake} an algorithm is given for the computation of a circular arc snake, whose sweep fits a given freeform surface. 

Finally, it should be mentioned that there is huge body of literature on the fitting of surfaces by ruled (or even developable) surfaces. As this topic, which is a problem for its own, goes beyond the scope of this paper, a review is omitted.

\subsection{Profile-affine surfaces: motivation and review}\label{subsec:profile_affine}

In this paper we want to present an algorithm which fits so-called profile-affine surface to a give point cloud. These surfaces introduced by Sauer and Graf  \cite{graf} can be seen as a generalization of moulding surfaces in the way that at each time instant $t$ the instantaneous rotation about the cylinder ruling $\mathbf{g}(t)$ of $\Delta$ is replaced by an instantaneous stretch-rotation; i.e.\ a composition of a rotation about $\mathbf{g}(t)$ and an axial dilatation with stretch factor $\eta$. The latter yields an affine deformation of the profile curve orthogonal to $\mathbf{g}(t)$, which gives rise for its nomenclature. 
If $\Delta$ degenerates into a finite line we get as special cases (stretch-)rotational surfaces. 

We pick one of the planes orthogonal to the ruling direction, which also contains a point $\mathbf{x}$ of the profile curve, and tag it as {\it base plane} $\tau_0$. The intersection of $\Delta$ with  $\tau_0$ is called {\it directrix} $\mathbf{d}$. The path of $\mathbf{x}$ under the generating motion $\mu$ of $\Psi$ gives the so-called {\it trajectory curve} $\mathbf{t}:=\mathbf{x}(t)$.

Profile-affine surfaces are of interest as they allow an isometric deformation within this surface class. For rotational surfaces this was first discovered by Minding \cite{minding}. For moulding surfaces this result dates back to Peterson \cite{peterson} (see also \cite[page 300]{staeckel}). For stretch-rotation surfaces this property was proven by Lagally \cite{lagally}. 
The isometric deformation of general profile-affine surfaces can be followed from the isometric deformability of stretch-rotational surfaces according to \cite[page 139]{sauer1970differenzengeometrie}, were also an alternative proof for the result of Lagally is given. 

The dicretized and semi-discretized versions of profile-affine surfaces recently attracted attention in the context of transformable design of surfaces \cite{SNRT21,MSN22} and tubular structures \cite{SMN23,MSNP23}, respectively,  
as they allow an intuitive access to their spatial shape by three control polylines\footnote{In the semi-discrete case one or two of them are smooth}. 
The discrete analogs are also known as {\it T-hedra}, where the T is implied by the property that the faces, which are arranged in the combinatorics of a square grid, are  all trapezoidal. For reasons of unifications in notations profile-affine surfaces are also called {\it T-surfaces}. 
A comparison between T-hedra and T-surfaces was done in tabular form \cite[§ 19]{graf} and in detail in Sauer's book \cite{sauer1970differenzengeometrie}. 
A contemporary comparison between the polyhedral and the smooth situation was given recently in \cite{izmestiev2023isometric}, where also the corresponding formulas for the isometric deformations of T-hedra are derived. 

The surface class under consideration comprises a rich variety of shapes and has potential for applications in various fields (e.g.\ kinetic architecture, transportable structures). This reasons our interest in the problem of fitting these surfaces to a point cloud, which originate e.g.\ from the sketch of a designer \cite{RWK23}. Therefore we aim for an efficient algorithm which has the capability to give almost real-time feedback, which is needed to serve as an interactive tool within the design process; e.g.\ implemented within the sketching app "MR.Sketch" \cite{mrsketch}.

\subsection{Contribution and outline of the paper}

In Section \ref{sec:theory} we provide the theoretical fundamentals on T-surfaces needed for their reconstruction from point clouds. In fact we will approximate the point cloud $\mathcal{X}$ not by a T-surface but by a T-hedron, whose resolution is a user input. 
The algorithm itself consists of two major phases:
    Determination of a first approximation of $\mathcal{X}$ by a T-hedron (Phase 1; cf.\ Section \ref{sec:initial_guess}), which is used as initial guess for the global optimization (Phase 2; cf.\ Section \ref{sec:global_opt}).
As the latter phase is based on a gradient descent approach, its outcome heavily depends on the initial guess. Therefore a lot of effort is put into the computation of a good initial guess, which can be subdivided into the following steps: 
\begin{enumerate}[a)]
    \item 
    Computing the ruling direction of $\Delta$ (Section \ref{subsec:direction})
    \item 
    Computation of the trajectory polyline $\mathbf{t}$ (Section \ref{subsec:trajectory})
    \item 
    Computation of the profile planes $\delta(t)$ (Section \ref{subsec:p_planes})
    \item 
    Computation of the profile polyline $\mathbf{p}$ (Section \ref{subsec:p_curve})
\end{enumerate}
The overall algorithm is novel, but some of the individual steps clearly rely on already known approaches to which we refer in the corresponding passages. 
The algorithm is demonstrated on a set of examples given in Section \ref{sec:results} including a discussion of the obtained results. Finally, the paper is concluded in Section \ref{sec:conclusion}.


\section{Theoretical background}\label{sec:theory}

First we have a look at the concepts of $\beta$-evolutes and $\beta$-involutes (Section \ref{sec:beta_e_i}), which are needed for an alternative construction of T-surfaces  (Section \ref{sec:Tsurface}) beside the already given kinematic generation (cf.\ Section \ref{subsec:profile_affine}). Analogue considerations for the discrete case of T-hedra are done in Section  \ref{sec:thedra}.


\subsection{$\beta$-evolutes and $\beta$-involutes}\label{sec:beta_e_i}

\begin{figure*}[t!]
    \begin{subfigure}[b]{.24\columnwidth}
    \centering
        \includegraphics[height = 40 mm,trim={0mm 0mm 0mm 0mm},clip]{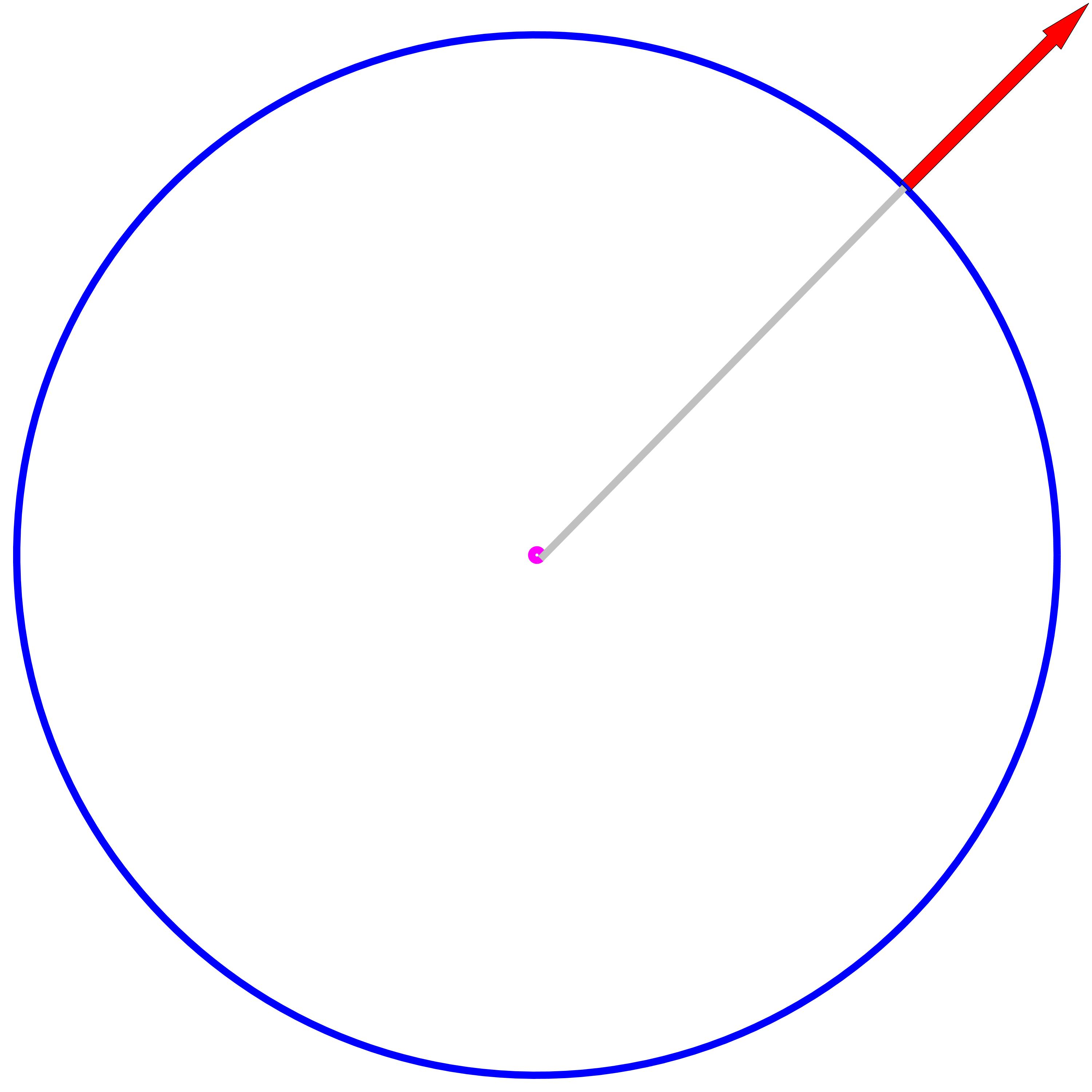}
        \caption{$\beta = 0$.}
        \label{fig:betaevol:AA}
    \end{subfigure}
    \hfill
    \begin{subfigure}[b]{.24\columnwidth}
    \centering
        \includegraphics[height= 43 mm ,trim={0mm 0mm 0mm 0mm},clip]{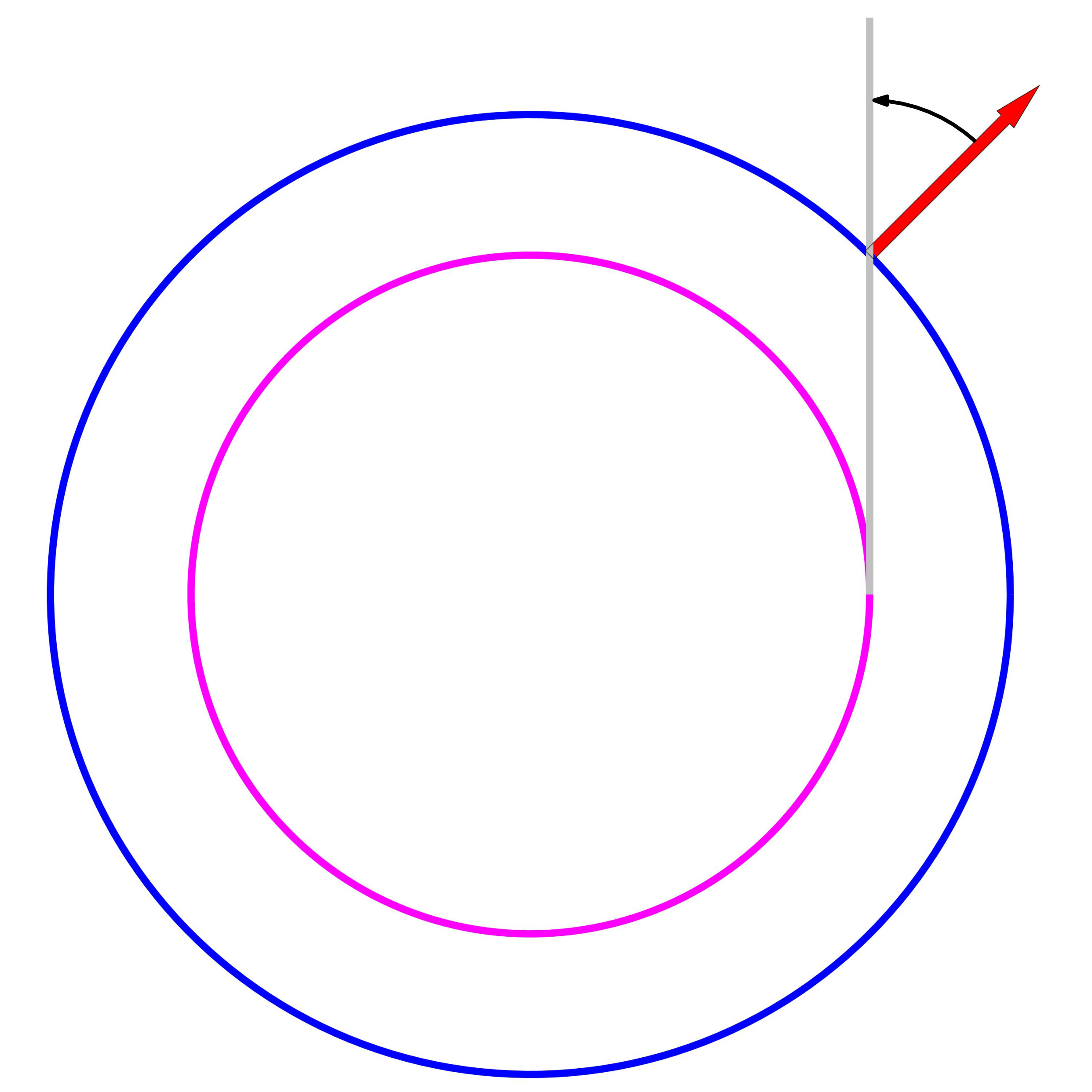}
        \caption{$\beta = \frac{\pi}{4}$.}
        \label{fig:betaevol:BB}
    \end{subfigure}
    \hfill
    \begin{subfigure}[b]{.24\columnwidth}
    \centering
        \includegraphics[grid,height= 41 mm,trim={0mm 0mm 0mm 0mm},,clip]{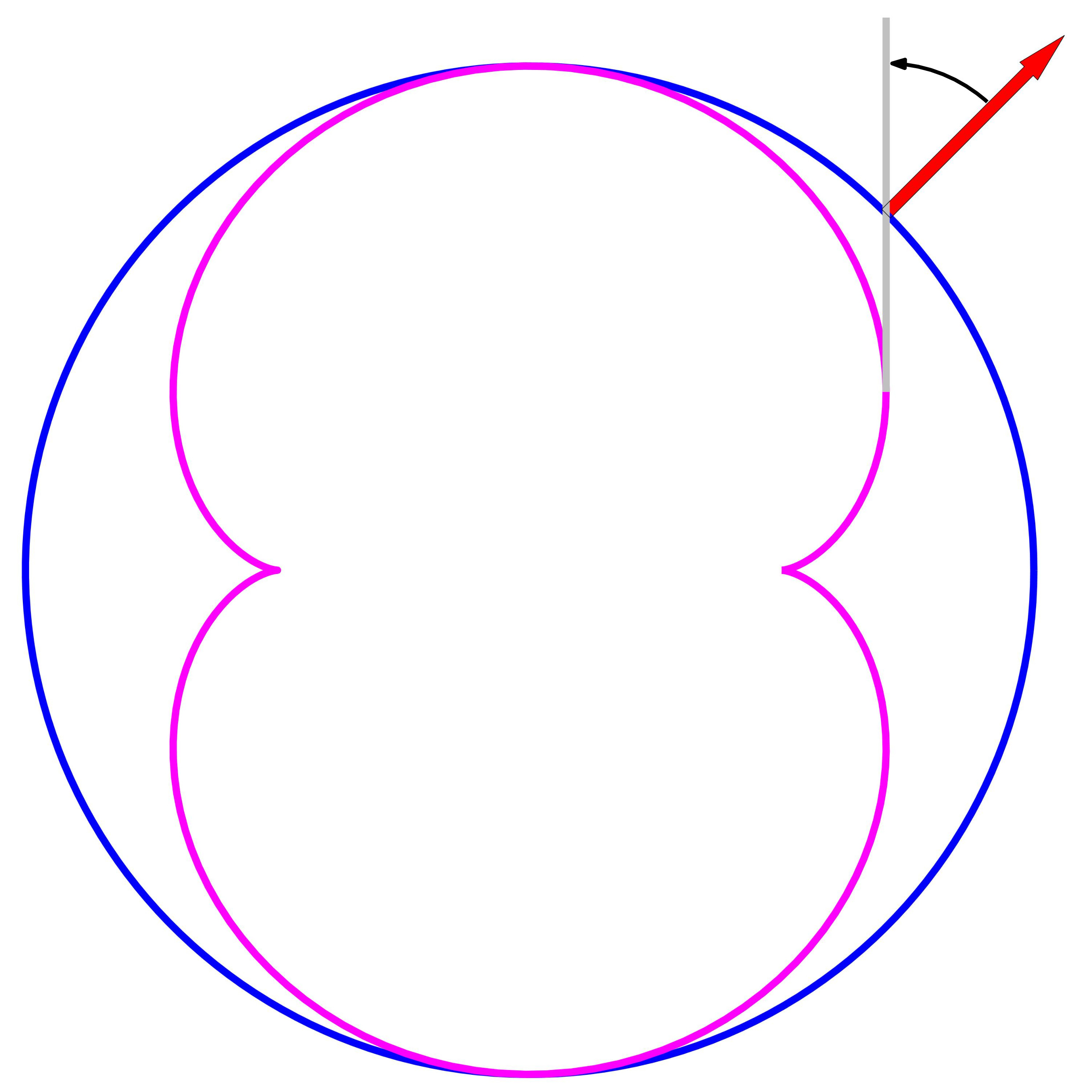}
        \caption{$\beta = t$.}
        \label{fig:betaevol:CC}
    \end{subfigure}
        \hfill
    \begin{subfigure}[b]{.24\columnwidth}
    \centering
        \includegraphics[grid,height= 43 mm,trim={0mm 0mm 0mm 0mm},,clip]{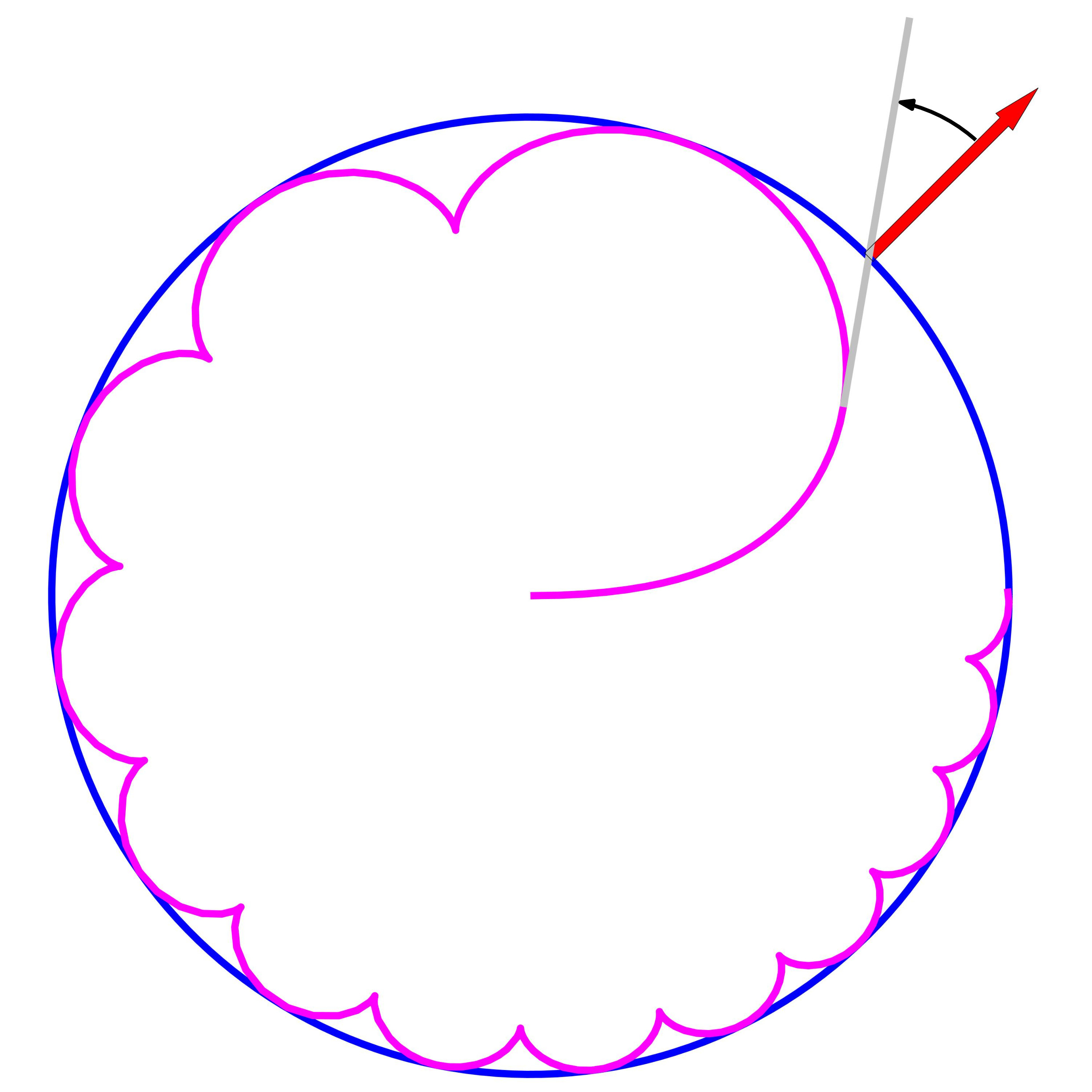}
        \caption{$\beta = t^2$.}
        \label{fig:betaevol:DD}
    \end{subfigure}
    \caption{Illustration of the $\beta$-evolutes of a unit circle for different values of $\beta$. In the case of $\beta(t) = t$ and $\beta(t) = t^2$, $\beta$-evolute is plotted for $t \in [0,2\pi]$. The red arrow shows the normal to the blue circle and the gray line shows the tangent to the $\beta$-evolute.
    }
    \label{fig:betaevol}
\end{figure*}

{The exploration of the $\beta$-involutes and $\beta$-evolutes done in this section is not relying on the \emph{intrinsic description}  \cite{apostol2010tanvolutes} defining curves by internal properties like arc length, tangential angle, and curvature, independent of their spatial orientation. Our approach diverges from this, focusing on integrating these curves as coordinate curves of T-surfaces.
As the studied smooth maps and vectors are functions of a single variable $t$, we will use the dot notation (e.g.\ $\dot{a}$) to denote derivatives with respect to this variable.}
\begin{definition}
Let $\mathbf{c} :I \rightarrow \mathbb{R}^2$ be a regular arc-length parameterized curve and $\beta : I \rightarrow \left(-\frac{\pi}{2},\frac{\pi}{2}\right)$ be a smooth function. Then the envelope of the rotated normals of $\mathbf{c}$ by the value $\beta$ about $\mathbf{c}$ (i.e.\, $\mathbf{R}\left(\beta\right).\mathbf{n}$ with $\mathbf{n}$ interpreted as the Frenet normals of $\mathbf{c}$) is called the $\beta$-evolute of $\mathbf{c}$ and is denoted by $\mathbf{c}_{*}(\beta(t))$. 
\label{def:beta:evolute}
\end{definition}
From above it is seen that in the case that $\beta = 0$ the $\beta$-evolute coincides with the classical notion of \emph{evolutes}. Naturally, as the evolutes can be interpreted as the curves obtained through the locus of centers of osculating circles of a given curve, a similar interpretation from $\beta$-evolutes is expected too. This interpretation is encoded into the following theorem.
\begin{theorem}
Let $\mathbf{c} :I \rightarrow \mathbb{R}^2$ be a regular arc-length parameterized curve and $\beta : I \rightarrow \left(-\frac{\pi}{2},\frac{\pi}{2}\right)$ be a smooth function. Then its $\beta$-evolute is
\begin{equation}
    \mathbf{c}_{*}(\beta(t)) = \mathbf{c}(t) + \left(\frac{\cos{\left(\beta(t)\right)}}{\kappa(t) + {\dot{\beta}(t)}}\right)\,\mathbf{R}(\beta(t)).\mathbf{n}(t),\quad\quad
    \text{where}\quad\quad
    \mathbf{R}(\beta(t)) = \left(\begin{array}{cc}
    \cos{(\beta(t))} & -\sin{(\beta(t))} \\
    \sin{(\beta(t))} &  \cos{(\beta(t))}
    \end{array}\right).
    \end{equation}
\label{theorem:beta:evolute}
\end{theorem}
\begin{proof}
Let us drop $(t)$ as pretty much everything is $t$ parameter dependent and let us abbreviate $\mathbf{c}_{*}(\beta(t))$ to $\mathbf{c}_{*}$. Since the $\beta$-evolute is the envelop of all the rotated normals of $\mathbf{c}$ it should be of the general form 
\begin{equation*}
    \mathbf{c}_{*} = \mathbf{c} + \lambda\,\mathbf{R}.\mathbf{n},
\end{equation*}
with $\lambda$ being the amount that one needs to travel along the rotated normal to reach  $\mathbf{c}_{*}$, such that $\dot{\mathbf{c}}_{*} \in \mathrm{span}\{\,\mathbf{R}.\mathbf{n}\,\}$. Henceforth, the only thing that remains to be indicated is $\lambda$. Expanding $\dot{\mathbf{c}}_{*}$ using Frenet-Serret formula for $\mathbf{c}$ (where $\mathbf{f}_t$ is the tangent vector, $\mathbf{f}_n = \mathbf{n}$ is the normal vector) gives
\begin{equation*}
    \dot{\mathbf{c}}_{*} = \mathbf{f}_t + \dot{\lambda}\,\mathbf{R}.\mathbf{f}_n + \lambda\,\dot{\mathbf{R}}.\mathbf{f}_n + \lambda\,\mathbf{R}\,\dot{\mathbf{f}}_n
                         = \mathbf{f}_t + \dot{\lambda}\,\mathbf{R}.\mathbf{f}_n + \lambda\,\dot{\mathbf{R}}.\mathbf{f}_n - \lambda\,\kappa\,\mathbf{R}.{\mathbf{f}_t},
\end{equation*}
where $\kappa$ is the curvature of $\mathbf{c}$. Now, by identifying $\mathbf{f}_t$ and $\mathbf{f}_n$ with $(1,0)^T$ and $(0,1)^T$ respectively, the linear dependency of $\dot{\mathbf{c}}_{*}$ and $\mathbf{R}.\mathbf{f}_n$ yields
\begin{equation}
    \det\left(\begin{array}{cc}
    -\sin{(\beta)}    & 1 - \dot{\lambda}\,\sin{(\beta)} - \lambda\,\dot{\beta}\,\cos{(\beta)} - \lambda\,\kappa\,\cos{(\beta)} \\
     \cos{(\beta)}    &     \dot{\lambda}\,\cos{(\beta)} - \lambda\,\dot{\beta}\,\sin{(\beta)} - \lambda\,\kappa\,\sin{(\beta)}
    \end{array}\right) = 0 \implies \lambda\,\dot{\beta} + \lambda\,\kappa - \cos{\beta} = 0
    \label{eq:nu:beta}
\end{equation}
which upon solving the last term for $\lambda$ gives the result.
\end{proof}
\begin{example}\label{ex:beta_const}
One of the special types of $\beta$-evolutes that are of importance in practical applications is the case where $\beta$ is constant. Assuming so simplifies our equations to 
\begin{equation}\label{eq:beta:evolute:const}
    \mathbf{c}_{*}(t) = \mathbf{c}(t) + \left(\frac{\cos{\beta}}{\kappa(t)}\right)\,\mathbf{R}(\beta).\mathbf{f}_n(t),\quad\quad
    \text{where}\quad\quad
    \mathbf{R}(\beta) = \left(\begin{array}{cc}
    \cos{\beta} & -\sin{\beta} \\
    \sin{\beta} &  \cos{\beta}
    \end{array}\right).
\end{equation}
One should note that, upon substituting $\beta = 0$ in Eq.\ (\ref{eq:beta:evolute:const}) one gets the classical formula for the evolutes. Fig.\,\ref{fig:betaevol} depicts $\beta$-evolutes for the four major scenarios where $\beta$ is constant zero (a), constant non-zero (b) and varying linearly (c) and non-linearly (d), respectively. 
\end{example}
\begin{definition}
Let $\mathbf{c}: I\rightarrow \mathbb{R}^2$ be a regular curve and $\beta: I \rightarrow \left(-\frac{\pi}{2},\frac{\pi}{2}\right)$ be a smooth function. Then the curves $\mathbf{c}^{*}_d : I \rightarrow \mathbb{R}^2$ whose rotated normals by the angle $\beta$ envelope $\mathbf{c}$ are called $\beta$-involutes of $\mathbf{c}$. 
\end{definition}


\begin{theorem}
Let $\mathbf{c} : I \rightarrow \mathbb{R}^2$ be a regular curve and $\beta: I \rightarrow \left(-\frac{\pi}{2},\frac{\pi}{2}\right)$ be a smooth function. Then the $\beta$-involute is
\begin{equation}
    \mathbf{c}^{*}_{d} =
    \mathbf{c} - \, \xi\,\left(\int_{\,0}^{\,t}\frac{\left|\,{\dot{\mathbf{c}}}\,\right|}{\xi}\,\mathrm{d}w + d\right)\,\frac{\dot{\mathbf{c}}}{\left|\,{\dot{\mathbf{c}}}\,\right|},
  \label{eq:beta:involute}
\end{equation}
where $\xi\,(t) := \exp\left( \int_{\,0}^{\,t}{\kappa(w)\,\tan\left(\beta(w)\right)\,\mathrm{d}w} \right)$ 
and $d$ is an arbitrary constant. 
\label{theorem:parametrization:beta:involute}
\end{theorem}
\begin{proof}
$\mathbf{c}$ being the envelope of rotated normals of $\mathbf{c}^{*}_{d}$ implies   
\begin{equation}
\mathbf{c}^{*}_{d} = \mathbf{c} - \lambda\,\frac{\dot{\mathbf{c}}}{\left|\,{\dot{\mathbf{c}}}\,\right|},
\label{eq:main}
\end{equation}
where $\lambda(t)$ is a function whose absolute value is the length of the connecting line from $\mathbf{c}$ to the curve $\mathbf{c}^{*}_{d}$ in such a way that its initial value $\lambda\,(0) = d$. Taking derivative with respect to $t$ gives: 
\begin{equation}\label{eq:diff:eq:1}
    \dot{\mathbf{c}^{*}_{d}} = \left(\,\left|\,\dot{\mathbf{c}}\,\right| - \dot{\lambda}\,\right)\,\frac{\dot{\mathbf{c}}}{\left|\,\dot{\mathbf{c}}\,\right|} - 
     \lambda \left(\frac{\ddot{\mathbf{c}}}{\left|\,\dot{\mathbf{c}}\,\right|} - \frac{\langle \ddot{\mathbf{c}}\,,\,\dot{\mathbf{c}}\rangle\,\dot{\mathbf{c}}}{\left|\,\dot{\mathbf{c}}\,\right|^3} \right)
\end{equation}
Now, writing the decomposition of $\dot{\mathbf{c}^{*}_{d}}$ in the Frenet frame 
$(\mathbf{f}_t,\mathbf{f}_n)$ of $\mathbf{c}$  and taking into account that the curvature $\kappa$ of a non-arc-length parametrized curve $\mathbf{c}$ is given by
\begin{equation*}
    \kappa = \left|\,\frac{\ddot{\mathbf{c}}}{\left|\,\dot{\mathbf{c}}\,\right|} - \frac{\langle\,\ddot{\mathbf{c}}\,,\,\dot{\mathbf{c}}\,\rangle\,\dot{\mathbf{c}}}{\left|\,\dot{\mathbf{c}}\,\right|^{\,3}}\,\right|
\end{equation*}
we end up with
\begin{equation*}
    -\left|\,\dot{\mathbf{c}^{*}_{d}}\,\right|\,(\cos\left(\beta\right)\,\mathbf{f}_n +\sin\left(\beta\right)\,\mathbf{f}_t) = \left(\,\left|\,\dot{\mathbf{c}}\,\right| - \dot{\lambda}\,\right)\mathbf{f}_t - {\lambda\,\kappa}\,\mathbf{f}_n.
\end{equation*}
The above equation then results in the following system of differential equations: 
\begin{equation*}
  \left|\,\dot{\mathbf{c}^{*}_{d}}\,\right|\,\cos\left(\beta\right) = {\lambda\,\kappa},\quad\quad\quad\quad
  \left|\,\dot{\mathbf{c}^{*}_{d}}\,\right|\,\sin\left(\beta\right) = \dot{\lambda} - \left|\,\dot{\mathbf{c}}\,\right|.
\end{equation*}
Having in mind that $\cos\left(\beta\right)$ never vanishes due to $\beta(t) \in \left(-\frac{\pi}{2},\frac{\pi}{2}\right)$ for all $t \in J$, one can solve the first equation for $\left|\,\dot{\mathbf{c}^{*}_{d}}\,\right|$ and substitute it into the second one and obtain the following \emph{linear first order differential equation}:
\begin{equation}
    \dot{\lambda} - \lambda\,\kappa\,\tan(\beta) = \left|\,\dot{\mathbf{c}}\,\right|,
    \label{eq:diff:eq:main}
\end{equation}
which upon solving yields 
\begin{equation}
    \lambda = \xi\,\left(\int_{\,0}^{\,t}\frac{\left|\,\dot{\mathbf{c}}\,\right|}{\xi}\,\mathrm{d}w + d\right),
    \label{eq:ell}
\end{equation}
where $1/\xi$ is the \emph{integrating factor} and $\xi\,(t) = \exp\left( \int_{\,0}^{\,t}{\kappa\,\tan\left(\beta\right)\,\mathrm{d}w} \right)$. Substituting Eq.\,(\ref{eq:ell}) into Eq.\,(\ref{eq:main}) gives the $\beta$-involutes. 
\end{proof}

\begin{remark}
    Note that, in Eq.\ (\ref{eq:beta:involute}), for every constant $d$ one gets a different $\beta$-involute. Therefore, having a given curve $\mathbf{c}$ results in having a family of $\beta$-involutes $\mathbf{c}^*_d$. \hfill $\diamond$
\end{remark}

At this juncture, we wish to underscore a significant aspect of the $\beta$-involute curves that is exploited later in the shape reconstruction process (see Section\,\ref{subsec:p_planes}):

\begin{property}\label{remark:beta:angle}
    The tangent lines of a curve intersects its $\beta$-involutes at corresponding points under equal angles.
\end{property}



\begin{figure*}[t!]
    \begin{subfigure}[b]{.24\columnwidth}
    \centering
        \includegraphics[height = 34 mm,trim={0mm 0mm 0mm 0mm},clip]{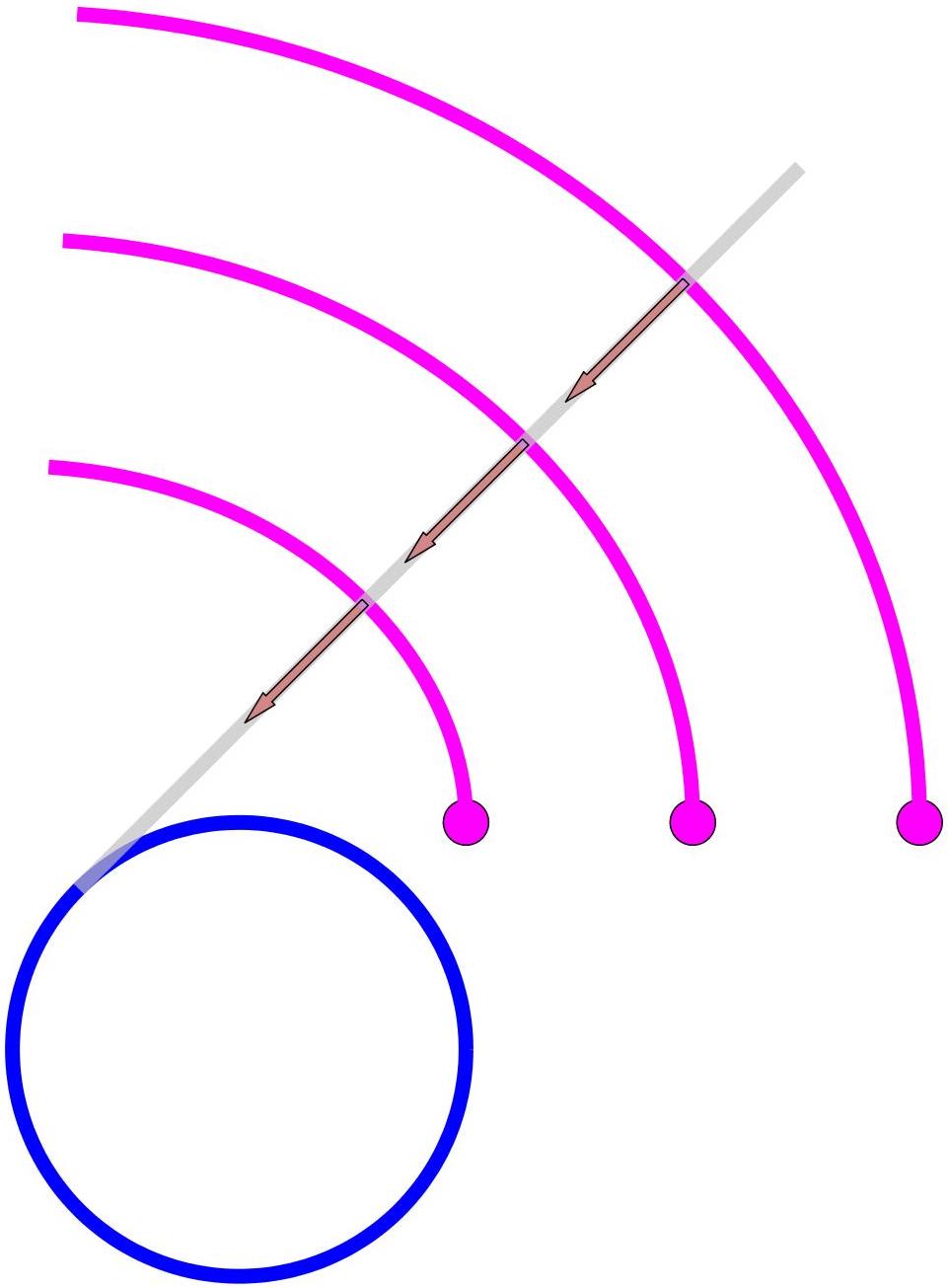}
        \caption{$\beta = 0$.}
        \label{fig:betainvol:AA}
    \end{subfigure}
    \hfill
    \begin{subfigure}[b]{.24\columnwidth}
    \centering
        \includegraphics[height= 52 mm ,trim={0mm 0mm 0mm 0mm},clip]{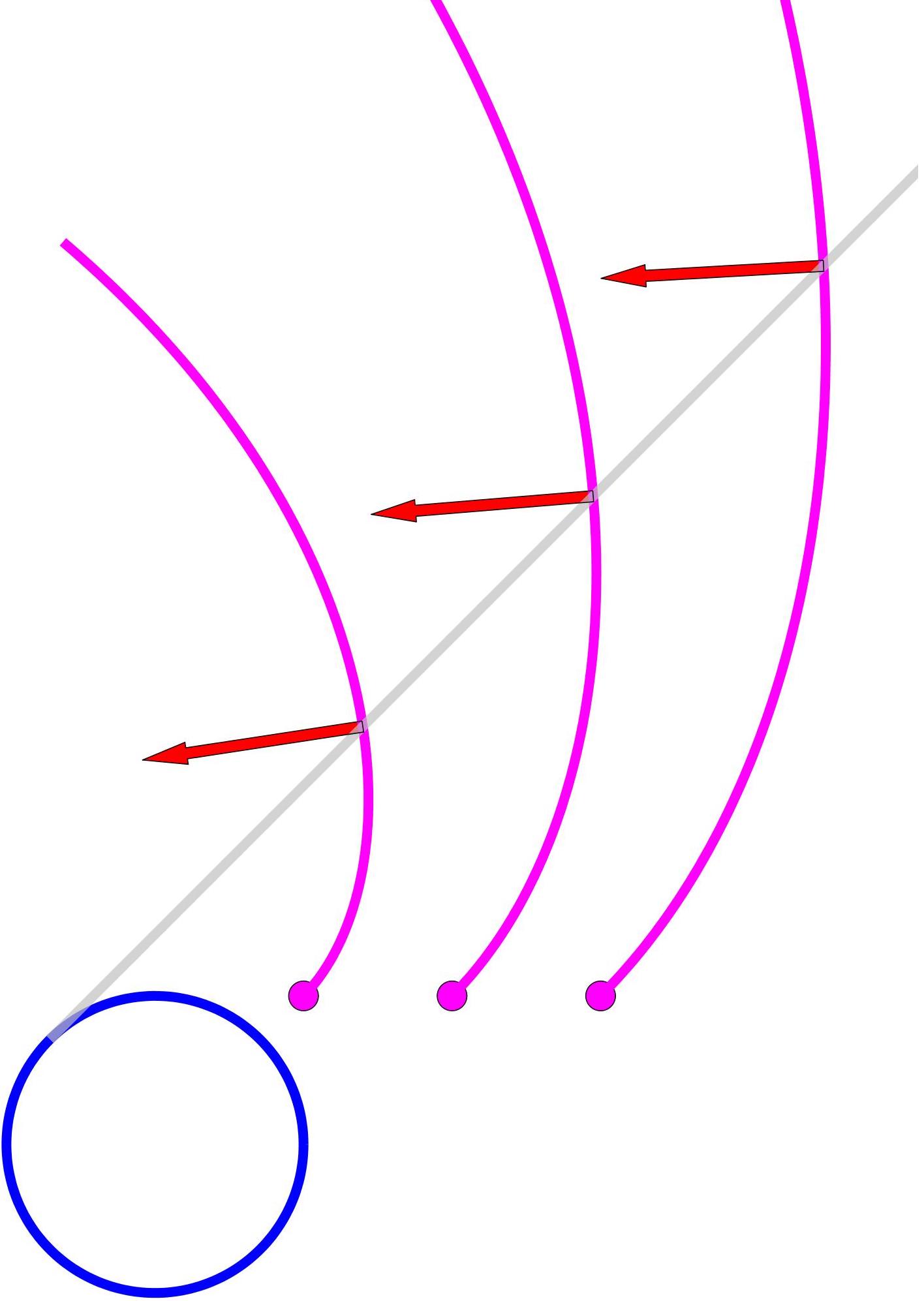}
        \caption{$\beta = \frac{\pi}{4}$.}
        \label{fig:betainvol:BB}
    \end{subfigure}
    \hfill
    \begin{subfigure}[b]{.24\columnwidth}
    \centering
        \includegraphics[grid,height= 45 mm,trim={0mm 0mm 0mm 0mm},,clip]{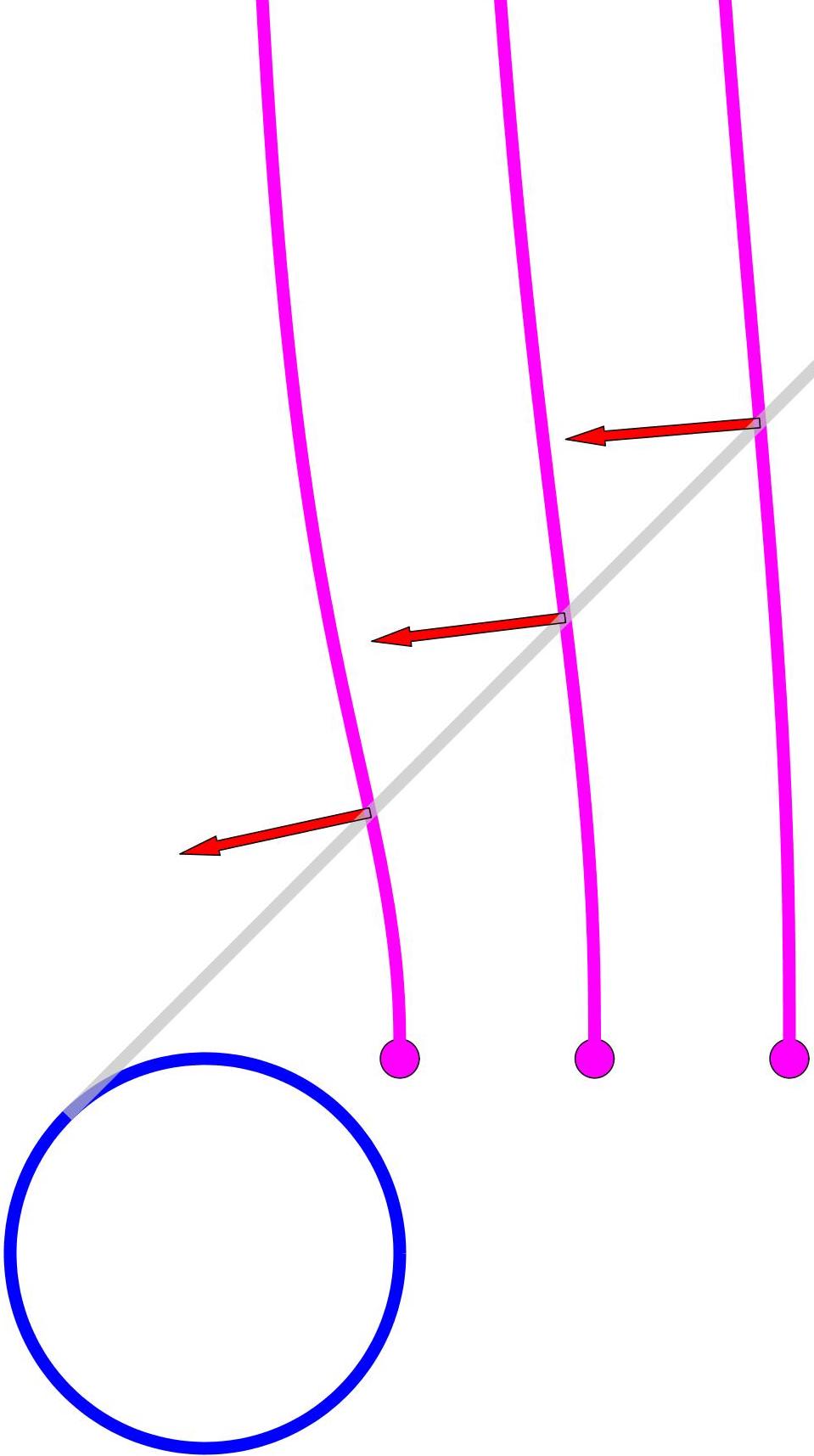}
        \caption{$\beta = t$.}
        \label{fig:betainvol:CC}
    \end{subfigure}
        \hfill
    \begin{subfigure}[b]{.24\columnwidth}
    \centering
        \includegraphics[grid,height= 44 mm,trim={0mm 0mm 0mm 0mm},,clip]{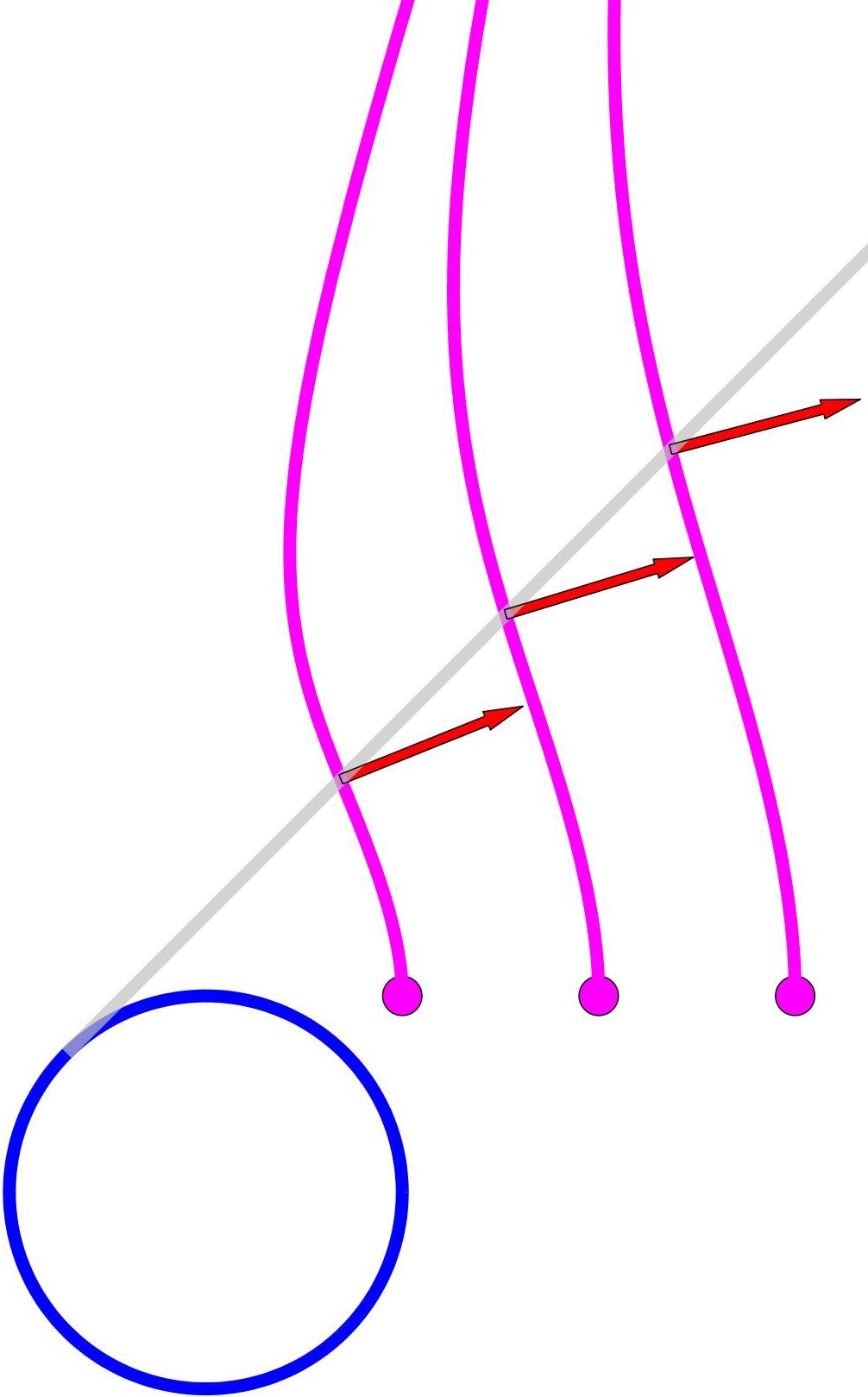}
        \caption{$\beta = t^2$.}
        \label{fig:betainvol:DD}
    \end{subfigure}
    \caption{Illustration of the $\beta$-involutes of a unit circle for different values of $\beta$ for the three values of $d=1$, $d=2$ and $d=3$. The $\beta$-involutes are plotted for $t \in \left[0,\frac{\pi}{2}\right]$.}
    \label{fig:betainvol}
\end{figure*}

\begin{example}
    Fig.\,\ref{fig:betainvol} depicts the $\beta$-involutes computed for a circle for the constant values of $d = 1$, $d = 2$ and $ d = 3$. 
    Additionally, it demonstrates that the tangent to the circle intersects the $\beta$-involutes under equal angles.
\end{example}

\subsection{T-surfaces}\label{sec:Tsurface}

According to \cite{izmestiev2023isometric} a T-surface can be defined as follows:
\begin{definition}
     A T-surface is defined as a parametrization  
     \begin{equation*}
         \mathbf{x}: [0,a]\times [0,b] \subset \mathbb{R}^2 \longrightarrow \mathbb{R}^3, \quad\quad (s,t) \longmapsto \mathbf{x}(s,t),\quad\quad a, b \in \mathbb{R}\setminus\{0\},
     \end{equation*}
     with two $1$-parameter family of planar coordinate curves where each of the carrier planes of the coordinate curves of one family is orthogonal to all the carrier planes of the other family. Furthermore, the aforementioned parametrization forms a conjugate system; i.e.\ $\det(\frac{\partial\mathbf{x}}{\partial t},\frac{\partial\mathbf{x}}{\partial s},\frac{\partial^2\mathbf{x}}{\partial s \partial t}) = 0$. 
\end{definition}

The orthogonality of the carrier planes imply that at least one family of these planes has to be consisted of parallel planes (see \cite{izmestiev2023isometric}). In this article, the $s$-coordinate curves (for fixed $t$) are referred to as \emph{profile curves}, denoted by $\mathbf{p}_t(s)$, whereas their $t$-coordinate counterparts are termed \emph{trajectory curves}, represented by $\mathbf{t}_s(t)$. Notably, the trajectory curves are situated on parallel carrier planes. The carrier planes corresponding to the trajectory curves are symbolized by $\tau(s)$, while those associated with the profile curves are designated by $\delta(t)$. Fig.\,\ref{fig:TrajProPlanes}-(a) illustrates a smooth T-surface along with its coordinate curves and carrier planes. 

In \cite{izmestiev2023isometric}, T-surfaces are \emph{locally} classified in three major sub-classes, namely, \emph{general case}, \emph{axial surface} and \emph{translational surface}\footnote{Graf and Sauer also acknowledge these classes in their works \cite{graf,sauer1970differenzengeometrie}. A similar classification system is adopted by the authors in \cite{SNRT21}.}. The method of generating T-surfaces outlined in \cite{izmestiev2023isometric} aligns well with the initialization of the system of PDEs underpinning the theory of smooth parametrization. However, for the objectives of this article, we adopt a parametrization grounded in kinematic descriptions.

We will prove in the forthcoming Theorem \ref{thm:T-invo} that a T-surface of the \emph{general case type} can be conceptualized as a collection of $\beta$-involutes, which, depending on the parameter $d$, are situated at varying heights on a planar curve. Thus, Theorem \ref{theorem:parametrization:beta:involute} provides the essential ingredients required for the parametrization of a general type T-surface. In \ref{appendix:subclasses} it is shown how other classes of the T-surfaces, with the exception of \emph{translational surfaces}, are obtained as ``special" sub-classes of this general case. Therefore T-surfaces of the translational type are not the subject of study in this article (see also Section \ref{sec:conclusion}). Consequently, our attention will be exclusively on the \emph{general case} going forward. For brevity, we will refer to this simply as the T-surface in the remainder of the paper. 

\begin{figure}[t!]
\begin{center}
    \begin{subfigure}[b]{.495\columnwidth}
        \centering
        \begin{overpic}[width=\textwidth]{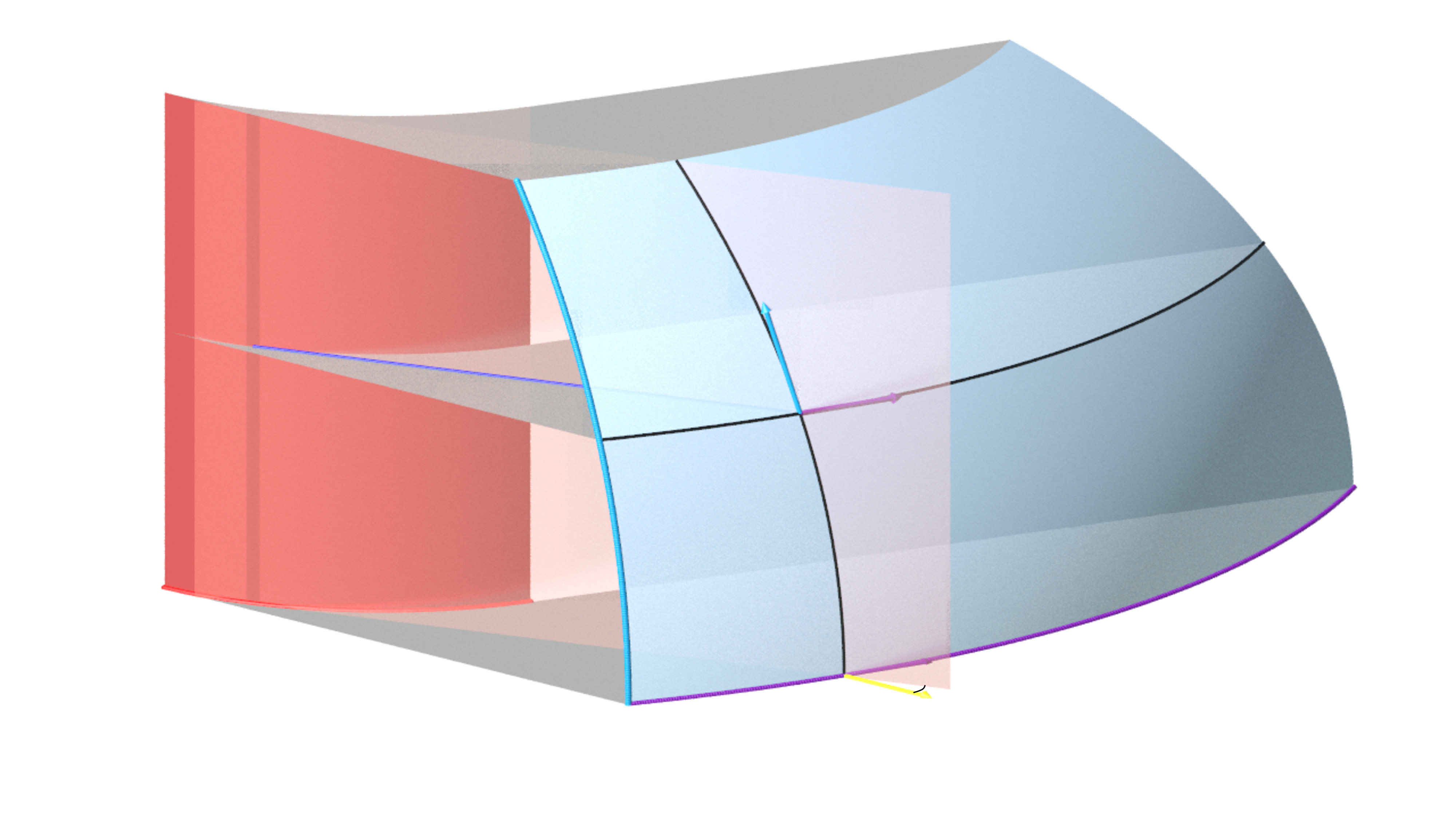}
            \put(40,55){$\mathbf{x}(s,0) = \mathbf{p}_0(s)$}
            \put(43,20){\color{black}\line(0,1){34}}
            \put(60,40){\color{black}\line(1,0){30}}
            \put(70,8){$\mathbf{x}(0,t) = \mathbf{t}_0(t)$}
            \put(91,39){$\delta(t)$}
            \put(33,11){$\tau(0)$}
            \put(33,29){$\tau(s)$}
            \put(20,20){$\Delta$}
            \put(33,45.5){$\tau(a)$}
            \put(41,4){$\mathbf{x}(0,0)$}
            \put(22,4){$\mathbf{d}(t)$}
            \put(25,14){\color{black}\line(0,-1){6}}
            \put(55,33){$\mathbf{x}_s$}
            \put(62,25){$\mathbf{x}_t$}
        \end{overpic}
        \caption{T-surface}
    \end{subfigure}
    \begin{subfigure}[b]{.495\columnwidth}
        \centering
        \begin{overpic}[width=\textwidth]{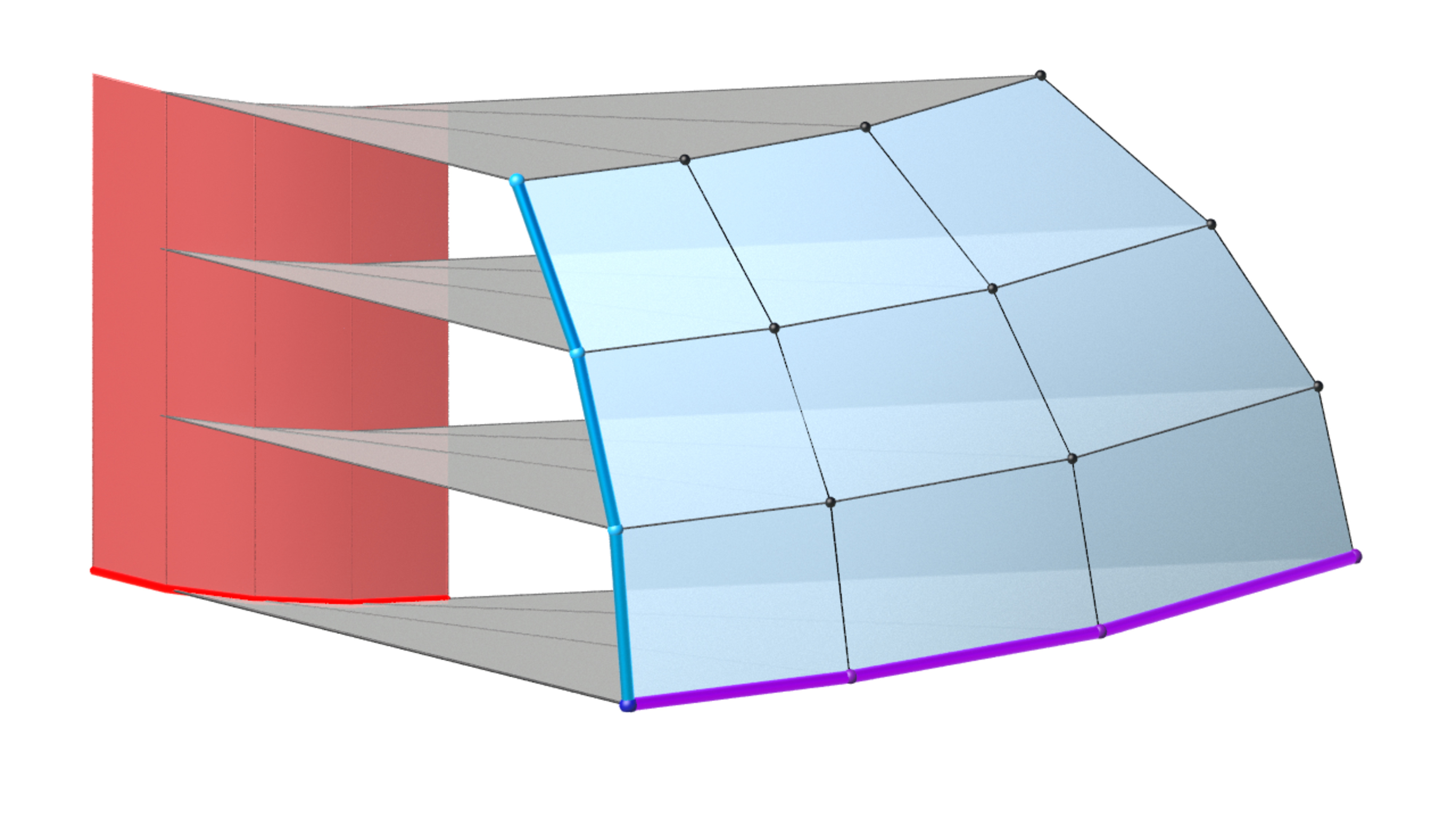}
            \put(70,8){$\mathbf{x}_{0,j} = \mathbf{t}_{0}(j)$}
            \put(43,25){$\mathbf{x}_{i,0} = \mathbf{p}_{0}(i)$}
            \put(67,39){$\mathbf{x}_{i,j}$}
            \put(7,42){${\delta}_{0}$}
            \put(13,42){${\delta}_{1}$}
            \put(19,42){${\delta}_{j}$}
            \put(24,42){${\delta}_{j+1}$}
            \put(33,12){${\tau}_{0}$}
            \put(33,23.5){${\tau}_{1}$}
            \put(33,35.5){${\tau}_{i}$}
            \put(33,46.5){${\tau}_{i+1}$}
            \put(41,4){$\mathbf{x}_{0,0}$}
            \put(15,20){$\Delta$}
        \end{overpic}
        \caption{T-hedron}
    \end{subfigure}
\end{center}
\caption{Illustration of a T-surface (a) and a T-hedron (b) their coordinate curves/polylines and their corresponding carrier planes.} 
\label{fig:TrajProPlanes}
\end{figure}
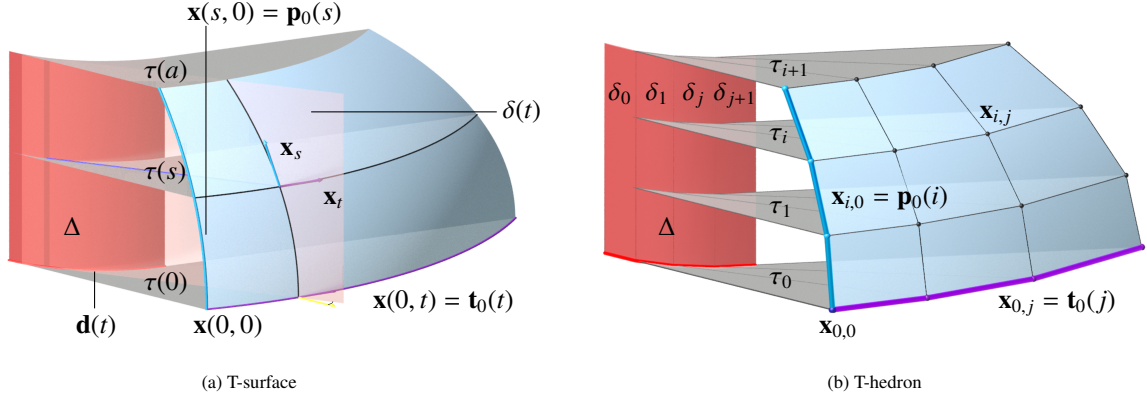
\begin{theorem}\label{thm:T-invo}
Up to a rigid motion every T-surface is uniquely determined by a planar profile curve 
\begin{equation*}
    \mathbf{p} := \mathbf{p}_0 :[0,a]\subset\mathbb{R}\rightarrow\delta(0)\subset\mathbb{R}^3,\quad\quad\quad\quad \mathbf{p}(s) = (p_x(s),0,p_z(s))^T,
\end{equation*}
living in $xz$-plane, a planar directrix curve 
\begin{equation*}
    \mathbf{d}:[0,b]\subset\mathbb{R}\rightarrow \tau(0)\subset\mathbb{R}^3,\quad\quad\quad\quad \mathbf{d}(t) = (d_x(t),d_y(t),0)^T,
\end{equation*}
living in $xy$-plane, with the curvature $\kappa$ and the initialization $\mathbf{d}(0) = (0,0,0)^T$, $\dot{\mathbf{d}}(0)$ in the direction of $x$-axis and a smooth function $\beta : [0,b]\subset\mathbb{R}\rightarrow (-\frac{\pi}{2},\frac{\pi}{2})$ as the angle that the tangent to the directrix makes with the normal to the trajectory curves. Under consideration of $\xi\,(t)$ from
Eq.\ (\ref{eq:beta:involute})
a T-surface supports the following parametrization:
\begin{equation}\label{eq:Tsurface:general:case}
    \mathbf{x}\left(s,t\right) =
    \left(\begin{array}{>{\displaystyle}c}
    d_x(t) - \,\xi(t)\,\left(\int_{\,0}^{\,t}\frac{\left|\,\dot{\mathbf{d}}(w)\,\right|}{\xi(w)}\,\mathrm{d}w + p_x(s)\right)\,\frac{\dot{d}_x(t)}{\left|\,\dot{\mathbf{d}}(t)\,\right|}\\[0.3cm]
    d_y(t) - \,\xi(t)\,\left(\int_{\,0}^{\,t}\frac{\left|\,\dot{\mathbf{d}}(w)\,\right|}{\xi(w)}\,\mathrm{d}w + p_x(s)\right)\,\frac{\dot{d}_y(t)}{\left|\,\dot{\mathbf{d}}(t)\,\right|}\\[0.3cm]
    p_z(s)
    \end{array}
    \right).
\end{equation}
\label{theorem:parametrization:type:III}
\end{theorem}

\begin{proof}
    Without loss of generality, choose the \emph{global coordinate system} in $\mathbb{R}^3$ in such a way that the plane $\tau(0)$ is the $xy$-plane and $\delta(0)$ is the $xz$-plane. Now, the point $\mathbf{x}(0,0)$ lies on the $x$-axis. Considering these simplifying assumptions, we have the trajectory plane $\tau(s_0) \supset \mathbf{x}(s_0,J)$ parallel to the $xy$-plane. This implies that 
    \begin{equation}\label{eq:x:general:form}
        \mathbf{x}(s,t) = \left(\begin{array}{c}
          \mathbf{c}(s,t)\\
           z(s)  
        \end{array}\right),
    \end{equation}
    for smooth maps $\mathbf{c}: I\times J \rightarrow \mathbb{R}^2$ and $z: I \rightarrow \mathbb{R}$. On the other hand, given that for all $t\in J$, $\delta(t)$ ought to be orthogonal to the $xy$-plane, the profile planes are characterized by one of three local configurations: they are either \emph{parallel to each other}, \emph{intersect along an axis perpendicular to the $xy$-plane}, or \emph{envelope a cylindrical surface}. Here, we restrict our attention to the last case\footnote{The profile planes being \emph{parallel to each other} is not an object of study here. Furthermore, the profile planes intersecting along an axis can be shown to be a special case of them enveloping a cylindrical surface (cf.\ \ref{appendix:subclasses}).} where the profile planes envelope a cylindrical surface $\Delta$ (see Fig.\,\ref{fig:TrajProPlanes}-(a)). Now, the trajectory planes slice the T-surface and $\Delta$.
    {
    Project the aforementioned slices along with the profile planes $\delta(t)$ onto the $xy$-plane. Denote the projection of the slices by $\mathbf{d}: J \rightarrow \mathbb{R}^2$ and the projection of the profile planes by $l$. Having in mind that Eq.\ \ref{eq:x:general:form} implies that $\mathbf{c}(s,t)$ are the projection of the trajectory curves, we obtain the following correspondence between $\mathbf{c}(s,t)$ and $\mathbf{d}(t)$ due to $l$ being tangent to $\mathbf{d}$ and intersecting $\mathbf{c}(s,t)$:
    \begin{equation*}
        \mathbf{c}(s,t) = \mathbf{d}(t) - \lambda(s,t)\,\frac{\mathrm{d}}{\mathrm{d} t}{\mathbf{d}}(t),
    \end{equation*}
    for some smooth map $\lambda: I\times J \subset \mathbb{R}^2 \rightarrow \mathbb{R}$. Our assertion is that the set of curves defined by $\{\mathbf{c}(s_0,t)\}_{s_0\in I}$ represents the $\beta$-involutes of the curve $\mathbf{d}(t)$. To substantiate this claim, it suffices to demonstrate that for any two points $s_1$ and $s_2$ in the interval $I$, the derivative $\frac{\partial \mathbf{c}}{\partial t} (s_1,t)$ is parallel to $\frac{\partial \mathbf{c}}{\partial t} (s_2,t)$. In other words, we need to show that $\frac{\partial^2 \mathbf{c}}{\partial s\partial t}$ is parallel to $\frac{\partial \mathbf{c}}{\partial t}$ for all pairs of admissible $(s,t)$. 
    Now, since $\mathbf{x}(s,t)$ is a conjugate parameterization we have 
    \begin{equation}\label{eq:conjugate}
        \frac{\partial^2 \mathbf{x}}{\partial s\,\partial t} = \alpha\,\frac{\partial \mathbf{x}}{\partial s} + \beta\,\frac{\partial \mathbf{x}}{\partial t},
    \end{equation} 
    for some smooth functions $\alpha, \beta : I\times J \rightarrow \mathbb{R}$. Taking respective derivatives of Eq.\ \ref{eq:x:general:form} and substituting it back into Eq.~(\ref{eq:conjugate}) reveals that $\alpha = 0$ and consequently\footnote{Theoretically, the function $\beta(s,t)$ may become zero over the domain $I \times J$. However, this scenario leads to $\frac{\partial^2 \mathbf{x}}{\partial s \, \partial t} = 0$. Solving this condition implies that $\mathbf{x}$ represents the parametrization of the previously excluded case (translation surface) with parallel profile planes.} $\frac{\partial^2\mathbf{x}}{\partial s\partial t} = \beta\,\frac{\partial\mathbf{x}}{\partial t}$ meaning that $\frac{\partial^2\mathbf{x}}{\partial s\partial t} \parallel \frac{\partial\mathbf{x}}{\partial t}$ for all admissible $(s,t)$. Writing $\frac{\partial^2\mathbf{x}}{\partial s\partial t}$ and $\frac{\partial\mathbf{x}}{\partial t}$ according to Eq.\ (\ref{eq:x:general:form}) gives
    \begin{equation*}
        \frac{\partial\mathbf{x}}{\partial t} = \left(\begin{array}{c}
           \frac{\partial\mathbf{c}}{\partial t}\\
           0  
        \end{array}\right),\quad\quad\quad\quad
        \frac{\partial^2\mathbf{x}}{\partial s\,\partial t} = \left(\begin{array}{c}
           \frac{\partial^2\mathbf{c}}{\partial s\partial t}\\
           0  
        \end{array}\right).
    \end{equation*}
    Therefore, $\frac{\partial^2\mathbf{x}}{\partial s\partial t} \parallel \frac{\partial\mathbf{x}}{\partial t}$ implies  $\frac{\partial^2 \mathbf{c}}{\partial s\partial t} \parallel \frac{\partial \mathbf{c}}{\partial t}$.
    %
    }
    Therefore, the trajectories of the T-surface are the $\beta$-involutes of the horizontal slices of $\Delta$. Consequently, we can think of a T-surface as a collection of $\beta$-involutes. The only thing that remains is to identify the constant $d$ that appears in each $\beta$-involute parametrization (see Eq.\ (\ref{eq:ell})). To achieve this, by applying Theorem,\ref{theorem:parametrization:beta:involute}, we obtain the following relation:
    \begin{equation}\label{eq:lambda:arc:length}
        \lambda(s,t) =  \xi(t)\,\left(\int_{\,0}^{\,t}\frac{1}{\xi(w)}\,\mathrm{d}w + d(s)\right),
    \end{equation}
    Now, with the initialization $\mathbf{d}(0) = (0,0,0)^T$, $\dot{\mathbf{d}}(0) = (-1,0,0)^T$ and $\mathbf{x}(s,0) = (p_x(s),0,p_z(s))^T$, we get
    \begin{equation*}
        \left(\begin{array}{c}
         p_x(s)    \\
         0     
        \end{array}\right) = \mathbf{c}(s,0) = \left(\begin{array}{c}
         -\lambda(s,0)    \\
         0     
        \end{array}\right)\quad\quad \implies\quad\quad \lambda(s,0) = -p_x(s).
    \end{equation*}
    Combining the above result with Eq.\ (\ref{eq:lambda:arc:length}) gives $d(s) = p_x(s)$.  Finally, with the illumination of the previous explanation implies that any T-surface parametrization is a collection of $\beta$-involutes living on different elevations $z = p_z(s)$ with the constants being $d = p_x(s)$. The aforementioned parametrization with $\mathbf{d}(t)$ as its directrix curve would be Eq.\ (\ref{eq:Tsurface:general:case}).
\end{proof}

\subsection{T-hedra}\label{sec:thedra}
According to \cite{izmestiev2023isometric}, the T-hedra can be defined quite similarly to their smooth counterparts as follows:
\begin{definition}\label{def:Thedron}
    A T-hedron is defined as a planar quad parametrization with the combinatorics of a grid of the form 
    \begin{equation*}
        \mathbf{x}: \bar{I}\times \bar{J} \subset \mathbb{Z}^2 \rightarrow \mathbb{R}^3,\quad\quad \bar{I} := \{0,1,\ldots,m\}\quad\text{and}\quad\bar{J} := \{0,1,\ldots,n\},\quad \mathbf{x}(i,j) =: \mathbf{x}_{i,j},
    \end{equation*}
    with two $1$-parameter family of planar coordinate polylines where each of the carrier planes of the coordinate polylines of one family is orthogonal to all the carrier planes of the other family.
\end{definition}

Once again the orthogonality of the carrier planes results in one of the families being consisted of parallel planes (see \cite{izmestiev2023isometric} for details), which implies that the faces of a T-hedron are \emph{trapezoids} (see Fig.\,\ref{fig:TrajProPlanes}-(b)). Note that the conjugate net property results from the planarity condition of the quads (cf.\ Peterson \cite{peterson}).

Analogously to the smooth scenario, through dividing the quad surface upon necessity, one can categorize these surfaces to three major sub-classes, namely, \emph{general case}, \emph{axial surface} and \emph{translational surface}. 

In this article, our goal is to assign a T-hedron to a given point cloud. We choose the \emph{general} type of a T-hedron as the other cases (with exception of the translational type) can be thought of as special sub-classes of such a T-hedron, making our choice suitable for applications. Therefore, from now on, by a T-hedron we are referring to a T-hedron of the \emph{general} type. 

The approach employed in the proof of Theorem \ref{theorem:parametrization:type:III} can be similarly applied to a T-hedron: the profile planes envelope a \emph{discrete cylindrical surface} $\Delta$ whose generators are obtained by the intersection of consecutive profile planes. Identically, every horizontal slice gives us a planar curve of $\Delta$ and a trajectory of the T-hedron. Since $\Delta$ is a cylindrical surface all such slices can be identified with a \emph{discrete directrix curve} $\mathbf{d}(j)$ living in $\tau_0$. The trajectories and the directrix curve resemble the same properties as the $\beta$-involutes and $\beta$-evolutes as in the smooth scenario. However, in the discrete scenario, to construct the T-hedron it is easier to resort to a set of numbers mimicking scalings of discrete stretch-rotations $\sigma_j$ instead of using discrete $\beta$-angles. {Using this approach every T-hedron can be generated in the following way according to \cite{SNRT21}:}




\begin{theorem}\label{theorem:T:hedron:III}
    Up to a rigid motion, every T-hedron is uniquely determined by two planar polylines, namely the profile curve 
    \begin{equation*}
        \mathbf{p} := \mathbf{p}_0: \bar{I} \subset \mathbb{Z}\longrightarrow {\delta}_0 \subset \mathbb{R}^3,\quad\quad\quad\quad \mathbf{p}_i = (p_{i,x},0,p_{i,z})^T,
    \end{equation*}
    and a directrix curve
    \begin{equation*}
        \mathbf{d} : \bar{J} \subset \mathbb{Z} \longrightarrow {\tau}_0 \subset \mathbb{R}^3,\quad\quad\mathbf{d}_0 = (-1,0,0)^T,\quad\quad\mathbf{d}_1 =(0,0,0)^T,\quad\quad\mathbf{d}_j =(d_{j,x},d_{j,y},0)^T\quad\text{for}\quad j\geq 2,
    \end{equation*}
    and a sequence of real numbers $\{\eta_{i}\}_{i \in I} \in \mathbb{R}\setminus\{0\}$, with $\eta_0 = 1$. 
    Iteratively, a T-hedron is built by a sequence of the following stretch-rotations $\sigma_j$:
    \begin{equation}\label{eq:scale:rotation}
    \sigma_j: \mathbf{x}_{i,j}\mapsto \mathbf{x}_{i,j+1} \quad\text{with} \quad
        \left(\begin{array}{c}\mathbf{x}_{i,j+1}\\ 1 \end{array}\right) = \mathbf{T}_j.\mathbf{S}_j.\mathbf{R}_j.\mathbf{T}^{-1}_j.\left(\begin{array}{c}\mathbf{x}_{i,j}\\ 1 \end{array}\right),\quad\quad \mathbf{x}_{i,0} = \mathbf{p}_i = (p_{x,i},0,p_{z,i})^T,
    \end{equation}
    where
    \begin{equation}\label{eq:T:R:S}
        \mathbf{T}_j = \left(\begin{array}{cccc}
        1  & 0 & 0 & d^x_j\\
        0  & 1 & 0 & d^y_j\\
        0  & 0 & 1 & 0\\
        0  & 0 & 0 & 1
        \end{array}\right)\quad\quad\quad
        \mathbf{S}_j = \left(\begin{array}{cccc}
        \left|\eta_{j}\right|  & 0                     & 0 & 0\\
        0                      & \left|\eta_{j}\right| & 0 & 0\\
        0                      & 0                     & 1 & 0\\
        0                      & 0                     & 0 & 1
        \end{array}\right)\quad\quad\quad
        \mathbf{R}_j = \left(\begin{array}{cccc}
        \cos{(\theta_{j})}  & -\sin{(\theta_{j})} & 0 & 0\\
        \sin{(\theta_{j})}  &  \cos{(\theta_{j})} & 0 & 0\\
        0                   & 0                   & 1 & 0\\
        0                   & 0                   & 0 & 1
        \end{array}\right),
    \end{equation}
    \begin{equation*}
    l_j := \left|\,\mathbf{d}_{j+1}-\mathbf{d}_{j}\,\right|,\quad\quad 
    \mathbf{e}_j:= \frac{\mathbf{d}_{j+1}-\mathbf{d}_j}{l_j},\quad\quad
    \textnormal{ and } \quad\quad
    \theta_j := \left\{ \begin{array}{ll}
    \sphericalangle (\mathbf{e}_{j-1},\mathbf{e}_j), & \textnormal{if } \, \eta_j>0 \\
    \sphericalangle (\mathbf{e}_{j-1},-\mathbf{e}_j),& \textnormal{if } \, \eta_j<0 
    \end{array} \right.. \nonumber
    \end{equation*}
\end{theorem}

\section{Determination of the initial guess}\label{sec:initial_guess}

Constructing a unique T-hedron necessitates the creation of a \emph{trajectory curve}, a \emph{profile curve}, and a set of real numbers to act as \emph{scaling factors} (cf.\ Theorem \ref{theorem:T:hedron:III}). But before all one must identify the axis which is pinpointing the normal of the family of parallel trajectory planes $P$.
Consequently, the subsequent subsections aim to address each of these components individually. The procedure is outlined as follows:

\begin{enumerate}
\item Determination of the axis direction.
\item Slicing the point cloud using a pair of parallel planes orthogonal to the axis direction yield two trajectory curves.
\item Leveraging the continuous theory of T-surfaces, we pinpoint the profile planes by seeking directions that intersect two trajectories at equal angles (the $\beta$ angles).
\item With the profile planes identified, their envelope is computed to derive the discrete $\beta$-evolute.
\item Transitioning back to the discrete framework, the $\beta$-evolute and trajectories enable the calculation of the scaling factors.
\item The point cloud is sliced using the profile planes to establish the profile curves.
\item Employing the T-hedron parameterization from Theorem \ref{theorem:T:hedron:III}, the T-hedron is reconstructed iteratively.
\end{enumerate}

While the above steps provide a simplified overview, it's essential to note that, due to the theoretical considerations highlighted in \ref{sec:singularities}, the procedure is highly sensitive to noise. Put simply, even minor inconsistencies in the point cloud can impede several of the steps outlined. As such, the upcoming subsections will delve deeper into the practical execution of these steps, highlighting instances where certain phases necessitate embedded optimization strategies.

To ensure readers gain a comprehensive understanding of the methodology, we illustrate the algorithm's operation through a detailed example. This demonstration begins in Example \ref{example:axis} and extends across subsequent subsections, providing a step-by-step elucidation of the process.


\subsection{Computing the ruling direction}\label{subsec:direction}

We recap in the next section the results of \cite{boris} needed to understand our procedure for computing the axis direction $\mathbf{a}$, which is parallel to the rulings $\mathbf{g}(t)$ of the directing surface $\Delta$.

\subsubsection{Linear complexes of line-elements}
It is well known \cite{PW} that the Pl\"ucker coordinates of a line $\sf l$ are given by $(\mathbf{l},\mathbf{\overline{l}})\mathbb{R}$ where $\mathbf{l}\neq \mathbf{o}$ is a vector parallel to $\sf l$ and $\mathbf{\overline{l}}:=\mathbf{x}\times \mathbf{l}$. 
According to \cite[Def.\ 1]{boris} they 
can be extended to a line-element $({\sf l,X})$ with  ${\sf X}\in{\sf l}$ by the triple $(\mathbf{l},\mathbf{\overline{l}},\lambda)\mathbb{R}$ where
 $\lambda:=\langle\mathbf{x}, \mathbf{l}\rangle$.

The set of line-elements fulfilling a line equation
\begin{equation}
    \langle\mathbf{a}, \mathbf{\overline{l}}\rangle + 
    \langle\mathbf{\overline{a}}, \mathbf{l}\rangle + \alpha\lambda = 0
\end{equation}
is called a {\it linear complex of line-elements} with 
coordinates $(\mathbf{a},\mathbf{\overline{a}},\alpha)\mathbb{R}$. Moreover, the following theorem holds:

\begin{theorem}\cite[Thm.\ 1]{boris}
    The path-normal-elements under uniform equiform motions equals a linear complex of line-elements. 
\end{theorem}

Note that a line-element $({\sf l,X})$ with  $\langle\mathbf{l}, \mathbf{v}\rangle=0$ is a path-normal-element, where $\mathbf{v}$ denotes the velocity vector of the point $\sf X$ under an uniform equiform motion.

\subsubsection{Planar equiform kinematics}

According to the given kinematic generation of T-surfaces in Section \ref{subsec:profile_affine}
the complete geometric information beside the shape of the profile $\mathbf{p}$ can be extracted from the normal projection  in direction $\mathbf{a}$ of the cylinder rulings $\mathbf{g}(t)$, which is indicated by a prime.  

Then $\Delta^{\prime}$ equals its directrix $\mathbf{d}$ and $\delta(t)^{\prime}$ is a line tangent to $\mathbf{d}$ in the point $\mathbf{g}(t)^{\prime}$. At each time instant $t$ the line $\delta(t)^{\prime}$ performs an instantaneous planar equiform motion, which equals a stretch-rotation with center point $\mathbf{g}(t)^{\prime}$. The corresponding uniform motion is a uniform planar spiral motion and the logarithmic spiral is the corresponding invariant curve. For a review on planar equiform kinematics we refer to \cite[pages 458-480]{bottema} and the references given therein. Moreover, the following theorem holds:

\begin{theorem}\cite[Thm.\ 3]{boris}
    The line-elements of a linear complex of line-elements, which are contained in a plane, are the path-normal-elements of an uniform planar equiform motion. 
\end{theorem}

\subsubsection{Algorithm}

Let us assume that ${\sf X}_i\in \mathcal{X}$ and ${\sf n}_i$ is the estimated surface normal \cite{mitra} in this point for $i\in \left\{1,\ldots, q \right\}$ with $q:=\#\mathcal{X}$. 
Then $({\sf n}_i,{\sf X}_i)$ denotes the corresponding normal-element and its  orthogonal projection $({\sf n}_i^{\prime},{\sf X}_i^{\prime})$ onto the base plane yields a path-normal-element  of an uniform planar equiform motion. 

We determine the projection  direction $\mathbf{a}=(a_1,a_2,a_3)^T$ by exploiting this property as follows: Given is the position vector $\mathbf{x}_i$ of the point ${\sf X}_i$ and the unit-vector $\mathbf{n}_i$ parallel to the surface-normal ${\sf n}_i$. Then we can compute the projection ${\sf X}_i^{\prime}$ onto a plane  orthogonal to $\mathbf{a}$ containing the origin by:
\begin{equation}
    \mathbf{x}_i^{\prime}=\mathbf{x}_i-
    \tfrac{ \langle\mathbf{a}, \mathbf{x}_i\rangle}
    { \langle\mathbf{a}, \mathbf{a}\rangle} \mathbf{a}.
\end{equation}
In the same way we get:
\begin{equation}
    \mathbf{n}_i^{\prime}=\mathbf{n}_i-
    \tfrac{ \langle\mathbf{a}, \mathbf{n}_i\rangle}
    { \langle\mathbf{a}, \mathbf{a}\rangle} \mathbf{a}.
\end{equation}
Then we can complete the computation of the Pl\"ucker coordinates $(\mathbf{n}_i^{\prime},\mathbf{\overline{n}}_i^{\prime},\nu_i)\mathbb{R}$ of the line-element $({\sf n}_i^{\prime},{\sf X}_i^{\prime})$ by
$\mathbf{\overline{n}}_i^{\prime}= \mathbf{x}_i^{\prime} \times \mathbf{n}_i^{\prime} $ and $\nu_i=\langle \mathbf{x}_i^{\prime}, \mathbf{n}_i^{\prime}\rangle$.

Now let us consider three further surface-normal elements $({\sf n}_{i,j},{\sf X}_{i,j})$ for $j=1,2,3$ with the property that ${\sf X}_{i,j}\in\mathcal{X}$ is the $j$th closest point to ${\sf X}_i$ with respect to the Euclidean distance. Similar to the above procedure we can compute $(\mathbf{n}_{i,j}^{\prime},\mathbf{\overline{n}}^{\prime}_{i,j},\nu_{i,j})\mathbb{R}$. 
Now we can compute the linear complex of line-elements $(\mathbf{a},\mathbf{\overline{a}}_i,\alpha_i)\mathbb{R}$, which fits best the four line-elements $({\sf n}_i^{\prime},{\sf X}_i^{\prime})$ and $({\sf n}_{i,j}^{\prime},{\sf X}_{i,j}^{\prime})$ for $j=1,2,3$. 
From the three equations 
\begin{equation}
    \langle\mathbf{a}, \mathbf{\overline{n}}_{i,j}^{\prime}\rangle + 
    \langle\mathbf{\overline{a}}_i, \mathbf{n}_{i,j}^{\prime}\rangle + \alpha_i\nu_{i,j} = 0
\end{equation}
one can compute $\alpha_i$ and two out of three coordinates of $\mathbf{\overline{a}}_i=(\overline{a}_{i,1},\overline{a}_{i,2},\overline{a}_{i,3})$; i.e.\ $\overline{a}_{i,u}$ and $\overline{a}_{i,v}$ with pairwise distinct $u,v,w\in\left\{1,2,3\right\}$.
Plugging the obtained equations into:
\begin{equation}
    \langle\mathbf{a}, \mathbf{\overline{n}}_i^{\prime}\rangle + 
    \langle\mathbf{\overline{a}}_i, \mathbf{n}_i^{\prime}\rangle + \alpha_i\nu_i = 0
\end{equation}
gives in the numerator a quartic polynomial $Q_i(\mathbf{a})$ in the remaining unknowns\footnote{$Q_i(\mathbf{a})$ is independent of $\overline{a}_{i,w}$.} $a_1,a_2,a_3$ with 4536 terms, which has to be fulfilled. 

Now, we can built the cost function
$f(\mathbf{a}):=\sum_{i=1}^q Q_i(\mathbf{a})^2$, which we want to minimize under the side-condition $h=0$ with $h(\mathbf{a}):=\langle\mathbf{a}, \mathbf{a}\rangle-1$. This can be solved by using the Lagrange approach considering $F(\mathbf{a},a):=f(\mathbf{a})-a\,h(\mathbf{a})$. 
To ensure a global minimum\footnote{If there is a further local minimum $\mathbf{a}^-$ where $f(\mathbf{a}^-)$ is close to  $f(\mathbf{a}^*)$, then one can also run the remaining algorithm with respect to this direction and compare at the end the obtained results. \label{fn:more}} denoted by $\mathbf{a}^*$, the system of four partial derivatives of the Lagrangian $F(\mathbf{a},a)$ is solved by the numeric algebra system {\tt Bertini} \cite{BHSW06,bates2013numerically} as follows:
    With a separation of the variables into two groups $\left\{a_1,a_2,a_3 \right\}$ and $\left\{a \right\}$ the multi-homogneous Bezeout number of $686$ (total number of paths) is obtained. There tracking yields $114$ finite solutions. As the method of homotopy continuation allows the parallelization of computations the solution of this system can be obtained in real-time. 
    Alternatively the system of equations can also be solved by means of Gr\"obner bases (e.g.\ using {\tt Maple}) but it is computationally much more expensive. We only used this method to recheck the solutions obtained by {\tt Bertini} within our test implementation.

\begin{remark}\label{rem:adjust}
{\it Greedy optimization:} The quality of the axis direction can be improved by applying the following iterative method:
    Assume that the global minimum $\mathbf{a}^*$ is obtained by the above method. Then one can rerun the procedure by replacing ${\sf X}_{i,j}$  by the  points ${\sf Y}_{i,j}\in\mathcal{X}$ which are the three closest points to ${\sf X}_i$ with respect to the isotropic metric
   \begin{equation}
      \sqrt{ \|\mathbf{x}_i-\mathbf{y}_{i,j}\|^2-
\langle \mathbf{a}^*, \mathbf{x}_i-\mathbf{y}_{i,j} \rangle^2 }
   \end{equation} 
    where $\mathbf{a}^*$ is the isotropic direction. Note that this equals the Euclidean distance between the orthogonal projections of the involved points along $\mathbf{a}^*$. One can iterate this procedure as long there is no more change in the clustering of points. 
    
    We abstained from this greedy optimization as the final global optimization step (cf.\ Section \ref{sec:global_opt}) includes an adjustment of the direction of the axis with respect to the point cloud.\hfill $\diamond$
\end{remark}

The direction $\mathbf{a}^*$, either obtained my the ordinary optimization or by the greedy one, can then be used to compute for each point 
${\sf X}_{i}$ the associated axis-element $(\mathbf{e}_i,\mathbf{\overline{e}}_i,\epsilon_i)\mathbb{R}$  with $\mathbf{e}_i=\mathbf{a}^*$ of the associated instantaneous equiform motion according to \cite[Eq.\ (19)]{boris}. It can easily be seen that the axis do not depend on the free parameter $\overline{a}_{i,w}$ in contrast to the center point on it; i.e.\ $\overline{a}_{i,w}$ only implies a change of the center along the fixed axis. 
Theoretically, one could use the information of the axis to compute the directrix curve $\mathbf{d}$, but this procedure is very sensitive to noise as the following example shows. 
\begin{figure*}[t!]
	\centering
	\begin{subfigure}[b]{0.24\textwidth} 
        \centering
		\includegraphics[height=25mm, trim={0mm 0mm 0mm 0mm}, clip]{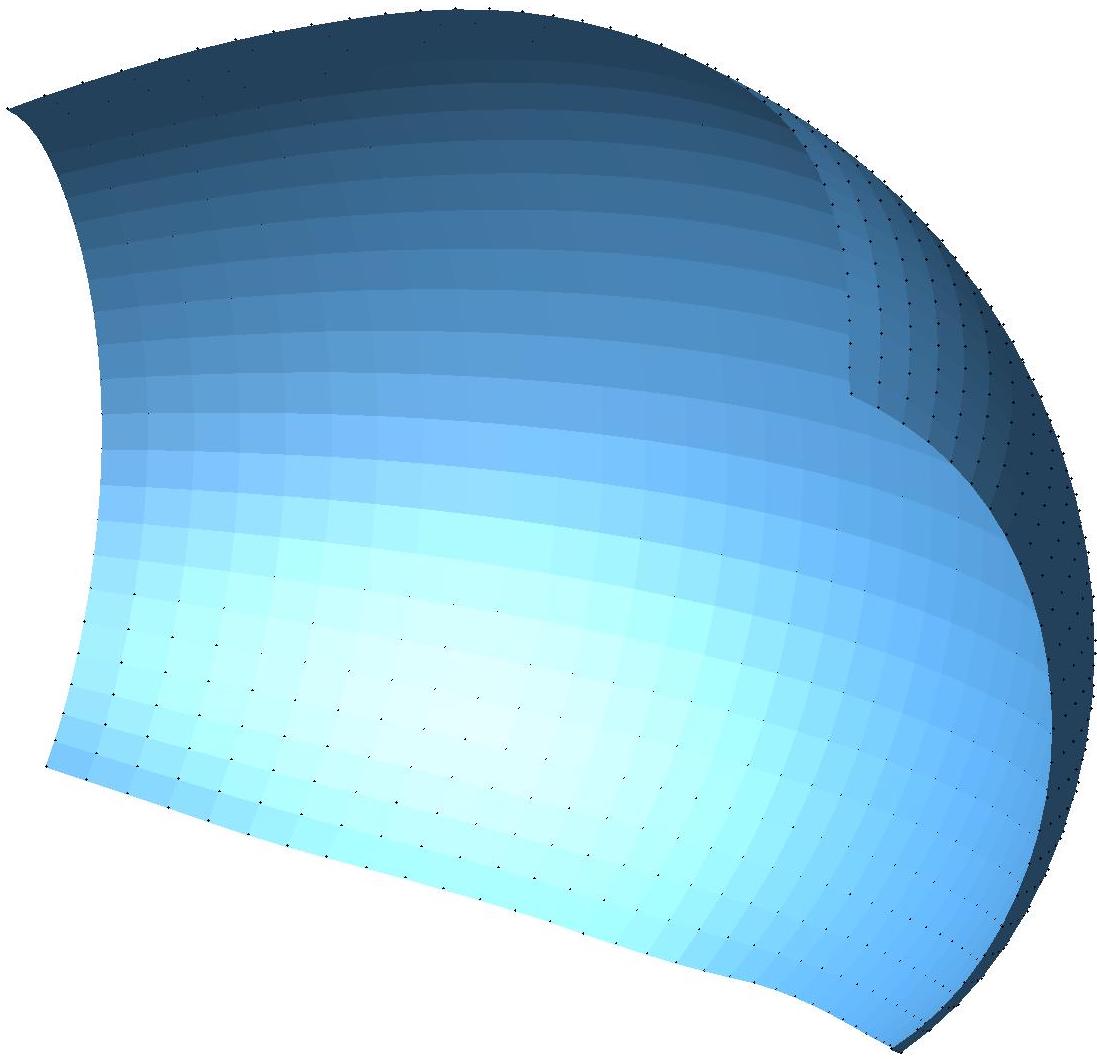}
		\caption{The non-T-hedron mesh.}
	\end{subfigure}
	\hfill
	\begin{subfigure}[b]{0.24\textwidth} 
        \centering
		\includegraphics[height=26mm, trim={0mm 0mm 0mm 0mm}, clip]{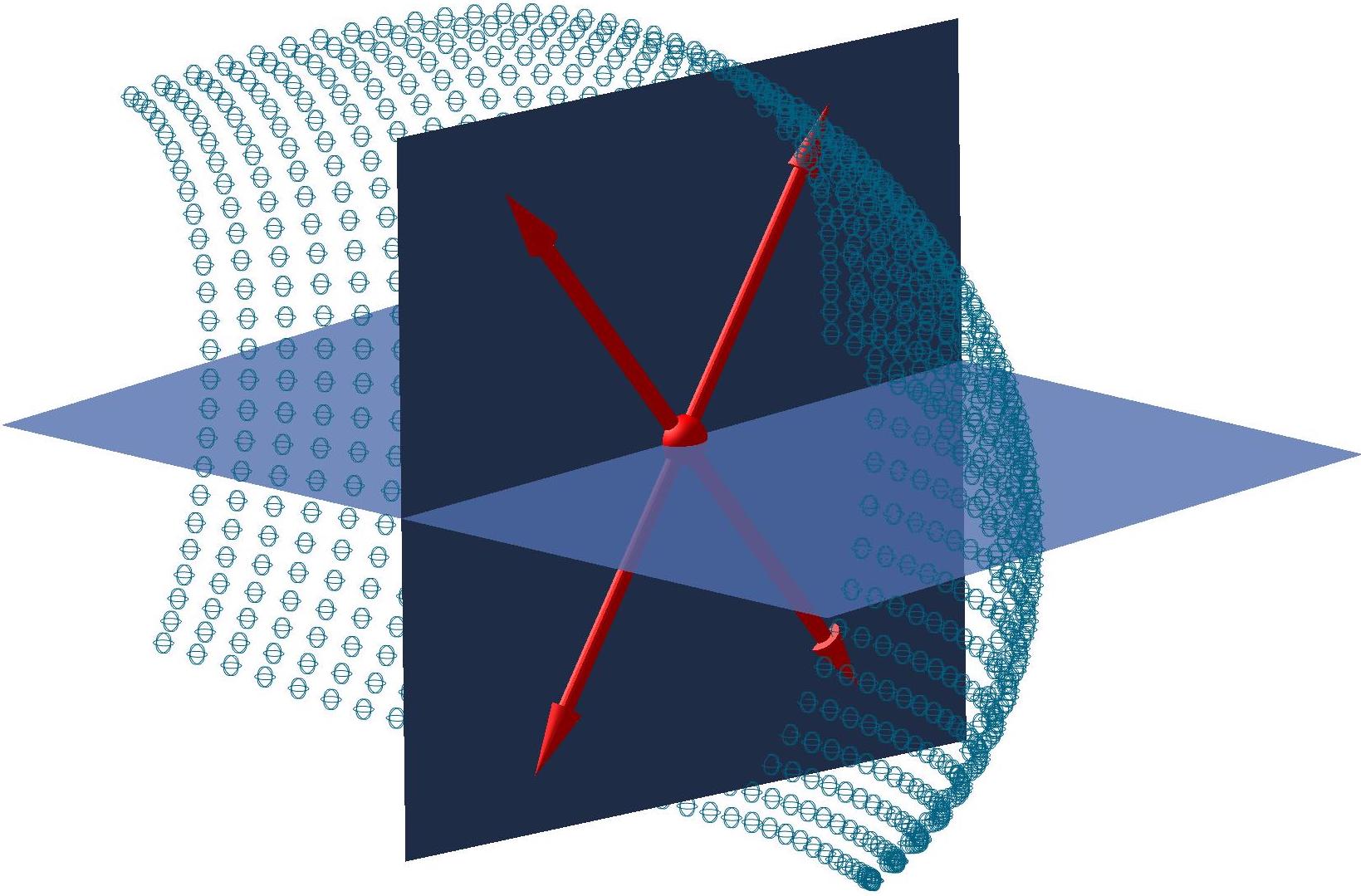}
		\caption{Symmetry planes.}
	\end{subfigure}
    \hfill
    \begin{subfigure}[b]{0.24\textwidth} 
        \centering
		\includegraphics[height=25mm, trim={0mm 0mm 0mm 0mm}, clip]{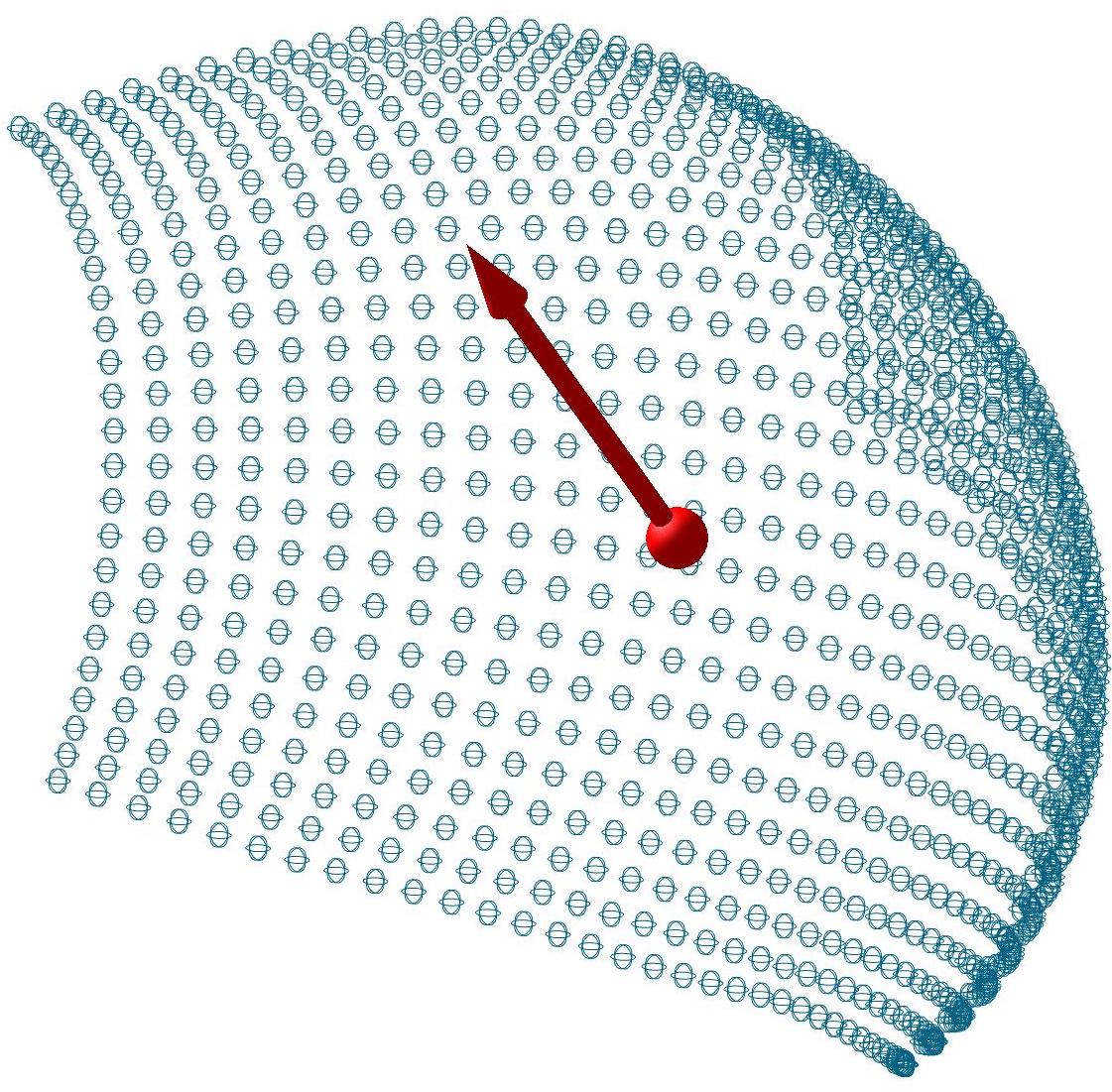}
		\caption{Choosing $\mathbf{a}^\ast$.}
	\end{subfigure}
    \hfill
    \begin{subfigure}[b]{0.24\textwidth} 
        \centering
		\includegraphics[height=26mm, trim={0mm 0mm 0mm 0mm}, clip]{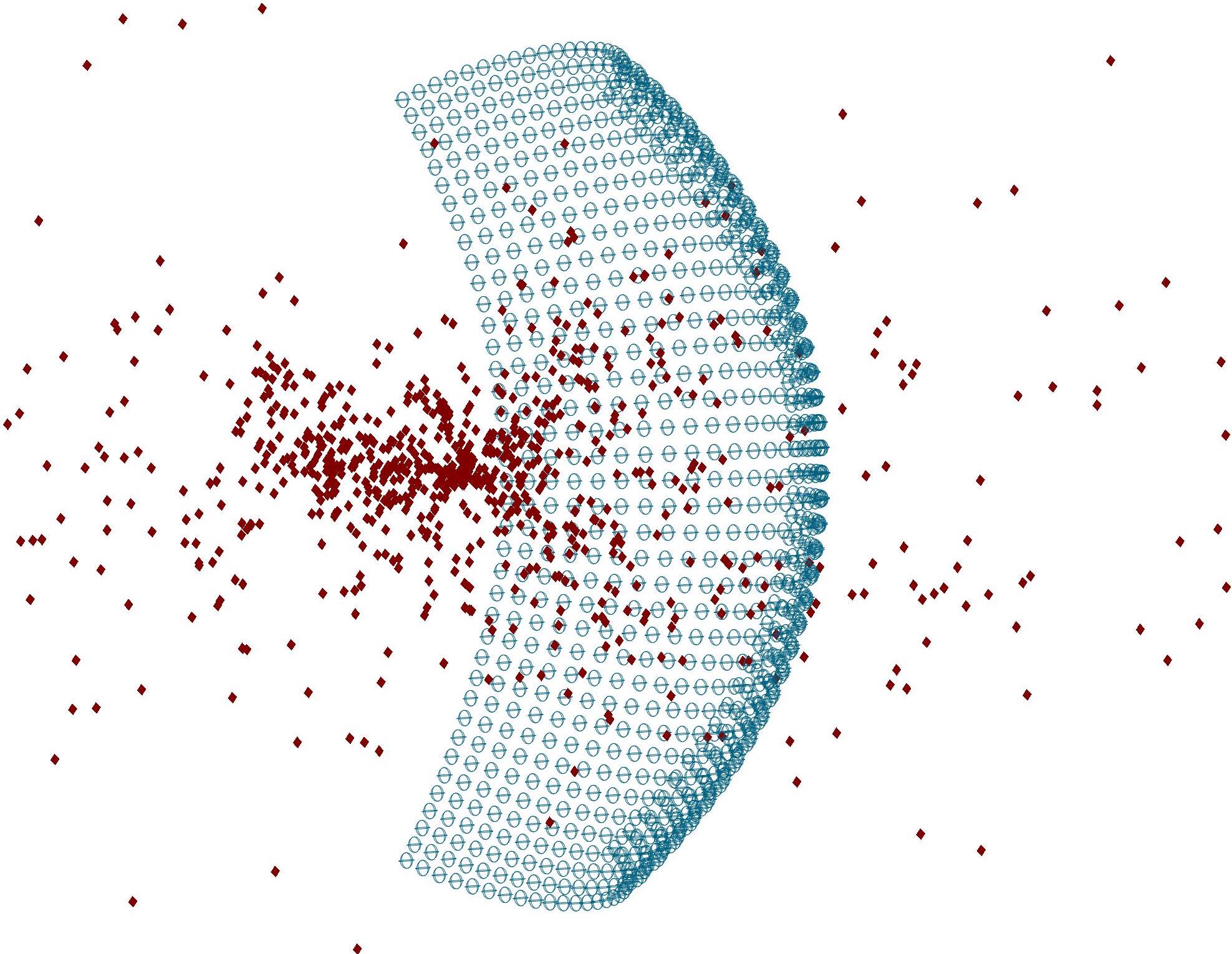}
		\caption{Top view with respect to $\mathbf{a}^\ast$.}
	\end{subfigure}
	\caption{Approximation of the direction of the axis for the point cloud $\mathcal{X}$. (a) Initially given non-T-hedron mesh. (b) The planes of symmetry  of $\mathcal{X}$ along with the four global minima direction candidates which are rendered at the origin. (c) Choosing one of the directions as $\mathbf{a}^*$. (d) The set of axes $(\mathbf{e}_i,\mathbf{\overline{e}}_i,\epsilon_i)\mathbb{R}$  with $\mathbf{e}_i=\mathbf{a}^*$, which should approximate the directrix curve $\mathbf{d}$ in the parallel projection along  $\mathbf{a}^*$.} 
	\label{fig:OBJ:axis}
\end{figure*}
\begin{example}\label{example:axis}
    As input of our example we consider the quad mesh illustrated in Fig.\,\ref{fig:OBJ:axis}-(a), which was computed in \cite[Fig.\ 11]{jiang} by an optimization technique to penalize an isometrically defomed surface with planar quads. 
 Throughout the rest of this section,  we demonstrate each step of the algorithm using this mesh. It is important to note, however, that the mesh is not a T-hedron; rather, we treat it solely as a point cloud (i.e., without considering the combinatorics) and denote it by $\mathcal{X}$.\\
    Now, we apply our algorithm to find the axis direction $\mathbf{a}^*$. As previously explained, we should seek solutions that are the global minima of our cost function $f(\mathbf{a})$. Using {\tt Bertini}, we identify four directions of the axis that, up to $10^{-5}$, serve as global minima of our cost function. Clearly, two of these four candidates are simply the negation of the other two. Hence, we are effectively left with only two vector candidates.\\
    Our point cloud $\mathcal{X}$ possesses two symmetry planes (see Fig.\,\ref{fig:OBJ:axis}-(b)), and the two candidate directions are on the one hand-side parallel to  one of these planes and on the other hand-side symmetric with respect to the other plane. Therefore, without loss of generality, we select one of them (see Fig.\,\ref{fig:OBJ:axis}-(c)).\\
    As mentioned earlier, from a theoretical standpoint, one should be able to reconstruct the directing cylinder $\Delta$ by  $(\mathbf{e}_i,\mathbf{\overline{e}}_i,\epsilon_i)\mathbb{R}$  with $\mathbf{e}_i=\mathbf{a}^*$. However, this process is highly susceptible to noise as illustrated in Fig.\,\ref{fig:OBJ:axis}-(d).
\end{example}

As this procedure for the computation of $\mathbf{d}$ is not suitable for practical application we proceed with the reconstruction of the trajectory polyline $\mathbf{t}$ in the next section. 

\begin{remark}
Following the idea of \cite{ANDREWS} and \cite{lin} one can also think of an alternative normalization condition $H(\mathbf{a})$ to $h(\mathbf{a}):=\langle\mathbf{a}, \mathbf{a}\rangle-1$ with
\begin{equation}
    H(\mathbf{a}):=\frac{1}{q}\sum_{i=1}^q\mathbf{v}(\mathbf{x}_i^{\prime})-1\quad \text{and} \quad \mathbf{v}(\mathbf{x}_i^{\prime})= \mathbf{a}\times \mathbf{x}_i^{\prime} + \alpha_i\mathbf{x}_i^{\prime} + \mathbf{\overline{a}}_i 
\end{equation}
where we select the center point within the base plane, which is obtained by choosing the free parameter $\overline{a}_{i,w}$ such that $\langle \mathbf{a}, \mathbf{\overline{a}}_i \rangle=0$ holds. In this way the velocity vector $\mathbf{v}(\mathbf{x}_i^{\prime})$ remains in the base plane and corresponds to the associated instantaneous planar equiform motion. Each velocity vector is in the numerator a polynomial of degree 8 in the coordinates of $\mathbf{a}$ and the denominator a squared polynomial of degree 3. By making the side-condition $H(\mathbf{a})$ free of denominators one can see that its degree rises to $8+6(q-1)$, thus this alternative normalization condition is not suitable for practical applications.
\hfill $\diamond$
\end{remark}


\subsection{Computation of the trajectory polyline}\label{subsec:trajectory}
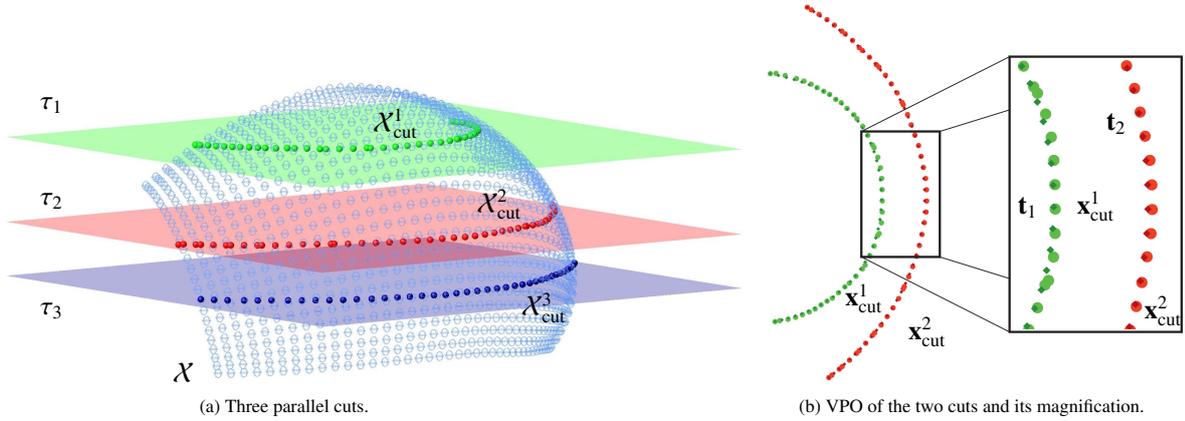
\begin{figure*}[t!]
	\centering
	\begin{subfigure}[b]{0.45\textwidth} 
        \centering
		\begin{overpic}[height=40mm, trim={0mm 0mm 0mm 0mm}, clip]{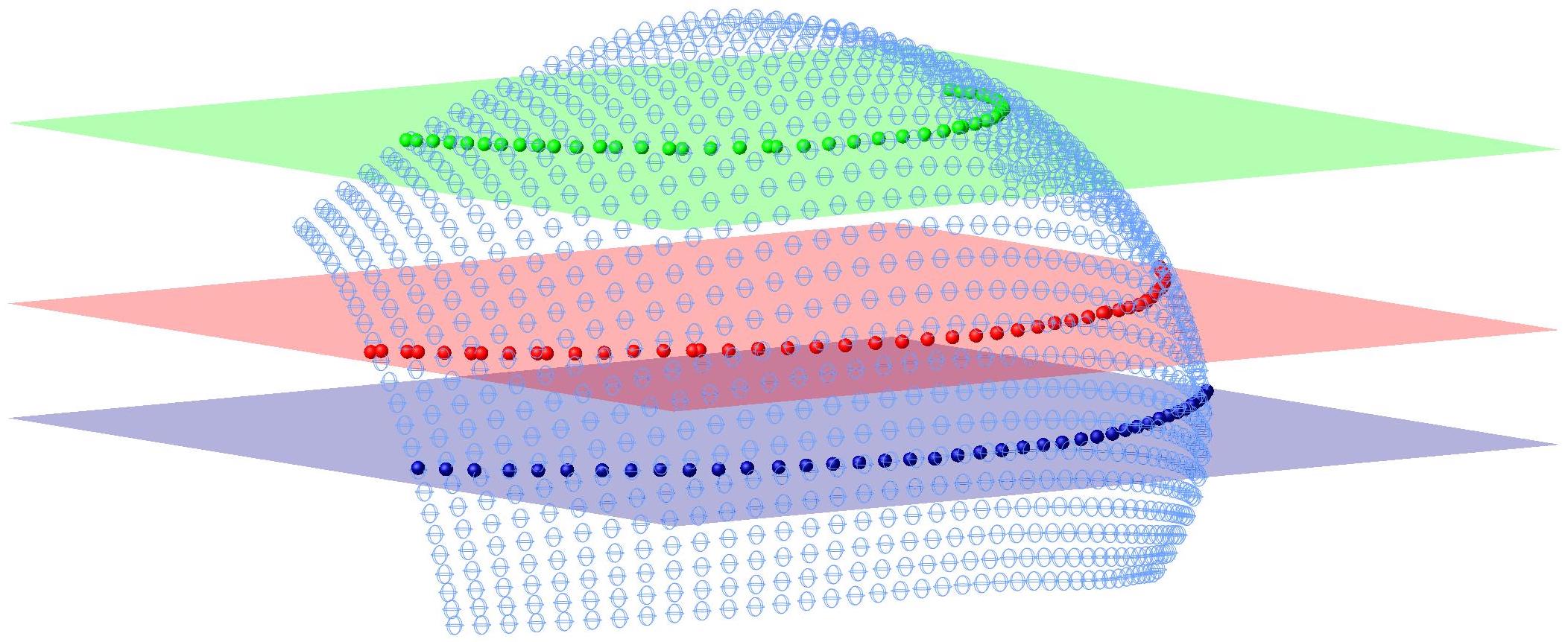}
           \put(5,9 ){$\tau_3$}
           \put(5,24){$\tau_2$}
           \put(5,37){$\tau_1$}
           \put(50,34){$\mathcal{X}_{\text{cut}}^1$}
           \put(64,23){$\mathcal{X}_{\text{cut}}^2$}
           \put(70,9){$\mathcal{X}_{\text{cut}}^3$}
           \put(23,0){$\mathcal{X}$}
        \end{overpic}
		\caption{Three parallel cuts.}
	\end{subfigure}
    \hfill
    \begin{subfigure}[b]{0.45\textwidth} 
        \centering
		\begin{overpic}[height=50mm, trim={0mm 0mm 0mm 0mm}, clip]{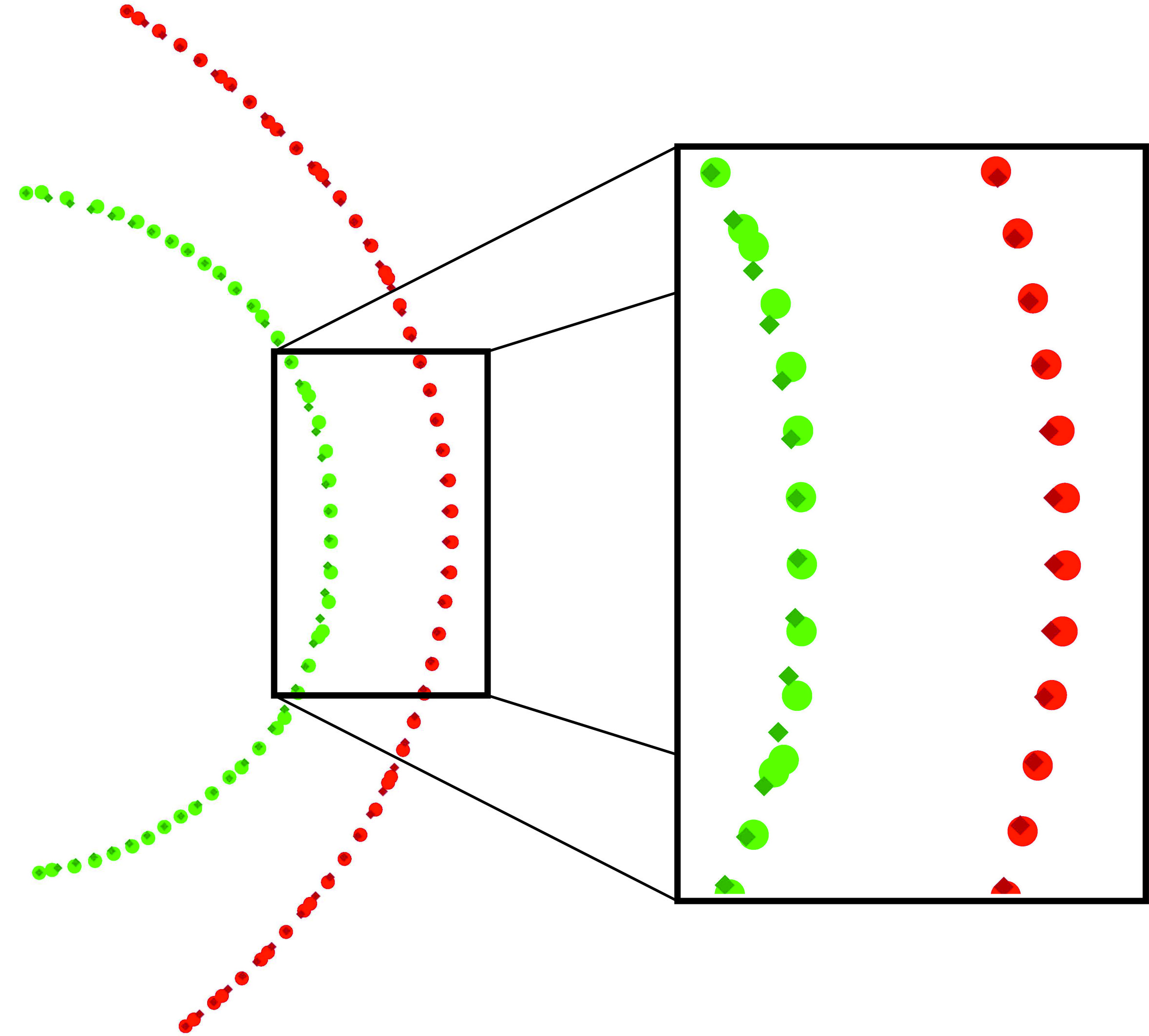}
           \put(75,40){$\mathbf{x}_{\text{cut}}^1$}
           \put(35,10){$\mathbf{x}_{\text{cut}}^2$}           
           \put(20,18){$\mathbf{x}_{\text{cut}}^1$}
           \put(91,15){$\mathbf{x}_{\text{cut}}^2$}           
           \put(61,40){$\mathbf{t}_1$}
           \put(82,60){$\mathbf{t}_2$}                      
        \end{overpic}
		\caption{VPO of the two cuts and its magnification.}
	\end{subfigure}
	\caption{Illustration of cutting the point cloud $\mathcal{X}$ with a set of parallel planes (a) and the variational path optimization (VPO) of two of them (b). Subfigure (b) contains also a magnified view showing how the vertices of the trajectories are updated via the VPO.  The dark green ($\mathbf{t}_1$) and dark red ($\mathbf{t}_2$) show the trajectories $\mathbf{x}_{\text{cut}}^1$ and $\mathbf{x}_{\text{cut}}^2$ after the optimization process.}
	\label{fig:OBJ:cuts}
\end{figure*}
Let our unorganized point cloud be given by an $q \times 3$  matrix $\mathcal{X}$, where each row houses the coordinates of a distinct point. We use $\mathcal{X}_i$ to denote the $i$-th row. When referring to the $x$, $y$, and $z$-coordinates of a specific row, they are respectively symbolized by $\mathcal{X}_{i,1}$, $\mathcal{X}_{i,2}$, and $\mathcal{X}_{i,3}$.

Having the axis direction $\mathbf{a}^*$ is equivalent to knowing the family of trajectory planes $\{\delta_{j}\}_{j\in \bar{J}}$. Therefore, the trajectory polylines are contained in the intersection of the trajectory planes with the point cloud. Without loss of generality (we can do a rigid motion if necessary), let us assume that $\mathbf{a}^*$ is pointing toward the positive direction of the $z$-axis. 
Define two parallel planes, namely, $\tau_1$ and $\tau_2$ by the equations $z = h_1$ and $ z = h_2$ where {$h_1\neq h_2$}. Now, consider a threshold $\epsilon$ for each of these two planes. Then project the points $\mathcal{X}_{i}$ onto the plane 
$\tau_k$ if
\begin{equation*}
    |\,\mathcal{X}_{i,3}  - h_{k}\,| \leq \epsilon
\end{equation*}
for $1\leq i \leq m$ and $k = 1,2$. This is simply taking place by replacing $\mathcal{X}_{i,3}$ with $h_{k}$ depending on the plane $\tau_i$ that $\mathcal{X}_{i}$ is in the vicinity of. In this way we obtain two unorganized point clouds namely $\mathcal{X}_{\text{cut}}^1$ and $\mathcal{X}_{\text{cut}}^2$ which are supposed to serve as our trajectories. Fig.\,\ref{fig:OBJ:cuts}-(a) depicts such parallel cuts for the point cloud of Example\,\ref{example:axis}. However, before we can assign a polyline to these unorganized curve-point clouds, it is essential to establish an order for the vertices.

\subsubsection{Assigning polylines to $\mathcal{X}_{\text{cut}}^1$ and $\mathcal{X}_{\text{cut}}^2$}
\begin{figure}[t!]
\centering
	\begin{subfigure}[b]{0.32\textwidth} 
        \centering
        \begin{center}
        \begin{overpic}[height = 35 mm]{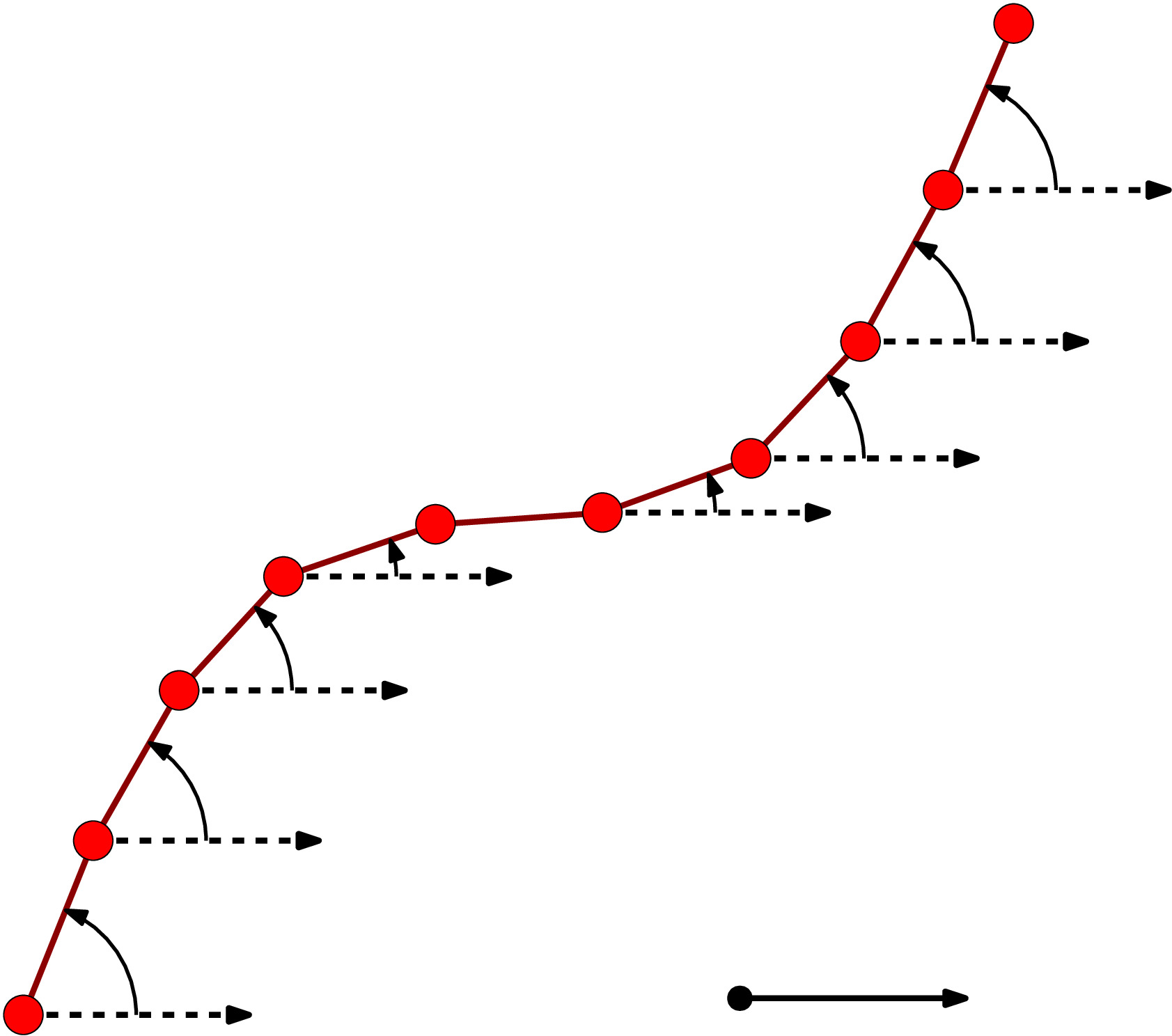}
        	\begin{small}
        	\put(63,7){chosen direction: $\mathbf{i}$}
            \put(12,7){$\theta_1$}
        	\put(19,22){$\theta_2$}
            \put(26,33){$\theta_3$}
            \put(0,20){$-1$}
            \put(8,33){$-1$}
            \put(20,44){$-1$}
            \put(34,49){$-1$}
            \put(47,50){$+1$}
            \put(60,56){$+1$}
            \put(67,65){$+1$}
            \put(73,76){$-1$}
        	\end{small}
          \end{overpic}  
        \end{center}
        \caption{}
    \end{subfigure}
    \hfill
    \begin{subfigure}[b]{0.32\textwidth} 
        \centering
        \begin{center}
        \begin{overpic}[height = 35 mm]{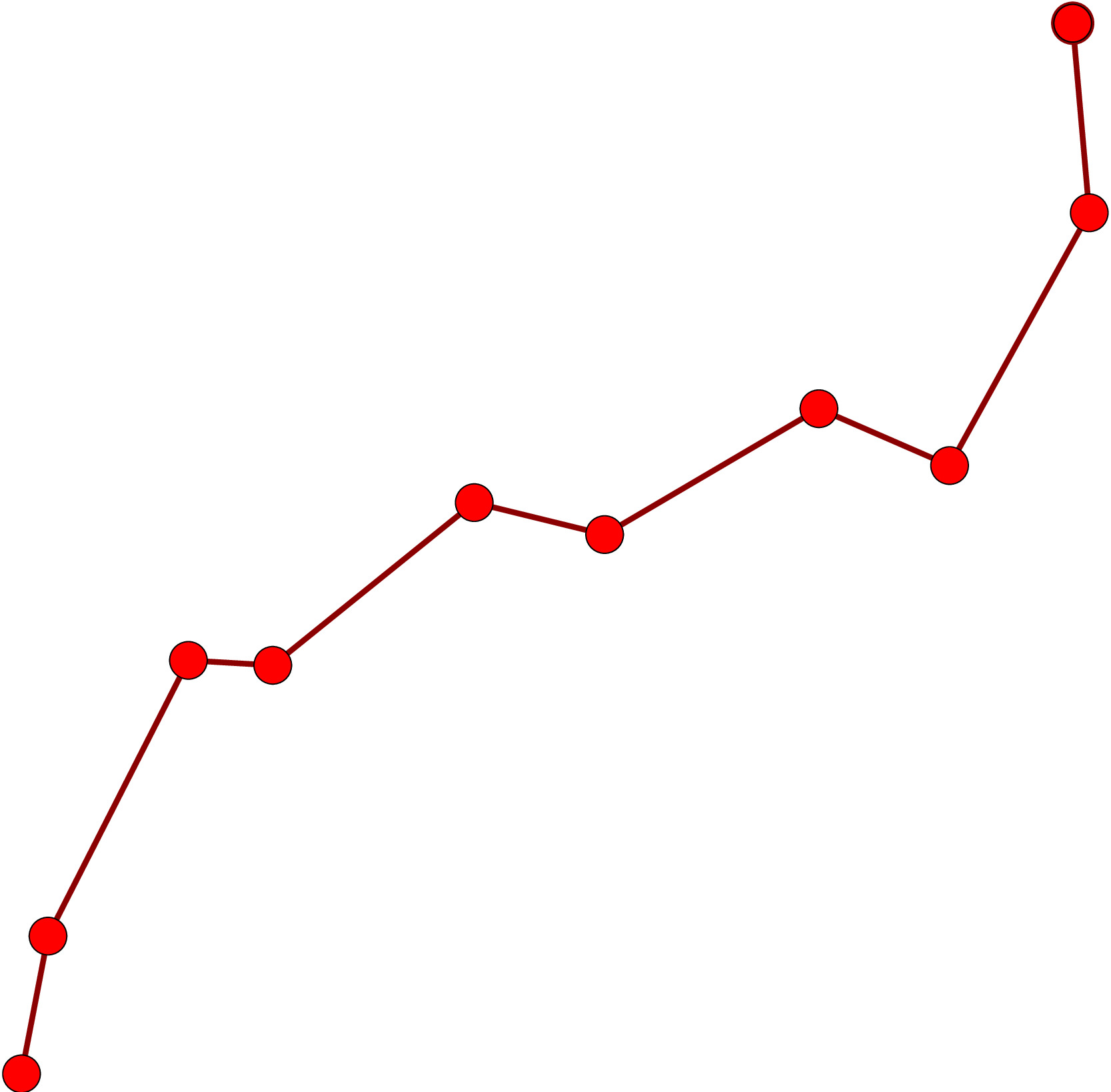}
        	\begin{small}
            \put(-5,20){\color{cyan}$-1$\color{black}}
            \put(3,40){\color{cyan}$-1$\color{black}}
            \put(20,44){\color{cyan}$+1$\color{black}}
            \put(34,57){\color{cyan}$-1$\color{black}}
            \put(47,38){\color{magenta}$+1$\color{black}}
            \put(68,65){\color{magenta}$-1$\color{black}}
            \put(85,45){\color{magenta}$+1$\color{black}}
            \put(85,80){\color{magenta}$+1$\color{black}}
        	\end{small}
          \end{overpic}  
        \end{center}
        \caption{}
    \end{subfigure}
    \hfill
    \begin{subfigure}[b]{0.32\textwidth} 
        \centering
        \begin{center}
        \begin{overpic}[height = 35 mm]{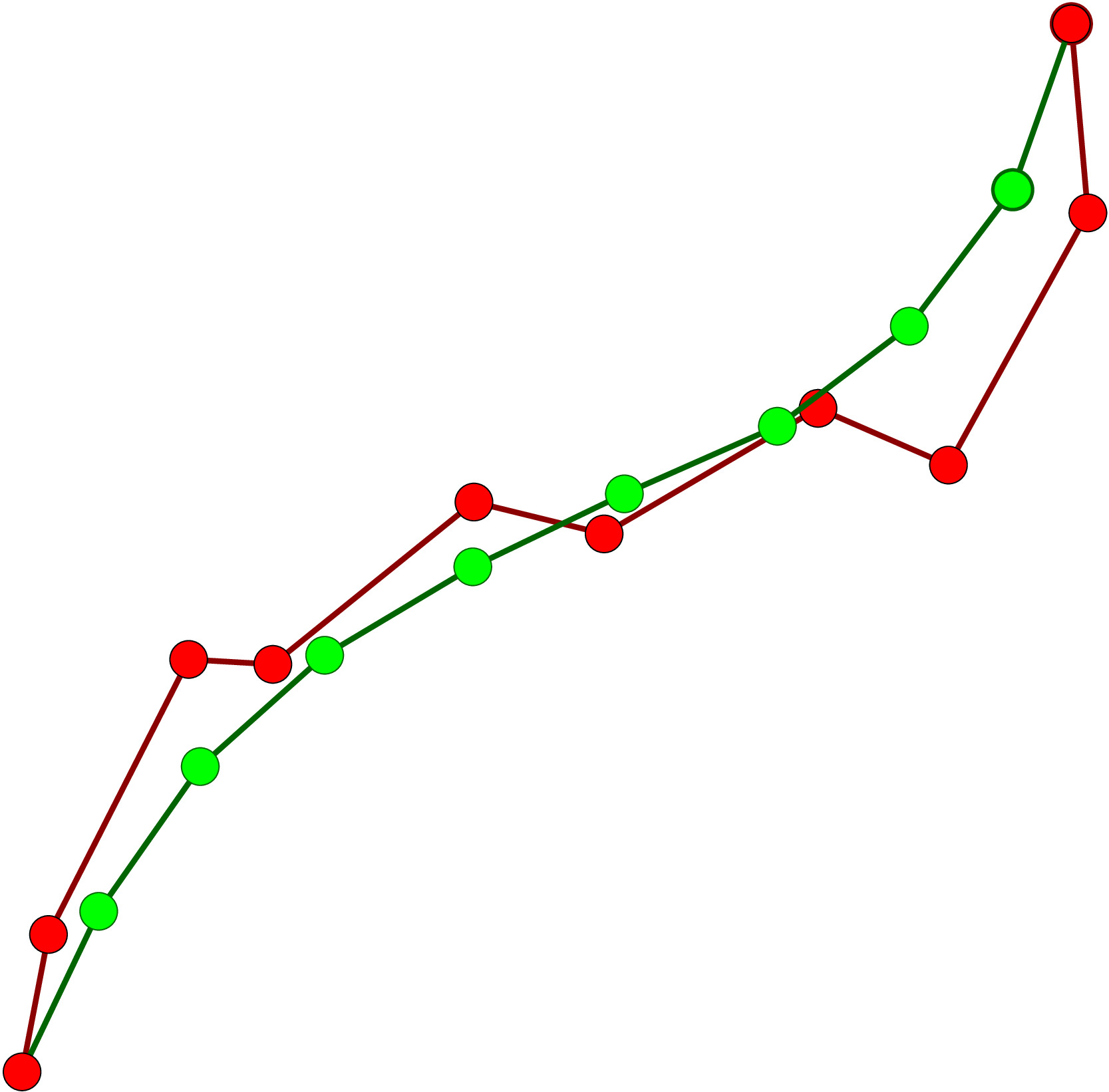}
        	\begin{small}
            \put(10,10){\color{cyan}$-1$\color{black}}
            \put(20,20){\color{cyan}$-1$\color{black}}
            \put(30,30){\color{cyan}$-1$\color{black}}
            \put(40,36){\color{cyan}$-1$\color{black}}
            \put(50,60){\color{magenta}$+1$\color{black}}
            \put(63,68){\color{magenta}$+1$\color{black}}
            \put(76,75){\color{magenta}$+1$\color{black}}
            \put(82,85){\color{magenta}$+1$\color{black}}
        	\end{small}
          \end{overpic}  
        \end{center}
        \caption{}
    \end{subfigure}
	\caption{Illustration of the (discrete) curvature sign of a polyline (a), a noisy version of it (b) and its smoothened version (in green) via VPO (c).}
	\label{fig:sign}
\end{figure}
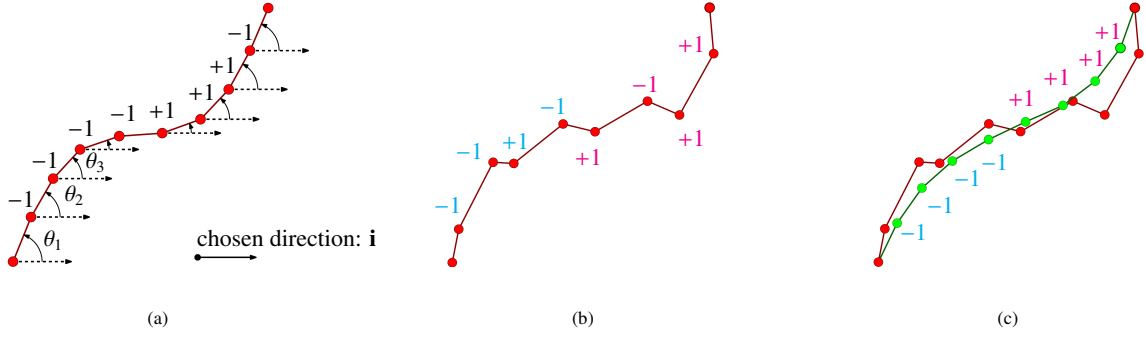

To address this challenge, we utilize an algorithm rooted in a modified version of moving least-squares, which details are given in \cite{lee2000curve}. This algorithm adeptly approximates a collection of unorganized points with a curve devoid of self-intersections, as illustrated in Fig.\ \ref{fig:OBJ:cuts}-(b). While the resulting curve bears a visual semblance to a smooth trajectory, it merely provides a preliminary foundation. Further refinement and manipulation are still necessary. Consequently, we name the now ordered versions of $\mathcal{X}_{\text{cut}}^1$ and $\mathcal{X}_{\text{cut}}^2$ by $\mathbf{x}^{1}_{\text{cut}}$ and $\mathbf{x}^{2}_{\text{cut}}$, respectively.\\
At this stage, our immediate objective is to apply the continuous theory of T-surfaces to derive the $\beta$-angles from our two ordered trajectories, $\mathbf{x}^{1}_{\text{cut}}$ and $\mathbf{x}^{2}_{\text{cut}}$, as outlined in Property \ref{remark:beta:angle}. To achieve this, it is crucial that our trajectory curves exhibit consistent (discrete) curvature signs. Without this consistency, it may not be possible to accurately determine such $\beta$-angles. This step, aimed at ensuring curvature sign consistency, can be viewed as a precautionary measure to reduce noise in the results. Therefore, we take a detour to introduce some discrete objects and then deal with the (discrete) curvature signs.

Let $\mathbf{x}^{k}_{\text{cut}} : \{0,...,m^k_{\text{cut}}\}\rightarrow\mathbb{R}^2$ for $k = 1,2$ be the ordered and smooth-like curve obtained from $\mathcal{X}^{k}_{\text{cut}}$ by the algorithm of \cite{lee2000curve} (see Fig.\ \ref{fig:OBJ:cuts}-(b)). Define the edge unit vectors of the aforementioned trajectory curves by 
\begin{equation*}
    \mathbf{e}^{k}_{\text{cut}}(i) := \frac{\mathbf{x}^{k}_{\text{cut}}(i+1) - \mathbf{x}^{k}_{\text{cut}}(i)}{\left|\,\mathbf{x}^{k}_{\text{cut}}(i+1) - \mathbf{x}^{k}_{\text{cut}}(i)\,\right|},\quad\quad\text{where}\quad\quad k =1,2.
\end{equation*}   
Let's assign a (discrete) normal $\mathbf{n}^{k}_{\text{cut}}(i)$ to $\mathbf{x}^{k}_{\text{cut}}(i)$ by rotating $\mathbf{e}^{k}_{\text{cut}}(i)$ clock-wise by $90^\circ$ in the plane for any $i \in \{0,...,m^k_{\text{cut}}-1\}$ (note that, in this way we assign no normal to the last vertex since there is no edge to rotate).

Intuitively, in the plane, the \emph{discrete curvature sign} acts similarly to its continuous counterpart where the positive sign means that the curve is turning ``left" while the negative one means that it is turning ``right". To compute the aforementioned sign of a polyline such as $\mathbf{x}^k_{\text{cut}}$, choose a predefined unit direction in the plane (e.g.\,$\mathbf{i} = (1,0)^T$). Now, compute the angle between $\mathbf{e}_{\text{cut}}(i)$ and the predefined direction by 
\begin{equation*}
    \theta_i = \arccos{\left(\langle\mathbf{e}^k_{\text{cut}}(i), \mathbf{i}\rangle\right)},
\end{equation*}
Finally, define the signum of the poyline by
\begin{equation*}
    \mathrm{sgn}(i) = \sign{\left({\theta}_i - {\theta}_{i-1}\right)}. 
\end{equation*} 
Therefore, one can observe that the curve is going to the right $\mathrm{sgn}(i) = -1$ and whenever the curve is moving left we have $\mathrm{sgn}(i) = +1$. 
An example is depicted in Fig.\,\ref{fig:sign}-(a). Naturally, when dealing with polylines we assign no sign to the start and end vertices. Finally, note that such a choice is independent of the orientation of the polyline.\\
If our trajectory curves exhibit excessive noise, characterized by the (discrete) curvature signs frequently oscillating between $-1$ and $+1$ (cf.\ Fig.\,\ref{fig:sign}-(b)), the existence of a consistent $\beta$-angle cannot be guaranteed. In other words, it may not be possible to rotate the normals of the first trajectory curve in such a manner that the lines they span consistently maintain the same angle with the normals of the second curve at all points. 
To circumvent this unfavorable scenario, we initially identify the discrete curvature signs at each vertex along the trajectories. 
Then we categorize the signs into groups designated as $-1$ and $+1$. This can be easily done using \emph{density based spatial clustering (dbs)} algorithms. Fig.\,\ref{fig:sign}-(b) shows such a grouping of signs with the colour labels of {magenta} and {cyan} for $+1$ and $-1$, respectively. Subsequently, we employ a \emph{variational path optimization} algorithm.
\begin{remark}
    {Variational path optimization (VPO)} is an algorithm via which one can ``smoothen" a curve (in any dimension) such that the vertices are distributed in a way that they would mimic an arc-length parameterization as much as possible depending on two weight factors $w_1$ and $w_2$. These weighted factors emphasize on the {geodesic energy} $E$ and {bending energy} $B$
    \begin{equation*}
        E := \sum_{i = 2}^{n} \left|\,\mathbf{c}_i -\mathbf{c}_{i-1}\,\right|^2,\quad\quad\quad\quad
        B := \sum_{i = 2}^{n-1} \left|\,\mathbf{c}_{i+1} - 2 \mathbf{c}_{i}  + \mathbf{c}_{i-1}\,\right|^2.
    \end{equation*}
    This algorithm was presented by Hofer and Pottmann \cite{hofer_pott} and a modified version\footnote{The weights being less dependent on the number of vertices.} was used by the authors \cite{rasoulzadeh2020variational}.
\hfill $\diamond$ 
\end{remark}
Here, we choose a certain arbitrary number for $w_1$ and gradually increase $w_2$ till the point where all the signs within the groups are consistent with respect to the sign assigned to the group. Fig.\,\ref{fig:sign}-(c) shows the result of applying VPO on the noisy curve, showing that the (discrete) curvature signs get consistent after such an application (i.e.\, the signs within the magenta and cyan groups are all becoming $+1$ and $-1$ respectively).
\begin{example}\label{example:trajectories}
    Coming back to our original example (see Example\,\ref{example:axis}) we apply the grouping algorithm to our case which implies that all our vertices should have the (discrete) curvature sign of $+1$ as this sign is strongly the dominant sign. Now, we applying the VPO algorithm results in the new curves as depicted in Fig.\,\ref{fig:OBJ:cuts}-(c). Both these curves have the sign $+1$ everywhere.
\end{example}
After applying the VPO algorithm we obtain the desired trajectories and therefore we change their naming from $\mathbf{x}^1_{\text{cut}}$ and $\mathbf{x}^2_{\text{cut}}$ to $\mathbf{t}_1$ and $\mathbf{t}_2$, respectively.
\subsection{Computation of the profile planes}\label{subsec:p_planes}
\begin{figure*}[t!]
	\centering
	\begin{subfigure}[b]{0.18\textwidth} 
        \centering
		\begin{overpic}[height=35mm, trim={0mm 0mm 0mm 0mm}, clip]{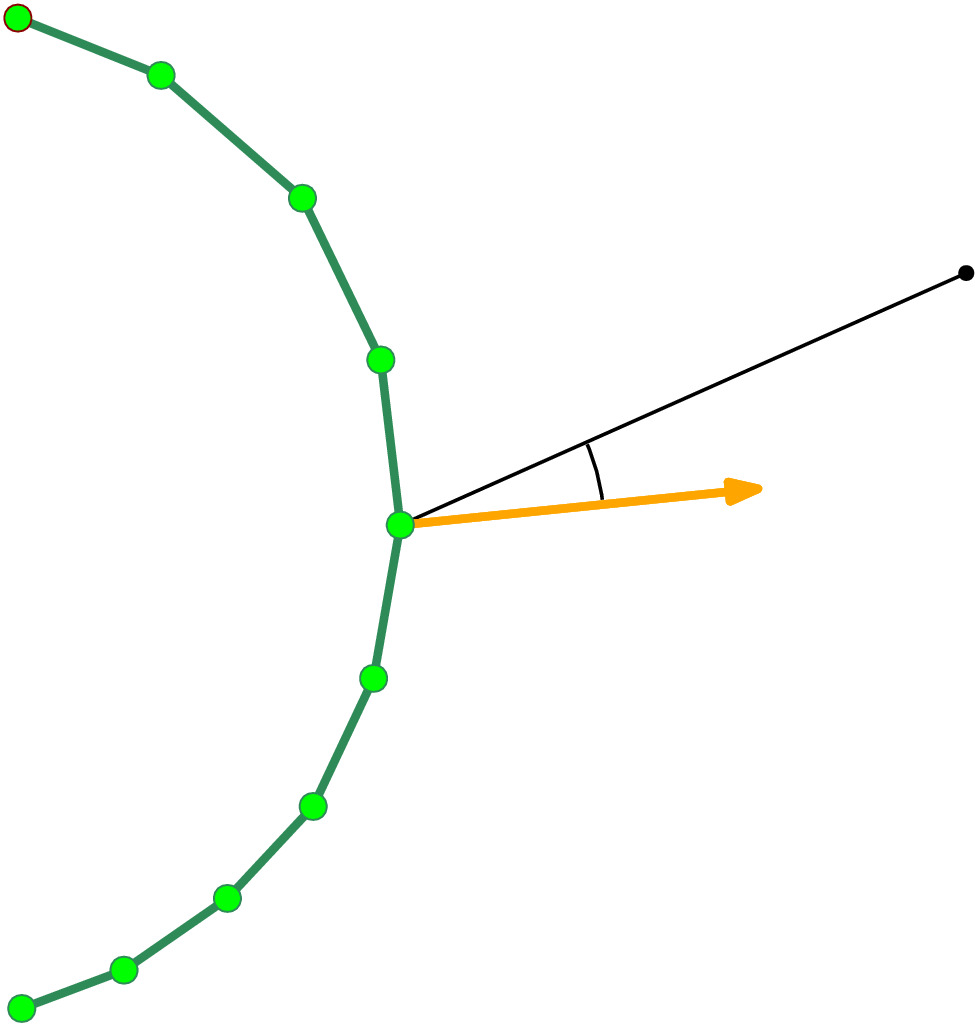}
           \put(18,46){$\mathbf{t}_1(i)$}
           \put(50,80){$j$-th line}
           \put(50,40){$\beta^1_{i,j}$}
        \end{overpic}
		\caption{Construction of $\beta^1$.}
	\end{subfigure}
	\hfill
	\begin{subfigure}[b]{0.22\textwidth} 
        \centering
		\begin{overpic}[height=40mm, trim={0mm 0mm 0mm 0mm}, clip]{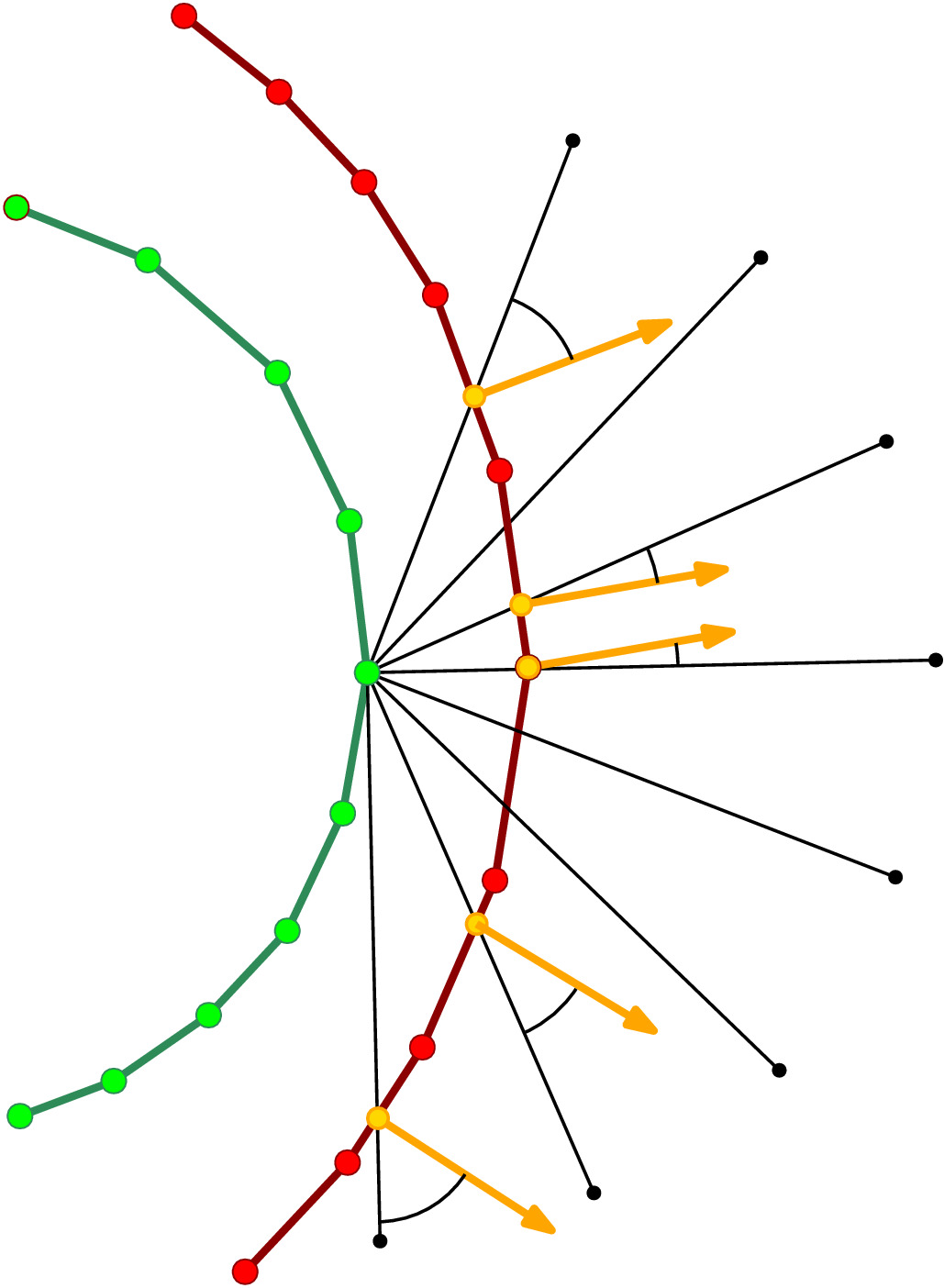}
             \put(10,46){$\mathbf{t}_1(i)$}
             \put(32,0){\tiny $\beta^2_{i,1}$}
             \put(45,15){\tiny $\beta^2_{i,2}$}
             \put(60,50){\tiny $\beta^2_{i,5}$}
             \put(60,58){\tiny $\beta^2_{i,6}$}
             \put(45,79){\tiny $\beta^2_{i,8}$}
        \end{overpic}
		\caption{Construction of $\beta^2$.}
	\end{subfigure}
    \hfill
	\begin{subfigure}[b]{0.22\textwidth} 
        \centering
        \begin{center}
		\begin{overpic}[height=35mm, trim={0mm 0mm 0mm 0mm}, clip]{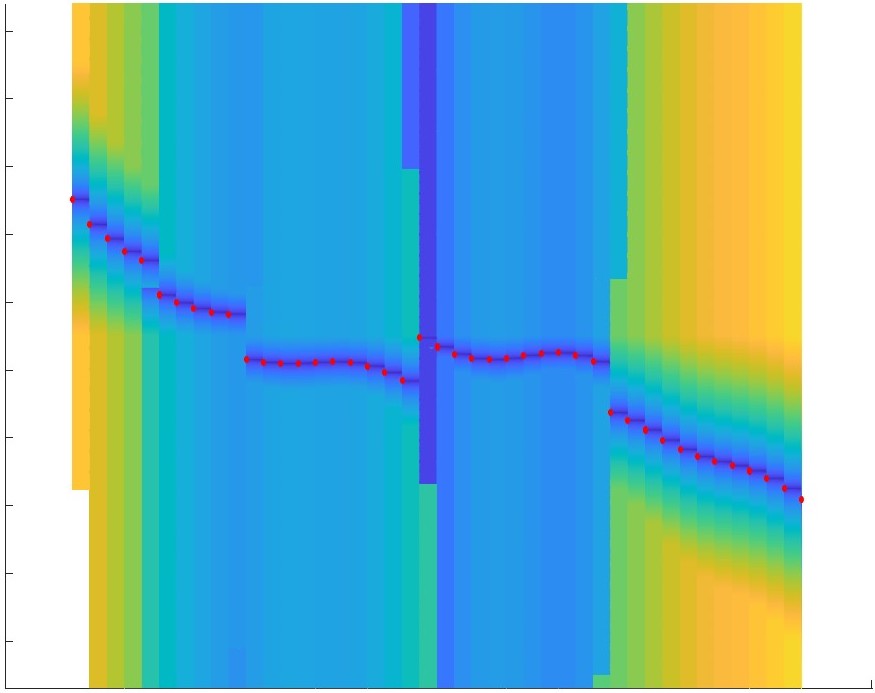}
            \put(50,-10){$i$}
            \put(9,-10){$4$}
            \put(90,-10){$46$}
            \put(-10,46){$j$}
        \end{overpic}
        \end{center}
		\caption{Construction of $\beta$-path.}
	\end{subfigure}
 	\hfill
	\begin{subfigure}[b]{0.22\textwidth} 
    \begin{center}
	\begin{overpic}[height = 40 mm,trim={0mm 0mm 0mm 0mm},clip]{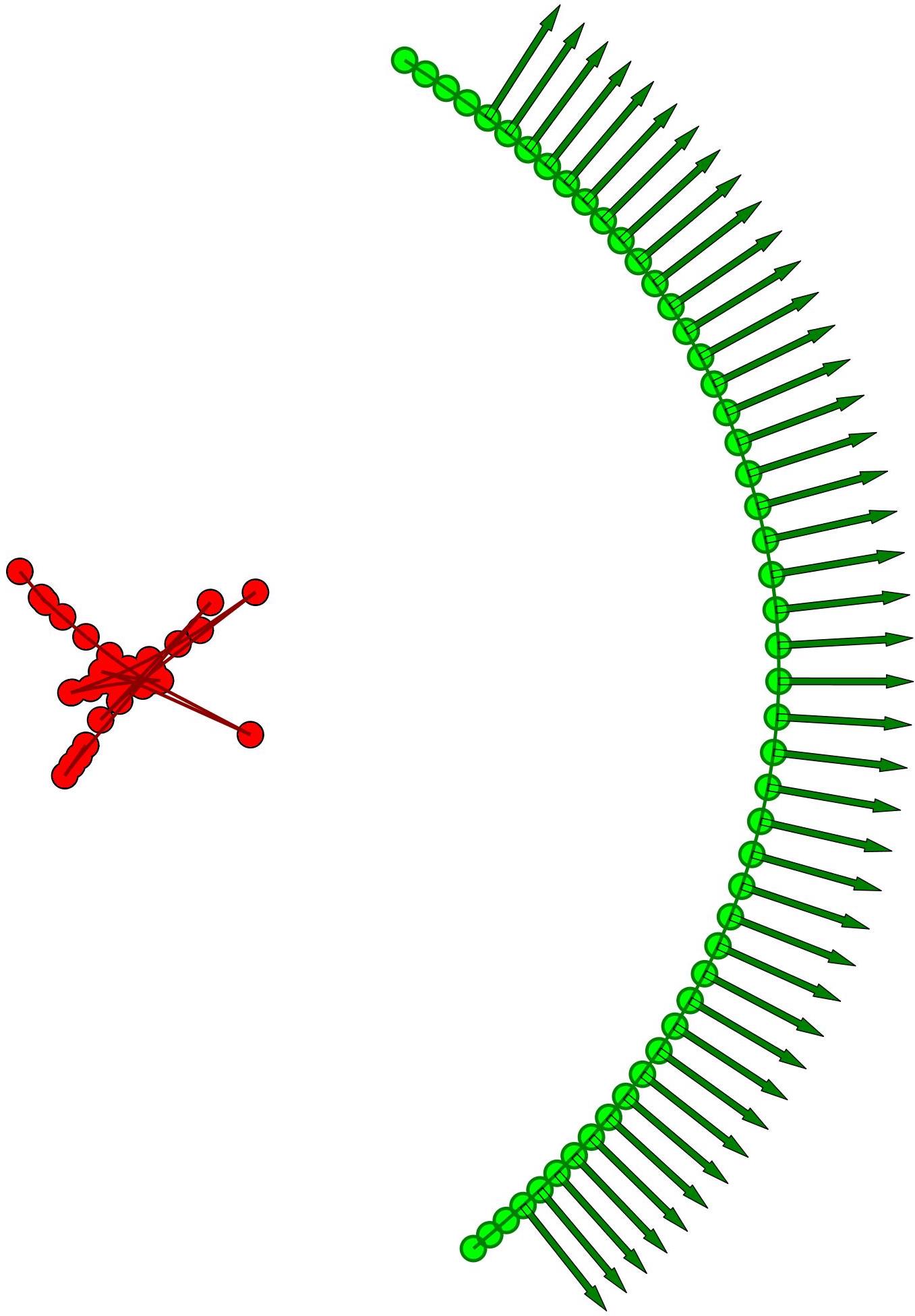}
        \put(50,40){$\mathbf{t}_1$}
        \put(60,0){$\mathbf{n}^{\,\beta}_1$}
        \put(10,60){$\mathbf{d}$}
    \end{overpic}
    \end{center}
    \caption{Construction of directrix curve.}
    \end{subfigure}
	\caption{(a,b): Illustration of a discrete pencil of line at each point of the trajectory curve $\mathsf{t}_1$ and its intersections with $\mathsf{t}_2$. Illustration of the discrete graph surface for determining the $\beta$-path (c) and the resulting construction of the directrix curve at $z = 0$ (d). }
	\label{fig:beta:path}
\end{figure*}
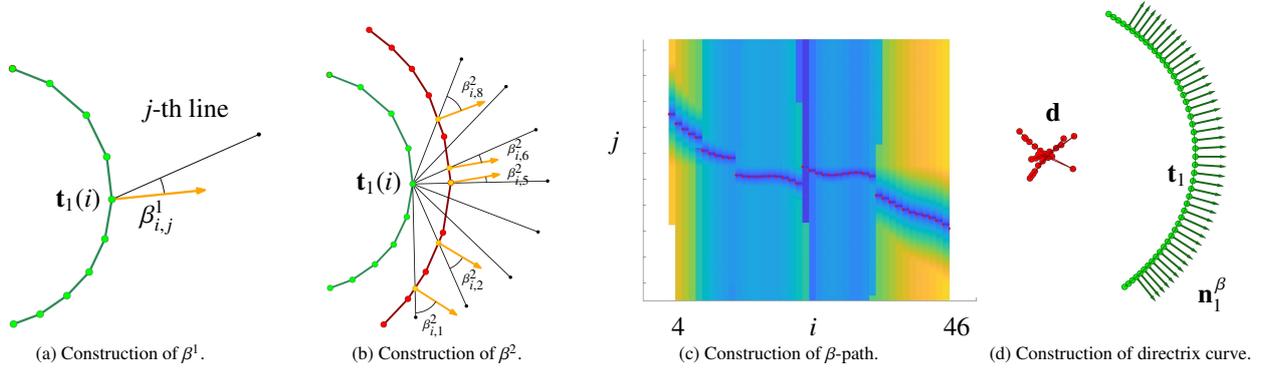
Having the trajectories $\mathbf{t}_1$ and $\mathbf{t}_2$ the goal is to estimate the profile planes. In order to do this we resort to an observation from the smooth theory that the 
profile planes of a T-surface intersect the trajectories under equal angles (cf.\ Property \ref{remark:beta:angle}). 
In another word, the $\beta$-angles that we desire to find are the angles that if we rotate the normals $\mathbf{n}_1$ of $\mathbf{t}_1$, the line that is spanned by it should make the same angle $\beta$ with the normals $\mathbf{n}_2$ of $\mathbf{t}_2$. To address this, we should identify lines that traverse the vertices of $\mathbf{t}_1$ and intersect with $\mathbf{t}_2$, whether at its vertices or along its edges. Crucially, the angles between the designated normals and edges at corresponding points must be consistent. To realize this, we formulate a discrete representation of a \emph{discrete pencil of lines} for each vertex of $\mathbf{t}_1$. Starting with an arbitrarily chosen line extended sufficiently around a vertex, we subsequently generate the remaining lines by rotating about that vertex. This rotation can be executed such that the pencil achieves desired density — for instance, having an angle of $1^\circ$ between consecutive lines. 

Consider $i$ as the index for the vertices of $\mathbf{t}_1$ and $j$ as the index for each line in the pencil. Let $\beta^1_{i,j}$ represent the angle formed between the $j$-th line of the pencil and the normal $\mathbf{n}_1(i)$. Similarly, let $\beta^2_{i,j}$ denote the angle between the $j$-th line and the normal at the point where this line intersects with $\mathbf{t}_2$ (Fig.\ \ref{fig:beta:path}-(a,b) show the concept for the less crowded version of our trajectories of Example\,\ref{example:trajectories}). Now, define 
\begin{equation*}
    \Delta\,\beta_{i,j} := \left|\,\beta^2_{i,j} - \beta^1_{i,j}\,\right|,
\end{equation*}
and consider the following \emph{discrete graph surface} 
\begin{equation}\label{eq:graf:surface}
    S := \left\{ \left( i, j, \Delta\,\beta_{i,j} \right) \colon 1 \leq i \leq M, \ 1^\circ \leq j \leq 180^\circ \right\}.
\end{equation}
Cut the graph surface at $z = 0$ with a certain threshold results in a path like shape. Via polynomial regression its possible to assign a curve to this path (one can even try to assign a high degree polynomial for safety). This results in a polyline, which is called $\beta$-path from now on, that shows us the changes of such $\beta$-angles along the trajectory curve $\mathbf{t}_1$. Let us denote the rotated normals of the polyline $\mathbf{t}_1$ about the angle $\beta$ by $\mathbf{n}^\beta_1$. This is all the information that we need, as each of our profile planes is characterised by the points $\mathbf{t}_1(i)$ and the vectors $\mathbf{n}^\beta_1(i)$.

The $\mathbf{n^\beta_1}$ vector field allows us to compute the directrix curve $\mathbf{d}$. As per Definition\,\ref{def:beta:evolute}, such a curve is the envelope of all the rotated normals which in our discrete setting means that it is simply the intersection of consecutive lines spanned by $\mathbf{n^\beta_1}(i)$ for all admissible $i$s. More accurately, let $\mathbf{n}^\beta_1(i) = \mathbf{R}(\beta_i).\mathbf{n}_1(i)$. Now, any two consecutive rotated lines can be presented as the set of points $\mathbf{d} := (x,y)^T$ satisfying $\langle \mathbf{d} - \mathbf{t}_1(i),\mathbf{J}.\mathbf{n}^\beta_1(i)\rangle = 0$ where $\mathbf{J} \in \mathrm{SO}(2)$ is a $90^\circ$ rotation matrix. Then each vertex of the directrix curve is computed by

\begin{equation}\label{eq:linear:system}
    \left(\begin{array}{c}
    (\mathbf{J}.\mathbf{n}^\beta_1(i))^T     \\
    (\mathbf{J}.\mathbf{n}^\beta_1(i+1))^T
    \end{array}\right).\mathbf{d} = 
    \left(\begin{array}{c}
    \langle \mathbf{J}.\mathbf{n}^\beta_1(i), \mathbf{t}_1(i)\rangle\\
    \langle \mathbf{J}.\mathbf{n}^\beta_1(i+1), \mathbf{t}_1(i+1)\rangle      
    \end{array}\right).
\end{equation}

\begin{remark}
    For utmost accuracy, one can compute the linear system of Eq.\ (\ref{eq:linear:system}) symbolically once and simply substitute the values inside at each $i$. \hfill $\diamond$
\end{remark}

\begin{example}\label{example:beta:eta:theta}
    Fig.\ \ref{fig:beta:path}-(a,b) illustrated a less crowded version of the trajectory curves $\mathbf{t}_1$ and $\mathbf{t}_2$ and describe how the pencil of lines is built over the trajectory $\mathbf{t}_1$. Fig.\,\ref{fig:beta:path}-(c) shows a top view of the graph surface of Eq.\ (\ref{eq:graf:surface}) that is computed for our original point cloud $\mathcal{X}$ with a gradient colouring. In this coloring the colder colors show depth while the warmer ones show heights.  In this specific example, the $\beta$-path is derived from the intersection of the graph surface with the plane where $z = 0$. 
    Now, with the $\beta$-path at hand, we assign the $\beta(i)$ angles to each vertex of $\mathbf{t}_1(i)$ with the exception of the last one (as no normal is assigned to it). In this way we can compute the directrix curve as the intersection of consecutive rotated normals by $\beta$ angles via Eq.\ (\ref{eq:linear:system}). The result is shown in Fig.\,\ref{fig:beta:path}-(d).
\end{example}

Finally, having the trajectory $\mathbf{t}_1$ and the directrix curve $\mathbf{d}$ we can compute the scaling factor and the rotation angles:
\begin{itemize}
    \item The sequence of the scaling factors $\eta$ is computed as follows: 
    \begin{equation}\label{eq:scaling}
        \eta_1 = 1,\quad\quad\quad\eta_i = \frac{\left|\,\mathbf{t}_1(i) - \mathbf{d}(i-1)\,\right|}{\left|\,\mathbf{t}_1(i-1) - \mathbf{d}(i-1)\,\right|}\quad\text{for}\quad i\geq 2.
    \end{equation}
    \item The sequence of the rotation angles $\theta$ can be thought of as the angle between the consecutive profile planes. More accurately, the angle $\theta_i$ is the angle enclosed by the planes $\delta(i)$ and $\delta(i+1)$. However, there is a subtlety  here. The aforementioned angle is an oriented angle. Therefore, its sign matters. Thankfully, as we have the directrix curve $\mathbf{d}$ and the trajectory curve $\mathbf{t}_1$ we can obtain such signs in  the following way
    \begin{equation*}
        \mathrm{sgn}(\theta_i) = \sign\left(\det\left(\mathbf{t}_1(i) - \mathbf{d}(i),\mathbf{t}_1(i+1) - \mathbf{d}(i) \right)\right),
    \end{equation*}
    where each of the vectors involved in the equation above are written as 2-dimensional vectors (since everything is planar). Fig.\,\ref{fig:prism:profile}-(a) illustrated the negative and positive version of $\theta_i$ in a schematic way.
\end{itemize}




\subsection{Computation of the profile polyline}\label{subsec:p_curve}
\begin{figure*}[t!]
	\centering
    \begin{subfigure}[b]{0.32\textwidth} 
    \begin{center}
	\begin{overpic}[height = 30 mm,trim={0mm 0mm 0mm 0mm},clip]{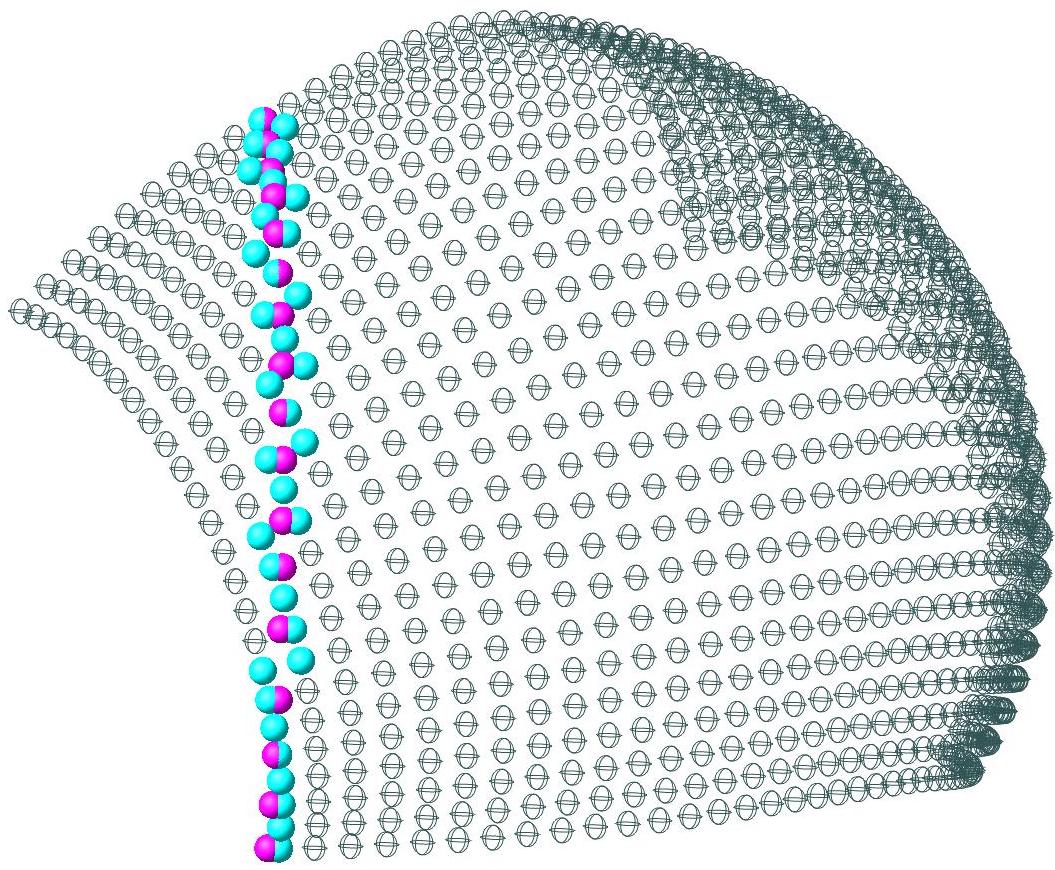}
    \end{overpic}
    \end{center}
    \caption{Front view.}
    \end{subfigure}
	\hfill
    \begin{subfigure}[b]{0.32\textwidth} 
    \begin{center}
	\begin{overpic}[height = 30 mm,trim={0mm 0mm 0mm 0mm},clip]{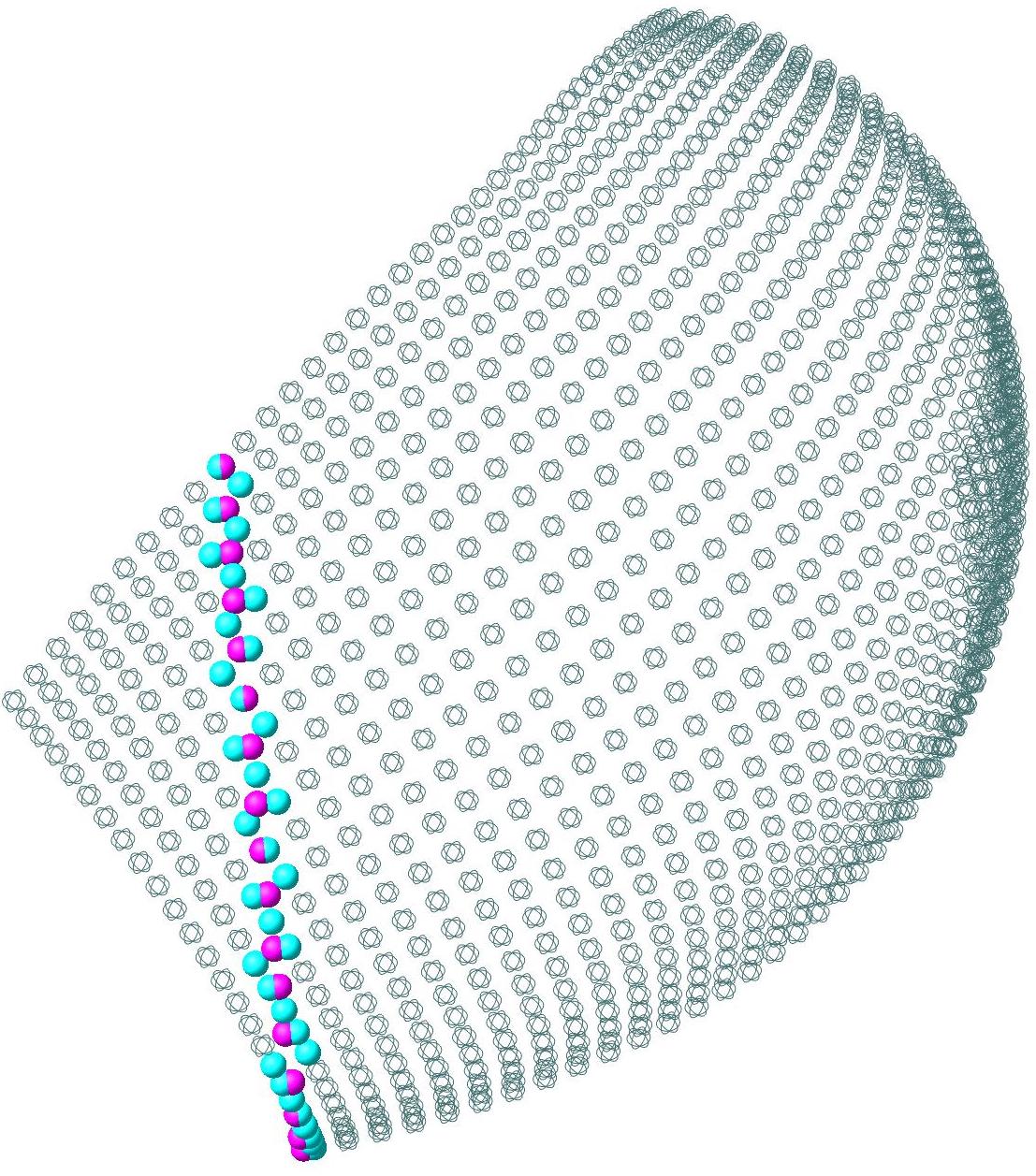}
    \end{overpic}
    \end{center}
    \caption{Rear view.}
    \end{subfigure}
    \hfill
    \begin{subfigure}[b]{0.32\textwidth} 
    \begin{center}
	\begin{overpic}[height = 27 mm,trim={0mm 0mm 0mm 0mm},clip]{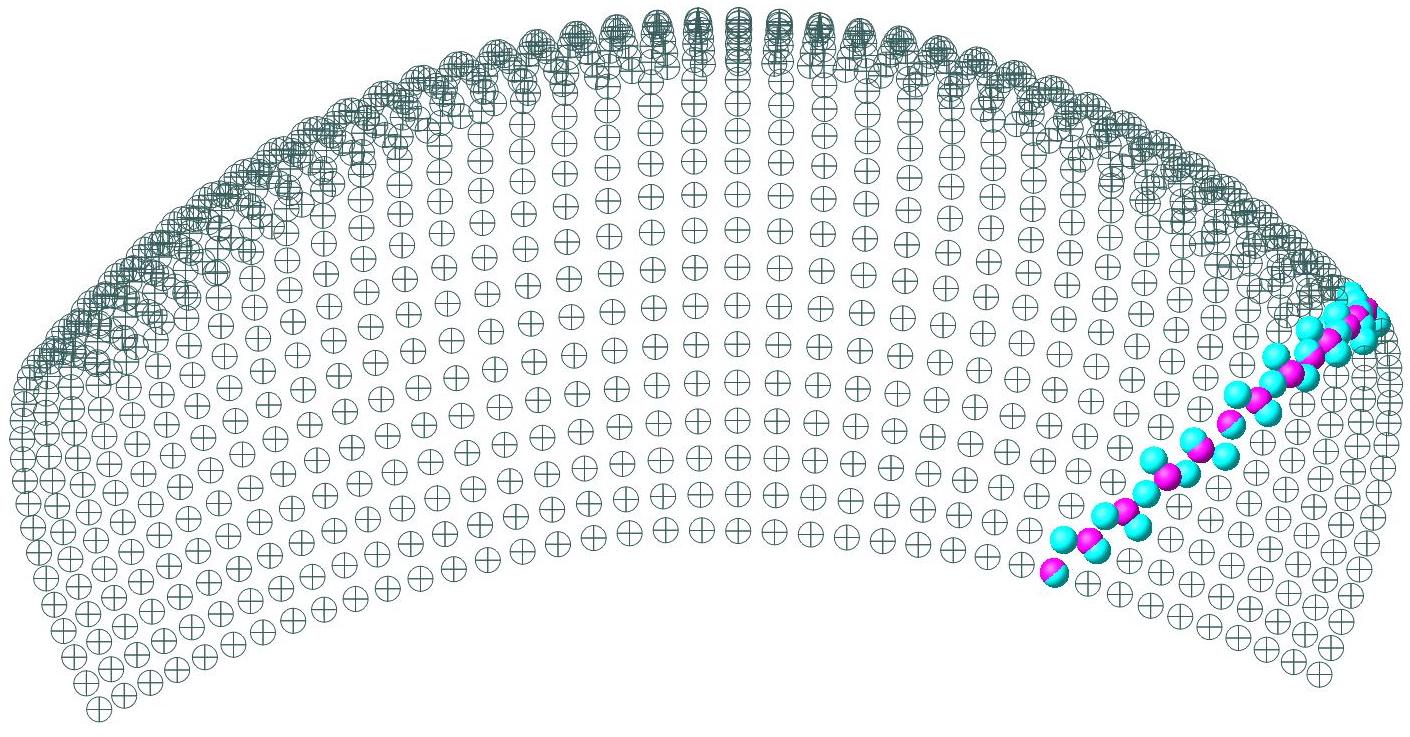}
    \end{overpic}
    \end{center}
    \caption{Top view with respect to the axis $\mathbf{a}^\ast$.}
    \end{subfigure}
	\caption{Illustration of the profile cut from three different angles. The cyan point cloud shows the cut while the magenta vertices stand for the ordered point cloud representing the profile polyline.}
	\label{fig:prism:profile}
\end{figure*}
The methodology presented in Section \ref{subsec:p_planes} yields the $\beta$-angles for the vertices of $\mathbf{t}_1$, enabling the construction of the profile planes $\delta$. These planes are defined by the points $\mathsf{t}_1(j)$ and the normals $\mathbf{a}^\ast\times \mathbf{n}^\beta_1(j)$ for all relevant indices $j$. This process generates a series of unorganized profile curves. Fig.\ \ref{fig:prism:profile}-(b,c) illustrate these intersections from both side and top views for the point cloud $\mathcal{X}$ in Example \ref{example:beta:eta:theta}.

At this stage, to organize these profile points into a coherent curve, we need not rely exclusively on the algorithm from \cite{lee2000curve}. Instead, an \emph{Euclidean minimum spanning tree} (EMST) algorithm can be effectively employed. To implement this, we first identify two points, $A$ and $B$, on the unorganized profile curve with the maximum and minimum $z$-coordinates, respectively. Then, using the EMST algorithm, we incrementally enlarge the radius of the neighborhoods around each vertex to establish the shortest path connecting $A$ and $B$. Although the resulting path may not be strictly planar, it is sufficiently close to a planar configuration. The availability of the profile planes $\delta(j)$ becomes beneficial at this point, as we can project the derived path onto these planes, ensuring alignment with the T-hedron structure.

\begin{remark}
    When dealing with a highly noisy point cloud, the aforementioned method may not yield an optimal curve. Specifically, it might result in a curve that closely resembles a straight line or exhibits an excessive number of zigzags. Under these circumstances, it is advisable to employ alternative curve-fitting techniques, such as polynomial regression or spline methods, which are typically more effective in accurately capturing the underlying pattern of the data despite the presence of significant noise. \hfill $\diamond$
\end{remark}

\begin{figure*}[t!]
	\centering
    \begin{subfigure}[b]{0.31\textwidth} 
    \begin{center}
	\begin{overpic}[height = 30 mm,trim={0mm 0mm 0mm 0mm},clip]{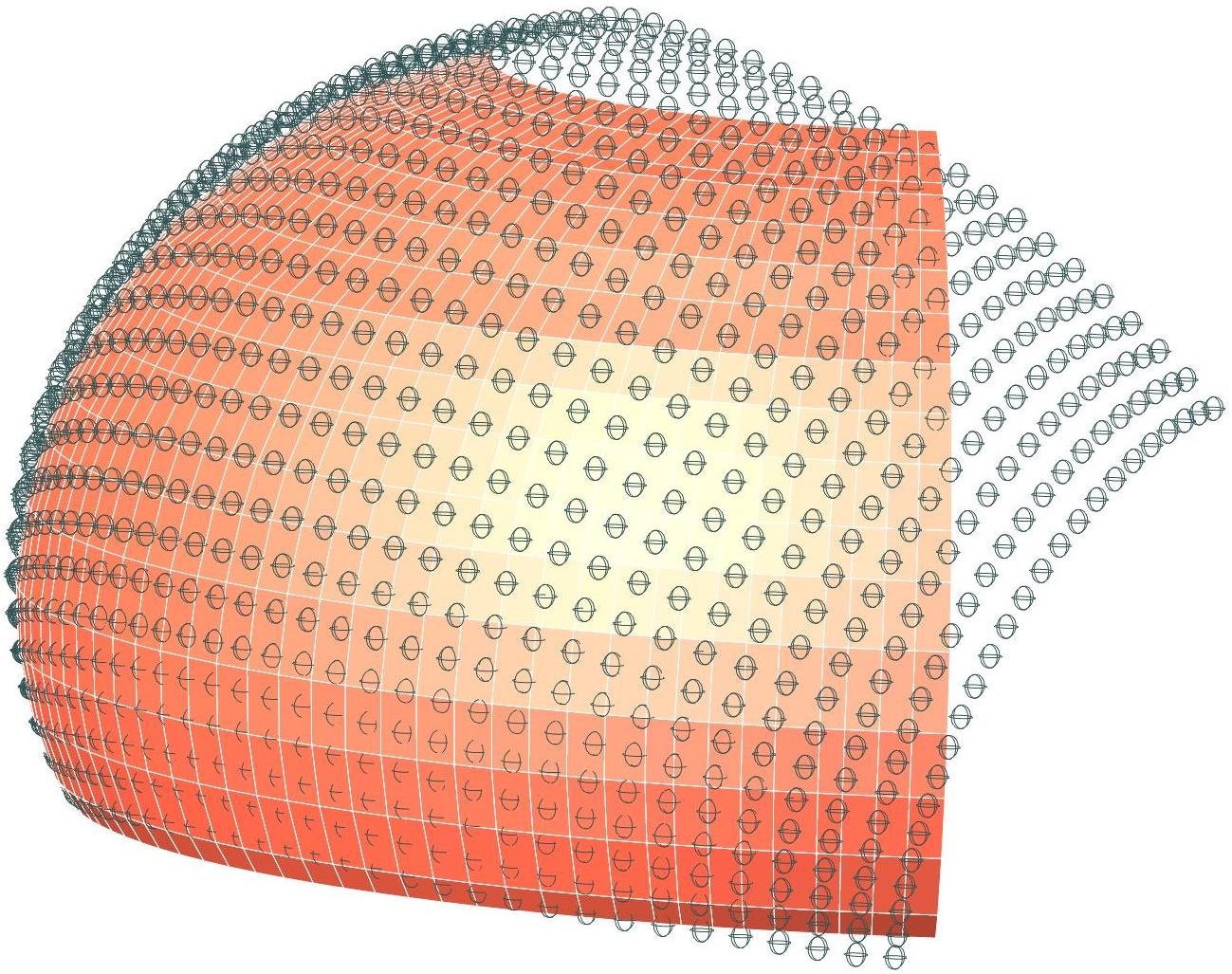}
    \end{overpic}
    \end{center}
    \caption{Front view.}
    \end{subfigure}
    \hfill
    \begin{subfigure}[b]{0.31\textwidth} 
    \begin{center}
	\begin{overpic}[height = 30 mm,trim={0mm 0mm 0mm 0mm},clip]{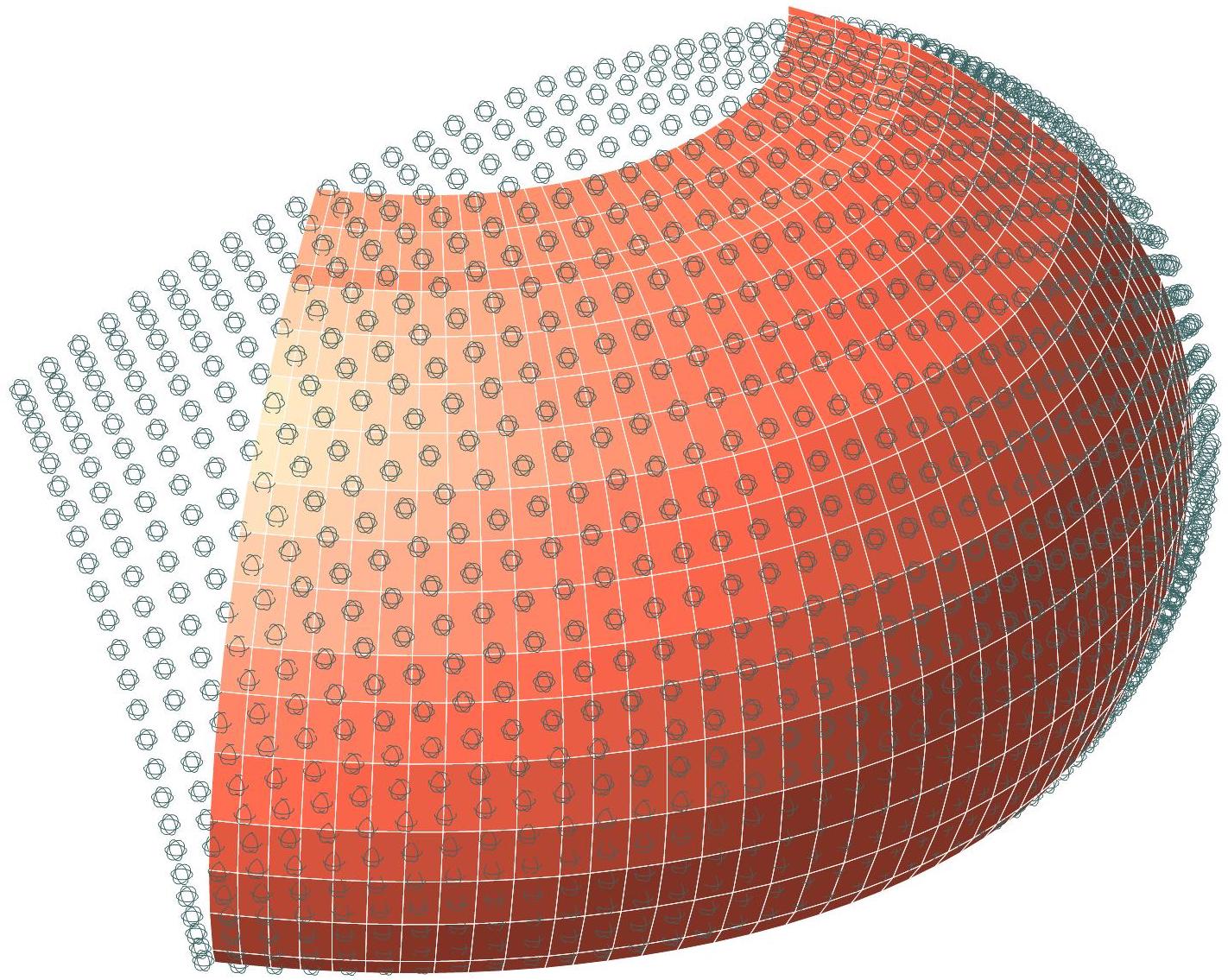}
    \end{overpic}
    \end{center}
    \caption{Rear view.}
    \end{subfigure}
    \hfill
    \begin{subfigure}[b]{0.31\textwidth} 
    \begin{center}
	\begin{overpic}[height = 27 mm,trim={0mm 0mm 0mm 0mm},clip]{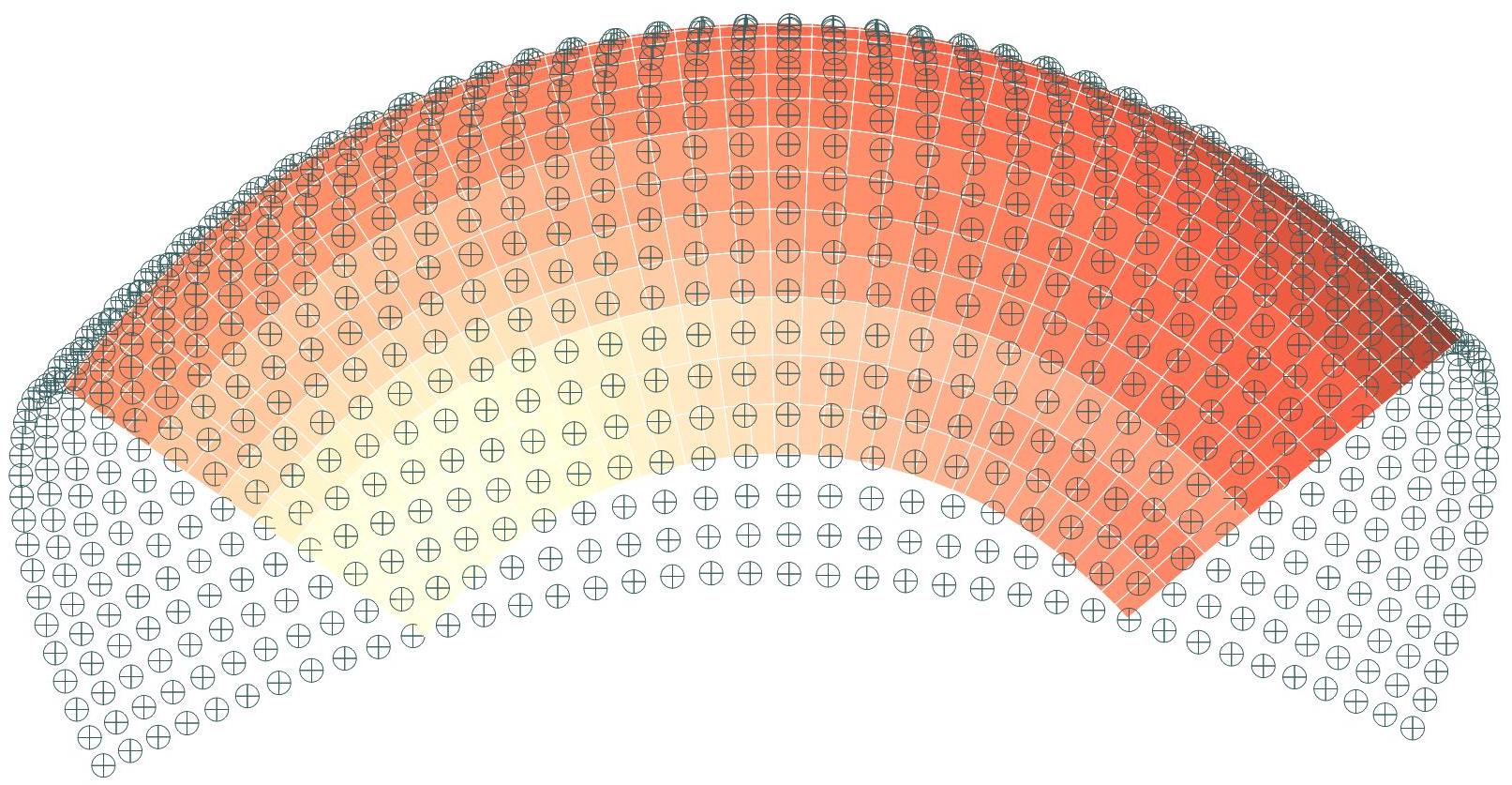}
    \end{overpic}
    \end{center}
    \caption{Top view with respect to the axis $\mathbf{a}^\ast$.}
    \end{subfigure}
	\caption{Illustration of the initial guess T-hedron with respect to the given point cloud.}
	\label{fig:mesh:initial}
\end{figure*}

Ultimately, equipped with the first profile polyline $\mathbf{p}_0$, the scaling factors $\eta_j$, and the rotation angles $\theta_j$, we possess all the elements required to construct $\mathbf{T}_j.\mathbf{S}_j.\mathbf{R}_j.\mathbf{T}_j^{-1}$ (cf.\ Eq.\ (\ref{eq:T:R:S})). 
By iteratively applying these transformations to each point of our profile polyline $\mathbf{p}_0(i)$, we can systematically construct the T-hedron. This process aligns with the principles set forth in Theorem \ref{theorem:T:hedron:III}, thereby enabling us to uniquely build our T-hedron. Fig.\ \ref{fig:mesh:initial} showcases a T-hedron mesh created using this methodology, aptly fitted to our unorganized point cloud $\mathcal{X}$.

\begin{remark}
    The availability of all the profile planes offers even further opportunities for analysis. Specifically, we can revert all the profile lines $\mathbf{p}_j$ on their respective profile planes $\delta_j$ back to the initial plane $\delta_0$. To accomplish this, we simply apply the inverse of the transformation matrix $\mathbf{T}_j.\mathbf{S}_j.\mathbf{R}_j.\mathbf{T}_j^{-1}$ to each profile line iteratively, excluding $\mathbf{p}_0$. This approach results in the superposition of all profile lines onto the plane $\delta_0$. Thus we can derive an optimal curve that best fits this amalgamation of profiles. This optimized curve can then be considered as our new $\mathbf{p}_0$. 
    \hfill $\diamond$
\end{remark}

\section{Global optimization}\label{sec:global_opt}

The intuitive plan for the global optimization is as follows. We start with the mesh $\mathbf{x_0}(i,j) := \mathbf{x}_{i,j}$ as our initial approximation of the point cloud $\mathcal{X}$. To achieve a more accurate representation of $\mathcal{X}$, we employ an optimization by using a gradient descent technique.
 Essentially, our aim is to update the vertices of our mesh so that they move ``closer" to the point cloud, while ensuring that the updated mesh remains a ``smooth enough" T-hedron. This can be accomplished by devising a cost function designed to guide the update vectors both in direction and magnitude, such that the resultant mesh conforms to the characteristics of a T-hedron.

To get what we require we resort to the method of \emph{Lagrange multipliers} with the \emph{Lagrange function} being 

\begin{equation}\label{eq:lagrange}
    {L}(\mathbf{x},\mathbf{a},\mathbf{b},\mathbf{c},\boldsymbol{\lambda^1},\boldsymbol{\lambda^2}) = F(\mathbf{x}) + \sum_{i=1}^{m-1}\sum_{j=1}^{n-1}\lambda^1(i,j)\,G^1_{i,j}(\mathbf{x}) + \sum_{i=1}^{m}\sum_{j=1}^{n}\lambda^2(i,j)\,G^2_{i,j}(\mathbf{x},\mathbf{a},\mathbf{b},\mathbf{c}),
\end{equation}
where 
\begin{itemize}
    \item $F(\mathbf{x})$ is a \emph{distance function} subject to minimization with variables stored in the vector $\mathbf{x}$,
    \item $G^1_{i,j}(\mathbf{x})$ and $G^2_{i,j}(\mathbf{x},\mathbf{a},\mathbf{b},\mathbf{c})$ are the constraints for all admissible $(i,j)$,
    \item $\lambda^1(i,j)$ and $\lambda^2(i,j)$ are the \emph{Lagrange multipliers} for all admissible $(i,j)$. 
\end{itemize}
Below, we explain each part of Eq.\ (\ref{eq:lagrange}) and the roles that these parts play.
\subsection{The cost function and the side-conditions}\label{sec:cost_side}


To update the mesh in such a way that it gets closer to the point cloud $\mathcal{X}$, we employ a projection method wherein each point $\mathbf{x}(i,j)$ of the mesh is mapped to the nearest point in $\mathcal{X}$, denoted as $\mathbf{y}(i,j)$. This proximity-based mapping leverages KD-tree algorithms, as detailed in \cite{muja_flann_2009}. A KD-tree algorithm efficiently partitions an $n \times k$ dataset by recursively subdividing the $n$ points in $K$-dimensional space, forming a binary tree structure. Upon constructing a \emph{KD Tree Searcher} model, the tree is traversed to identify the nearest neighbors in the point cloud for each query point. Consequently, this approach yields a set of closest points $\mathbf{y}(i,j)$ on the point cloud $\mathcal{X}$ corresponding to each point on the mesh $\mathbf{x}(i,j)$.

\begin{remark}
    It's important to note that the aforementioned step is highly dependent on the initial mesh $\mathbf{x_0}(i,j)$. Without an appropriate initial guess, such a projection process can get infeasible quickly. Consider an example where the initial guess includes a fold in such a way that both sides of the fold would be projected onto the same region of the point cloud. This overlap illustrates how deviations in the initial guess, such as folds, can lead to complexities in aligning the mesh with the point cloud.  
\hfill $\diamond$
\end{remark}
With the above explanations we define the cost function as 
\begin{equation}\label{eq:cost}
    F(\mathbf{x}) = \frac{1}{2} \sum_{i=1}^{m} \sum_{j=1}^{n} \left|\,\mathbf{x}(i,j) - \mathbf{y}(i,j)\,\right|^2.
\end{equation}
\begin{remark}
    An alternative approach to constructing the distance function involves considering the distance between $\mathbf{x}(i,j)$ and the projection of that point onto the tangent plane at the corresponding projection point $\mathbf{y}(i,j)$ in the point cloud. Mathematically, this distance function can be expressed as:
    \begin{equation*}
        F(\mathbf{x}) = \frac{1}{2} \sum_{i=1}^{m} \sum_{j=1}^{n} \langle\,\mathbf{x}(i,j) - \mathbf{y}(i,j),\mathbf{n}(i,j)\,\rangle^2,
    \end{equation*}
    where $\mathbf{n}(i,j)$ is the normal of the tangent plane to the point cloud at $\mathbf{y}(i,j)$. \hfill $\diamond$
\end{remark}

The mesh should fit into T-hedron criteria of Definition \ref{def:Thedron}. A T-hedron consists of trapezoids, which are quads where one pair of opposite edges is parallel. Since we would like the edges which are representing the trajectory polylines to be parallel, we can introduce this condition as 
\begin{equation}\label{eq:trapezoidal:faces}
    G_1(i,j) = \left|\,\mathbf{x}(i,j+1) - \mathbf{x}(i,j)\right)\times \left(\mathbf{x}(i+1,j+1) - \mathbf{x}(i+1,j)\,\right|^2.
\end{equation}
Therefore, whenever dealing with a T-hedron, $G_1(i,j)$ for all $1\leq i \leq m-1$ and $1\leq i \leq n-1$ should be zero.

The condition of planarity of the profile polylines can be incorporated in the following way. We write the implicit equation of the profile planes $\delta_j$ of the mesh $\mathbf{x}(i,j) = (x(i,j),y(i,j),z(i,j))^T$ for $j = 1,\ldots,n$ in the form 
\begin{equation}\label{eq:profile:planes}
   G_2(i,j) =  a(j)\,x(i,j) + b(j)\,y(i,j) + c(j).
\end{equation}
Whenever dealing with a T-hedron this condition should vanish. Note that, the planes characterized by $a(j)$, $b(j)$ and $c(j)$ are all orthogonal to the $xy$-plane.

\subsection{The gradient descent approach}\label{sec:gradient:descent}

The practical approach in obtaining a local minimum of the \emph{Lagrange function} of Eq.\ (\ref{eq:lagrange}) is to resort to \emph{gradient descent method}. The aforementioned method is a  first-order iterative optimization algorithm for finding a local minimum of a differentiable function. The idea is to take repeated steps in the opposite direction of the gradient (or approximate gradient) of the function at the current point, because this is the direction of steepest descent.

With this 
observation in mind, one starts with an initial guess $\mathbf{v_0}$ of a local minimum of $L$ where $\mathbf{v_{k}}$ is defined as:
\begin{equation*}
\mathbf{v_{k}} := (\mathbf{x_k},\mathbf{y_k},\mathbf{a_{k}},\mathbf{b_{k}},\mathbf{c_{k}},\boldsymbol{\lambda_{k}^1},\boldsymbol{\lambda_{k}^2}).    
\end{equation*}
Then we consider the sequence $\mathbf{v_0}$, $\mathbf{v_1}$, $\mathbf{v_2}$, ... which is obtained by the following procedure:
\begin{equation*}
      \mathbf{\hat{v}_{k}} := \mathbf{v_{k-1}} - \alpha\,\nabla L(\mathbf{v_{k-1}}),
\end{equation*}
with $\nabla L(\mathbf{v_{k-1}})$ being the gradient of the cost function $L$ at the point $\mathbf{v_{k-1}}$ and $\alpha \in \mathbb{R}^+$ being a small enough \emph{learning rate}. 
In general $\mathbf{\hat{x}_{k}}$ is not a T-hedron anymore due to the first order approximation. Therefore, in a final step $\mathbf{x_{k}}$ has to be obtained from $\mathbf{\hat{x}_{k}}$ by back-projection onto the set of T-hedra, which will be explained in detail in Section \ref{sec:backproj}. 
 From $\mathbf{x_{k}}$ we can derive the data $\mathbf{y_{k}}$ of $\mathbf{v_{k}}$. Moreover, we set  $\mathbf{a_{k}}:=\mathbf{\hat{a}_{k}}$, $\mathbf{b_{k}}:=\mathbf{\hat{b}_{k}}$ and $\mathbf{c_{k}}:=\mathbf{\hat{c}_{k}}$.   The determination of the values for the Lagrange multipliers 
$\boldsymbol{\lambda_{k}^1}$ and $\boldsymbol{\lambda_{k}^2}$ is discussed in Section \ref{sec:lagrange_mult}.

In the end, we would expect to get a monotone sequence $L(\mathbf{v_0})\geq L(\mathbf{v_1}) \geq L(\mathbf{v_2}) \geq ....$ which would eventually converge to a desired local minimum denoted by $\mathbf{v}^*$. 

\begin{remark}
    During the descent of $L$, there may be instances where, at a certain iteration $k$, we encounter $L(\mathbf{v_{k-1}}) \leq L(\mathbf{v_{k}})$. Given that the learning rate may vary from one iteration to the next, a practical response in such situations is to halve the learning rate and attempt a second iteration. This adjustment can be repeated until the sequence resumes a monotonically descending trajectory, or the algorithm may choose to terminate, concluding that the $(k-1)$-th step represented a local minimum.
    \hfill $\diamond$
\end{remark}

\subsubsection{Determination of the Lagrange multipliers}\label{sec:lagrange_mult}

For the completion of $\mathbf{v_k}$
we compute the Lagrange multipliers 
$\boldsymbol{\lambda_{k}^1}$ and $\boldsymbol{\lambda_{k}^2}$ in the following way:

To form the \emph{Lagrange system}, we take derivatives of $L$ with respect to $\mathbf{v}$ and substitute $\mathbf{v_k}$ into the resulting expressions, thereby obtaining an \emph{overdetermined linear system} in the variables $\boldsymbol{\lambda_{k}^1}$ and $\boldsymbol{\lambda_{k}^2}$.

To find solutions for this overdetermined system we can resort to the \emph{least square method}. Using the least squares method, we aim to approximate a solution that minimizes the sum of the squares of the residuals (i.e. the differences between the left and right sides of the equations). Our system, represented as
\begin{equation*}
\mathbf{A}\boldsymbol{\Lambda} = \mathbf{B},\quad\quad \text{with}\quad\quad \boldsymbol{\Lambda} = (\boldsymbol{\lambda_{k}^1},\boldsymbol{\lambda_{k}^2})^T,
\end{equation*}
involves $\mathbf{A}$ being an $M \times N$ matrix and $\mathbf{B}$ being an $M \times 1$ vector, where $M > N$ (Remembering that $m\times n$ was the size of the mesh, $M = 4mn + 2n -m +1$ and $N = mn + 2n - m +1$). The least squares solution seeks to minimize the residual norm $\left| \mathbf{A}\boldsymbol{\Lambda} - \mathbf{B}\right|^2$. By treating this residual norm as a cost function, the least squares method identifies the vector $\boldsymbol{\Lambda}$ that minimizes this cost function.

\begin{remark}
    Alternative methods, such as the pseudo-inverse, also exist for solving the aforementioned overdetermined system. The pseudo-inverse approach explicitly computes the solution using the formula $\boldsymbol{\Lambda} = (\mathbf{A}^T\mathbf{A})^{-1}\mathbf{A}^T\mathbf{B}$. However, this method is generally less numerically stable and is not typically recommended compared to approaches based on minimizing a cost function (for details see \cite{trefethen2022numerical}). 
    \hfill $\diamond$
\end{remark}


\subsubsection{Back-projection onto the set of T-hedra}\label{sec:backproj}

In this section we explain the back-projection of $\mathbf{\hat{x}_{k}}$ 
to the set of T-hedra. By this process we obtained the T-hedron $\mathbf{x_{k}}$. 
The set of profile planes $\delta_j$ ($j=0,\ldots ,n$) of $\mathbf{x_{k}}$ are already determined by the relations $\mathbf{a_{k}}:=\mathbf{\hat{a}_{k}}$, $\mathbf{b_{k}}:=\mathbf{\hat{b}_{k}}$ and $\mathbf{c_{k}}:=\mathbf{\hat{c}_{k}}$.
Therefore, we can project the respective profile polylines of $\mathbf{\hat{x}_{k}}$ orthogonally into these planes, which yields  $\mathbf{\hat{x}_{k}^{\perp}}$. These points are already contained in two set of orthogonal profile and trajectory planes, but the faces do not meet the planarity condition in general. Therefore we apply the following procedure. 

We map the plane $\delta_j$ back the plane $\delta_{j-1}$ by applying a stretch-rotation $\sigma_j^{-1}$  around the intersection line of  $\delta_j$ and $\delta_{j-1}$ about the angle $-\theta_j$ according to the notation of Theorem \ref{theorem:T:hedron:III}. The scaling factor $\eta_{j}^{-1}$ is computed in a way that the the sum of the squared distances of corresponding points of  $\sigma_j^{-1}(\mathbf{p}_j^{\perp})$ and $\mathbf{p}_{j-1}^{\perp}$ is minimized. 
In this way we can iteratively stretch-rotate back all profile polylines $\mathbf{p}_{j}^{\perp}$ to $\delta_0$, where the new profile $\mathbf{p}_{0}$ is computed as the polyline of $(m+1)$ barycenters of $(n+1)$ sets of corresponding points. By applying the sequence of stretch-rotations $\sigma_j$ to $\mathbf{p}_{0}$ we construct the aimed set of points $\mathbf{x_{k}}$.

\begin{figure*}[t!]
	\centering
	\begin{subfigure}[b]{0.24\textwidth} 
        \centering
		\includegraphics[height=31mm, trim={0mm 0mm 0mm 0mm}, clip]{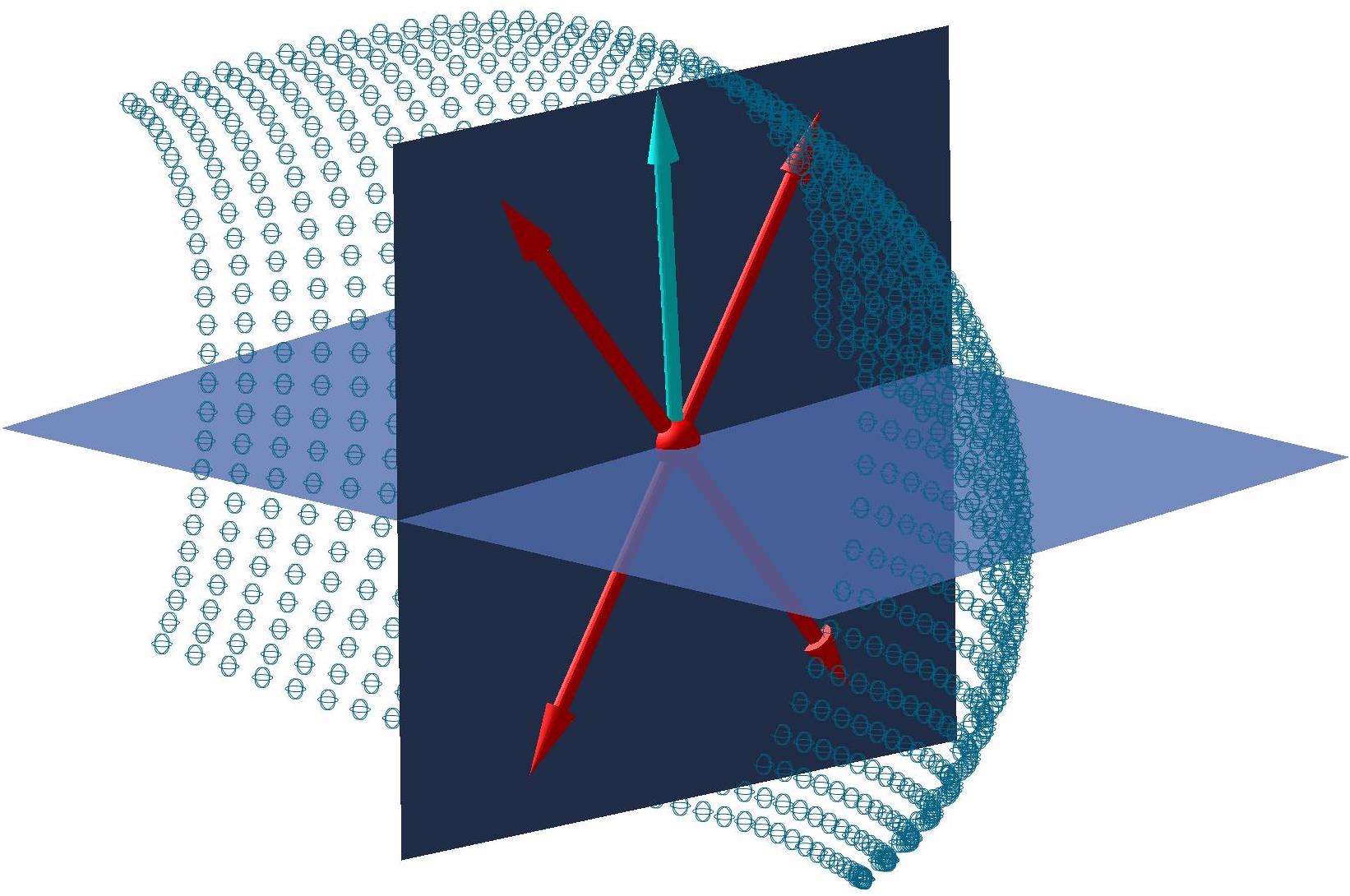}
		\caption{Symmetry planes and the axis.}
	\end{subfigure}
	\hfill
	\begin{subfigure}[b]{0.24\textwidth} 
        \centering
		\includegraphics[height=33mm, trim={0mm 0mm 0mm 0mm}, clip]{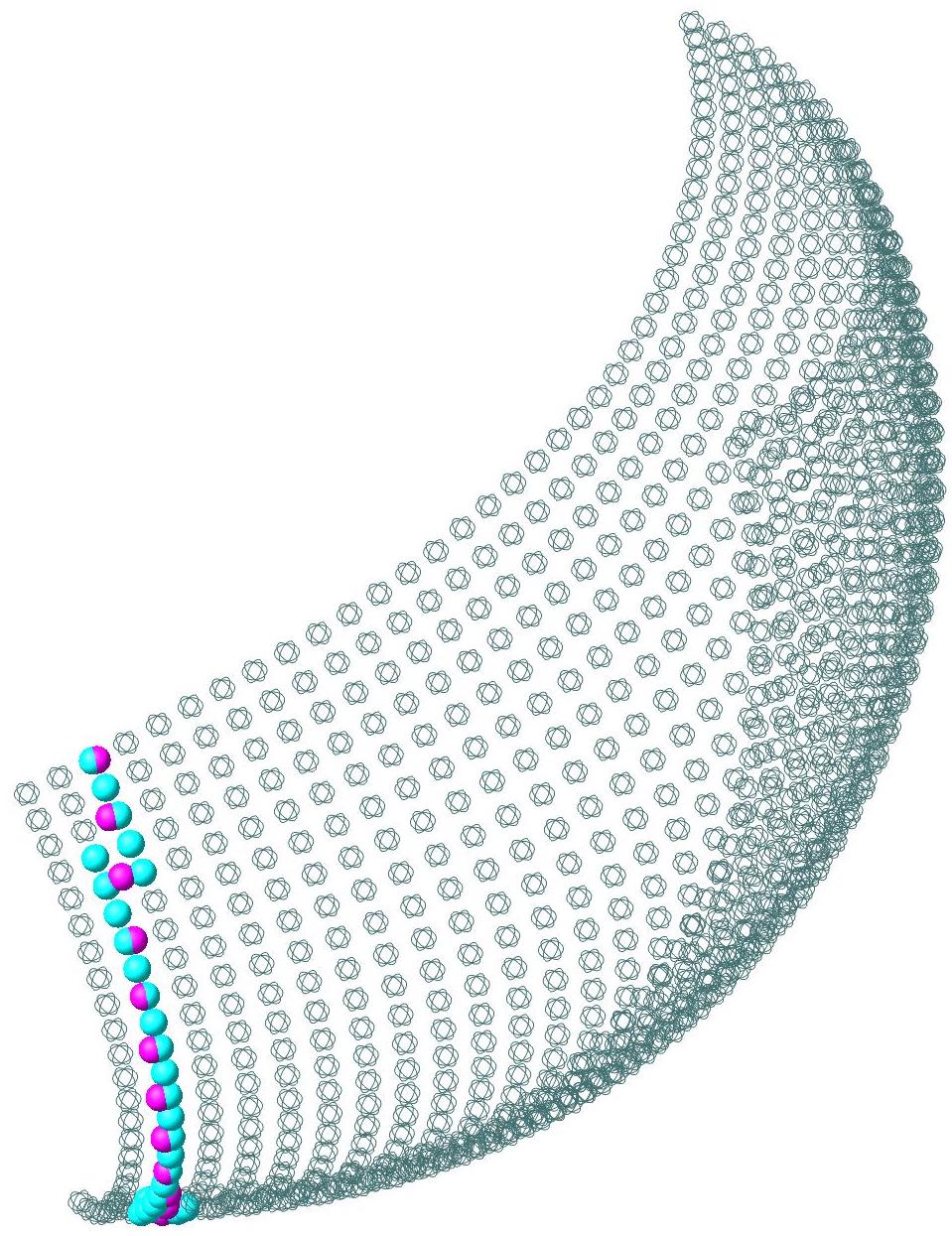}
		\caption{Profile cut.}
	\end{subfigure}
    \hfill
    \begin{subfigure}[b]{0.24\textwidth} 
        \centering
		\includegraphics[height=33mm, trim={0mm 0mm 0mm 0mm}, clip]{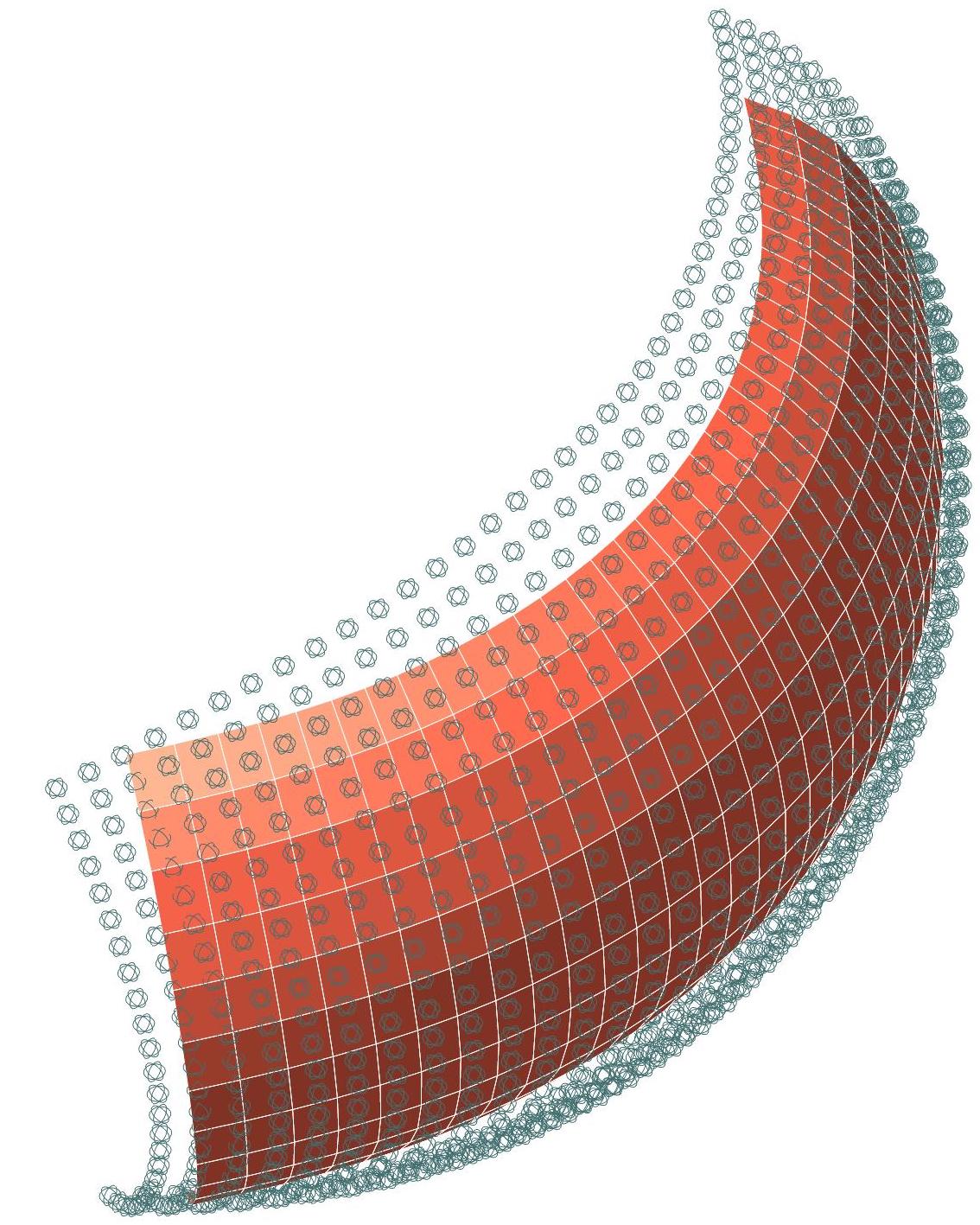}
		\caption{Initial guess.}
	\end{subfigure}
    \hfill
    \begin{subfigure}[b]{0.24\textwidth} 
        \centering
		\includegraphics[height=33mm, trim={0mm 0mm 0mm 0mm}, clip]{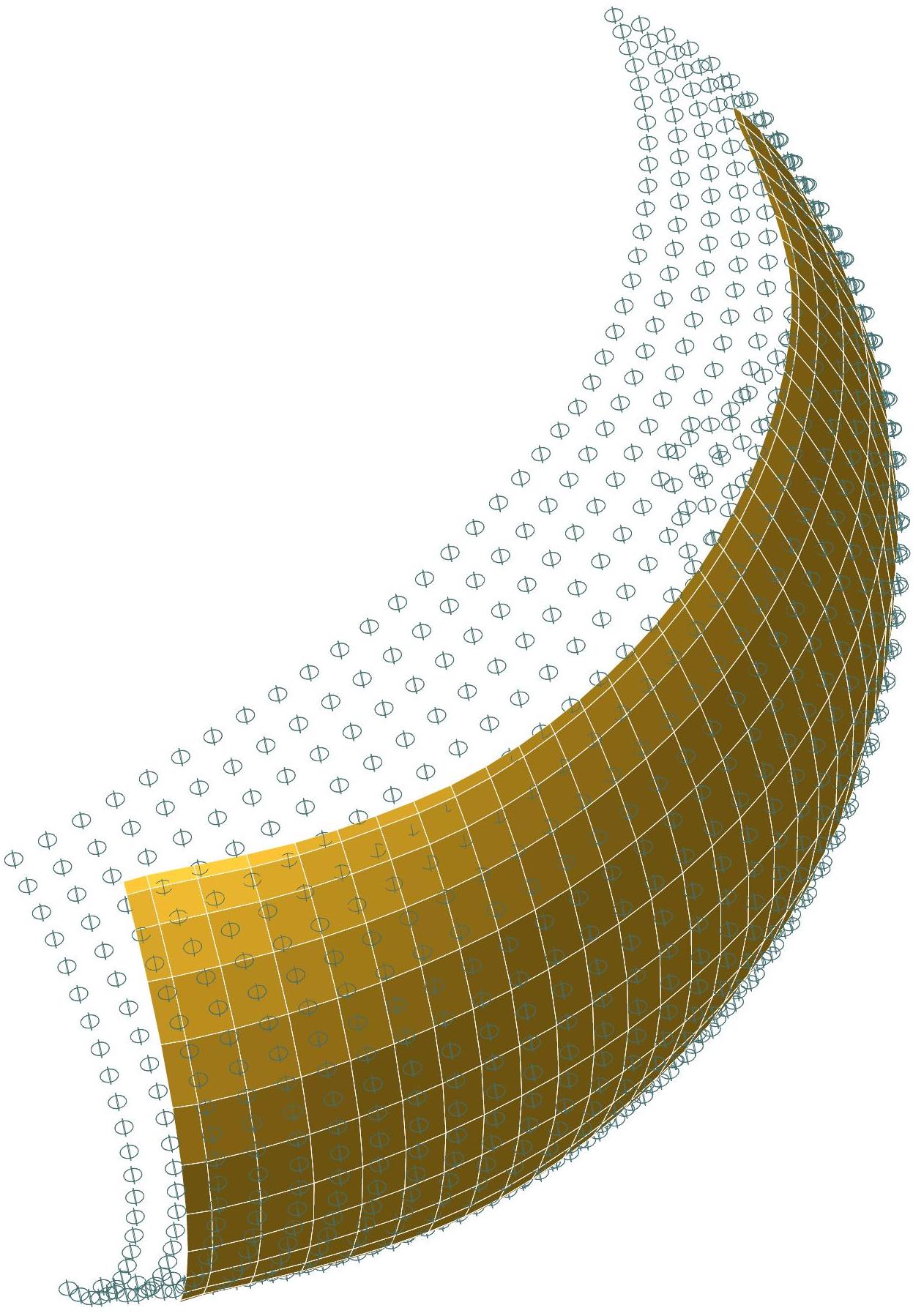}
		\caption{Final mesh.}
	\end{subfigure}
	\caption{Depiction of an optimization scenario with an initial inaccuracy. (a) Selection of a new axis (indicated by the cyan arrow). (b) Acquisition of the profile polyline through a sectional cut of the point cloud. (c) Construction of the initial guess, which does not completely align with the point cloud. (d) Creation of the final T-hedron using optimization technique described in Sections \ref{sec:cost_side} and \ref{sec:gradient:descent}.} 
	\label{fig:OBJ:axis:II}
\end{figure*}
\begin{example}
    To illustrate the effectiveness of our optimization approach, let us revisit Example \ref{example:axis}, but with an initially inaccurately chosen axis. Instead of using $\mathbf{a}^{\ast}$ (represented by one of the red arrows in Fig.\ \ref{fig:OBJ:axis}-(b)), we consider an alternative axis\footnote{This axis direction keeps up the symmetry of the point cloud during the T-hedral isometric deformation like the original surface of \cite[Fig.\ 11]{jiang}.}, depicted in cyan in Fig.\ \ref{fig:OBJ:axis:II}-(a).  
    Following the procedure outlined in our paper, we derive the directrix curve from a certain $\beta$-path. This step leads to the determination of the profile planes, which, when intersected with the point cloud, yield the profile polylines. Fig.\ \ref{fig:OBJ:axis:II}-(b) displays the unorganized point cloud cut (in cyan) alongside the corresponding organized profile polyline (in magenta).

    Utilizing the tools at our disposal and computing the scaling factors, we generate our first initial guess, as shown in Fig.\ \ref{fig:OBJ:axis:II}-(c). This initial mesh, however, does not properly align with the point cloud. Apart from its first and last profiles, none of the mesh's profile polylines coincide with the point cloud. At this juncture, we apply the optimization technique described in Sections \ref{sec:cost_side} and \ref{sec:gradient:descent}. This process allows the mesh to closely approximate the point cloud while maintaining its T-hedron characteristics, as depicted in Fig.\ \ref{fig:OBJ:axis:II}-(d).
\end{example}

\subsection{Adjustment of the axis direction}

In this phase of the global optimization algorithm the direction of the axis relatively to $\mathcal{X}$ is fine-tuned (cf.\ Remark \ref{rem:adjust}) in the following way.
Let as assume that we ended up in the local  minimum  $\mathbf{v}^*$ by applying the gradient descent approach of Section \ref{sec:gradient:descent}. 
Further let us compute $\mathbf{y}^*$ with respect to $\mathbf{x}^*$.
Now we can try to improve the cost function $F(\mathbf{x}^*)$ given in Eq.\ (\ref{eq:cost}) solving the registration problem of the point sets $\mathbf{x}^*$ and $\mathbf{y}^*$ with known correspondences, which can be solved explicitly in the following two steps:
\begin{enumerate}[$\bullet$]
    \item 
    We translate $\mathcal{X}$ in a way that the barycenter of $\mathbf{y}^*$ coincides with the  barycenter of $\mathbf{x}^*$. 
    \item 
    Then the repositioned point cloud is rotated about this barycenter, where the optimal orientation can be computed according to \cite{Horn}. 
\end{enumerate}
The output  $\mathcal{X}^*$ of this procedure is then  used together with $\mathbf{v}^*$  as new input for the gradient descent approach of Section \ref{sec:gradient:descent}. This process is iterated until the improvement of the cost function drops below a certain threshold.




\section{Results and discussion}\label{sec:results}

\begin{figure*}[t!]
    \centering
    \begin{subfigure}[b]{.32\textwidth}
        \includegraphics[height = 23mm]{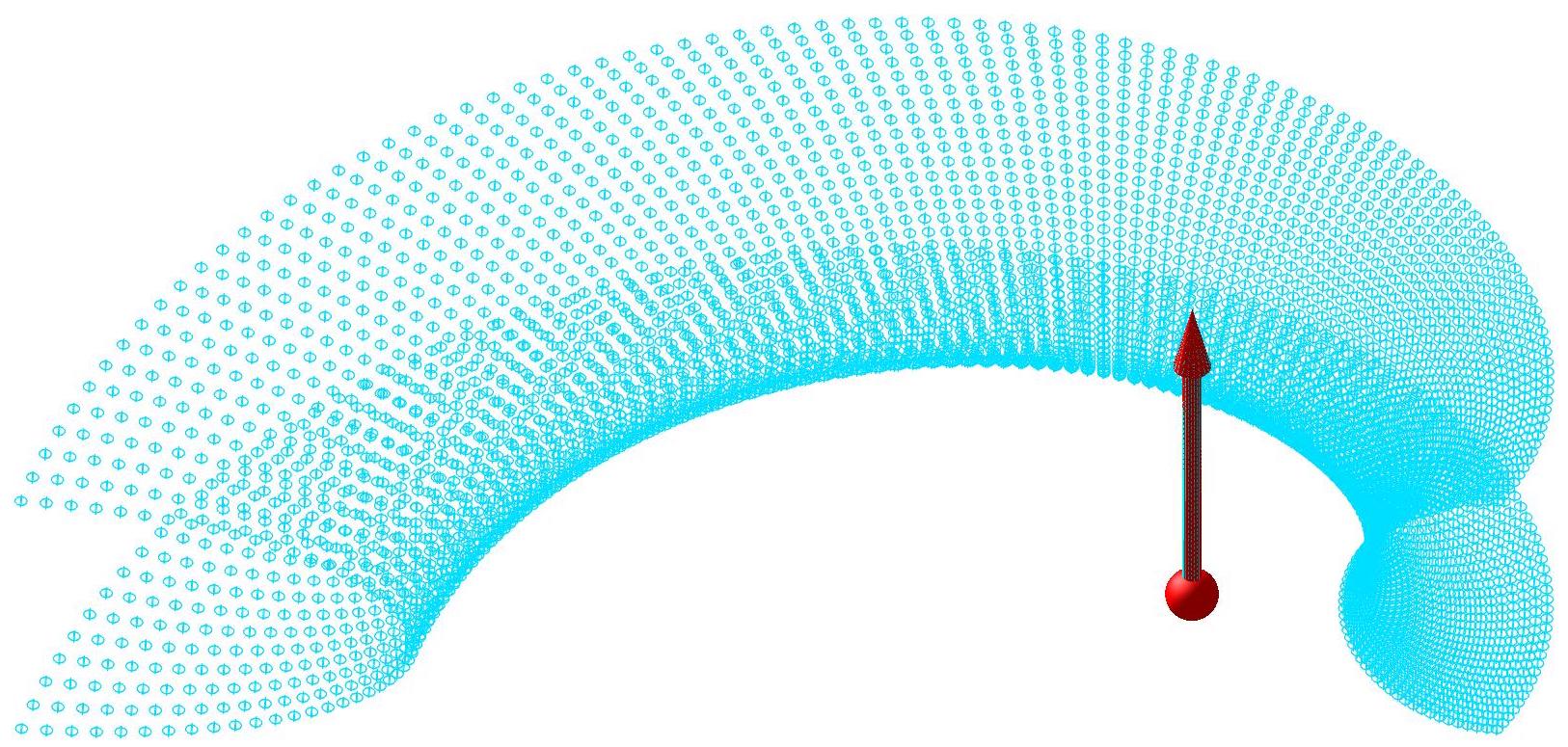}
        \caption{Axis $\mathbf{a}^\ast$ coinciding with the original axis.}
    \end{subfigure}\hfill
    \begin{subfigure}[b]{.32\textwidth}
        \includegraphics[height = 23mm]{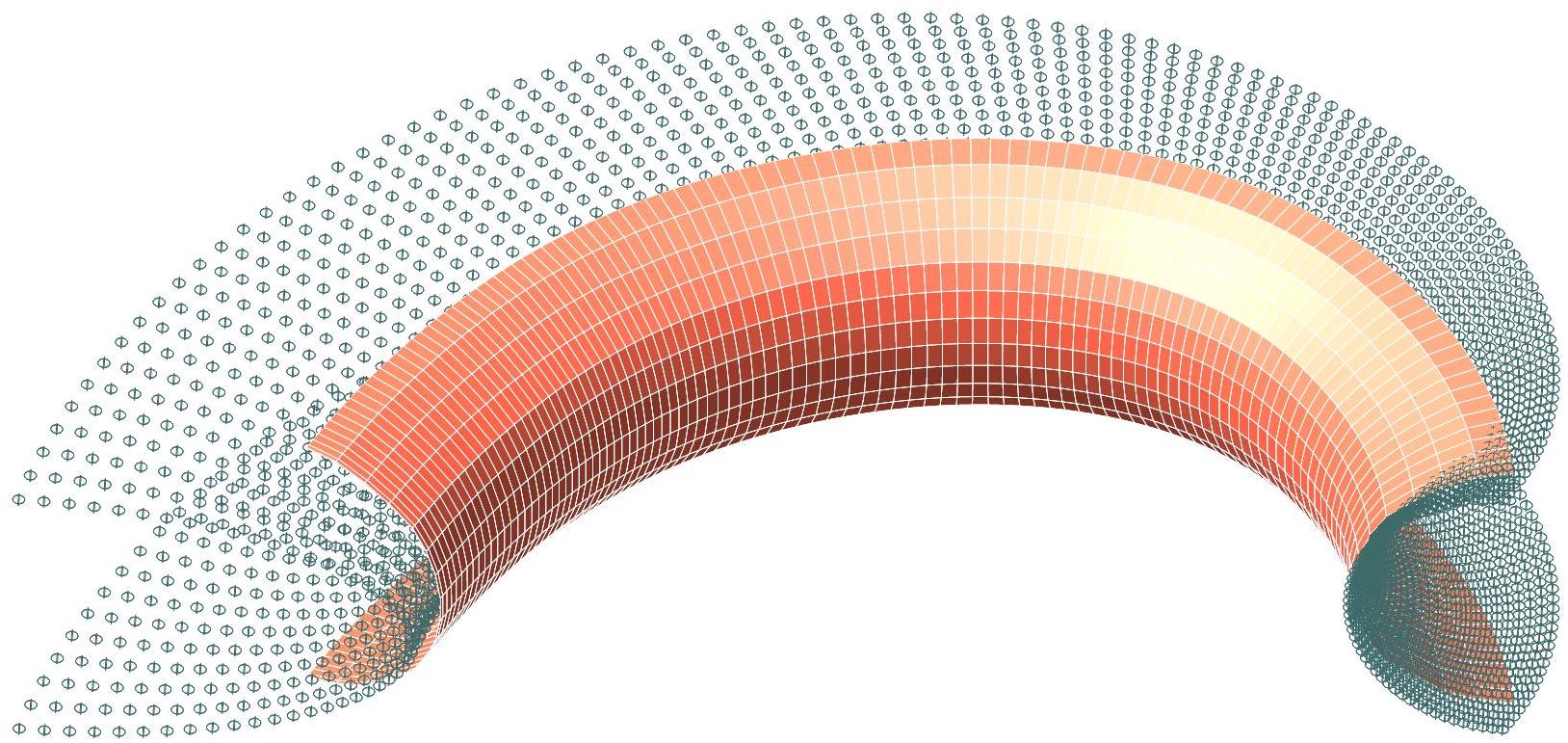}
        \caption{Initial guess.}
    \end{subfigure}\hfill
    \begin{subfigure}[b]{.32\textwidth}
        \includegraphics[height = 23mm]{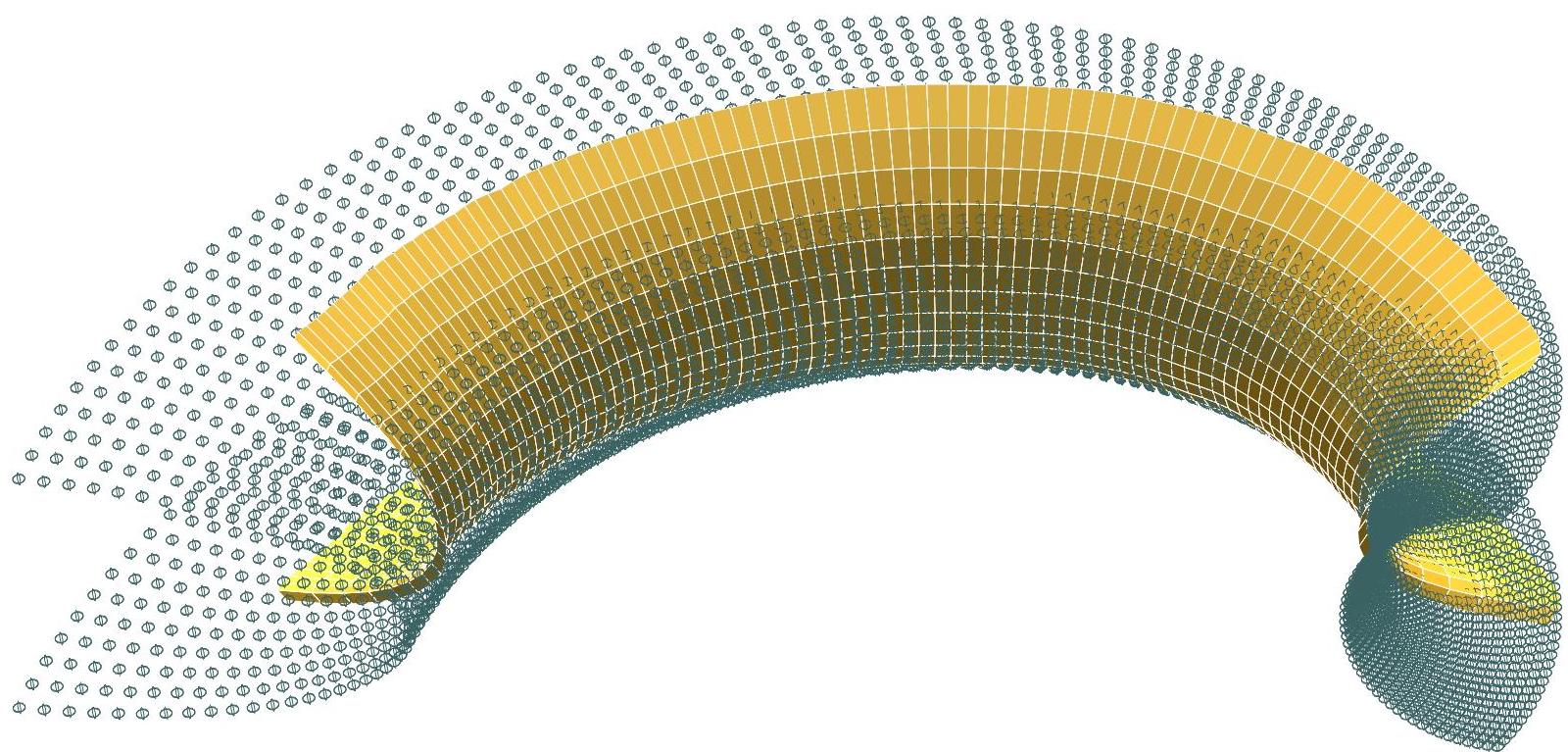}
        \caption{Final optimized mesh.}
    \end{subfigure}

    \begin{subfigure}[b]{.32\textwidth}
        \includegraphics[height = 23mm]{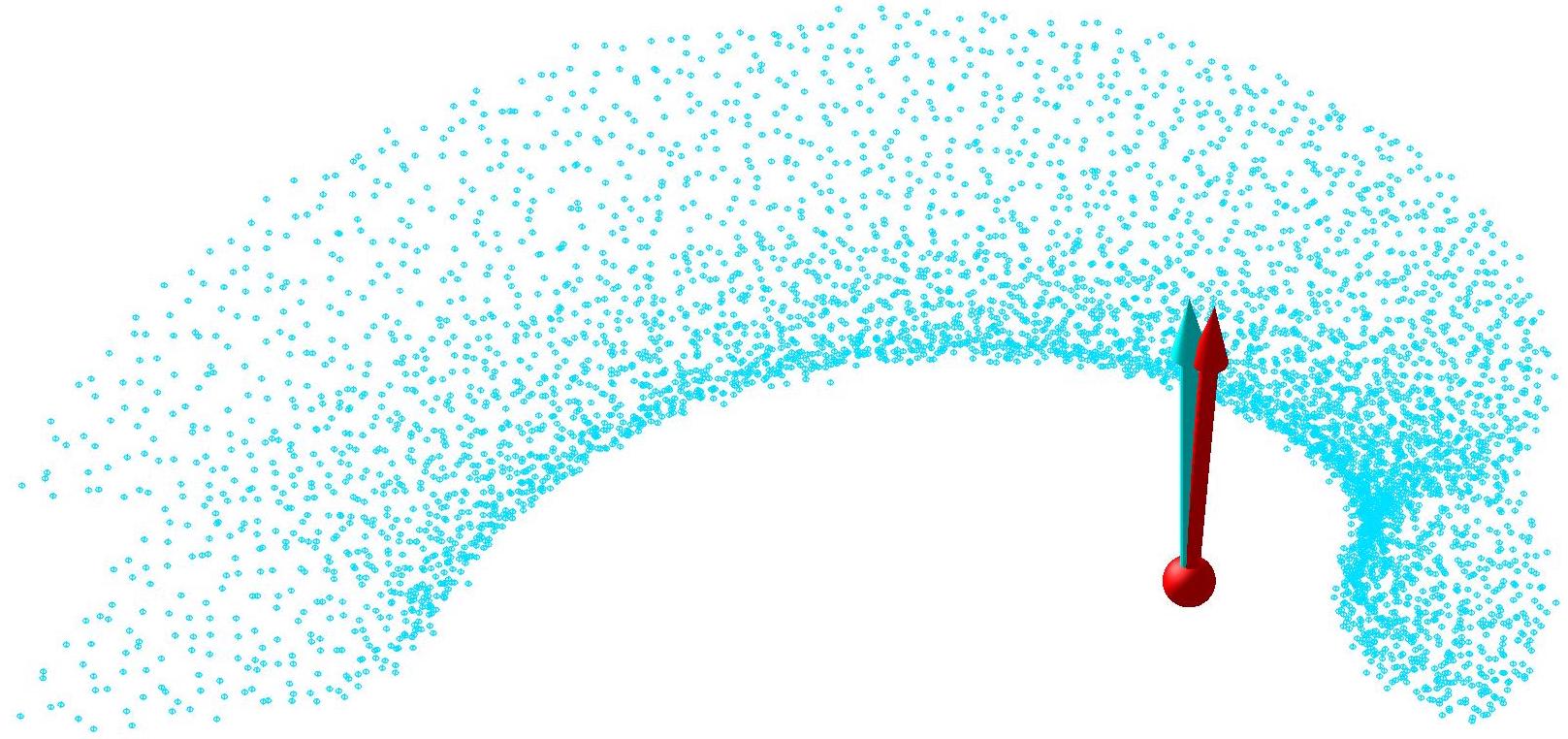}
        \caption{Axis $\mathbf{a}^\ast$ versus the original axis.}
    \end{subfigure}\hfill
    \begin{subfigure}[b]{.32\textwidth}
        \includegraphics[height = 25mm]{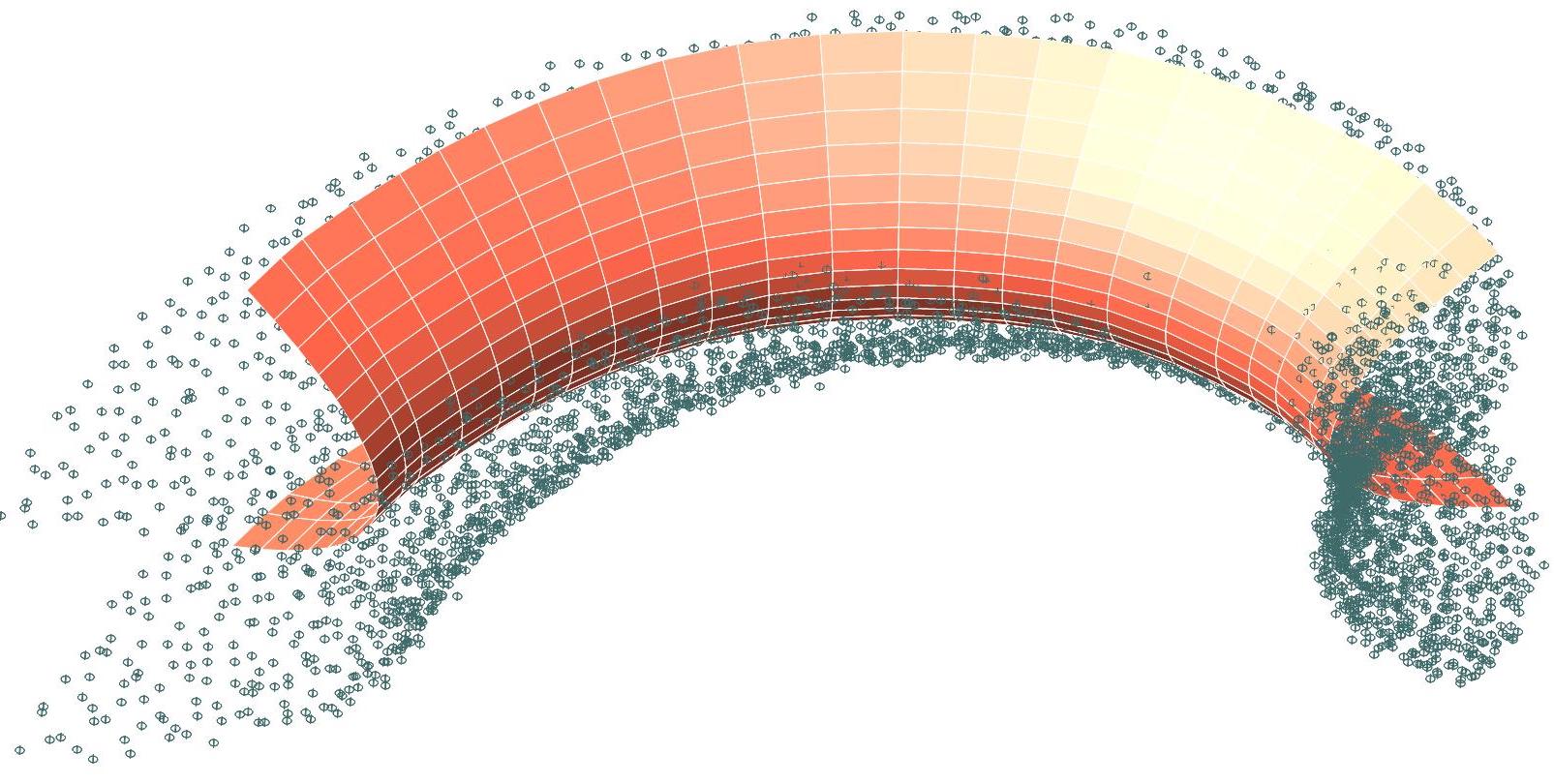}
        \caption{Initial guess.}
    \end{subfigure}\hfill
    \begin{subfigure}[b]{.32\textwidth}
        \includegraphics[height = 25mm]{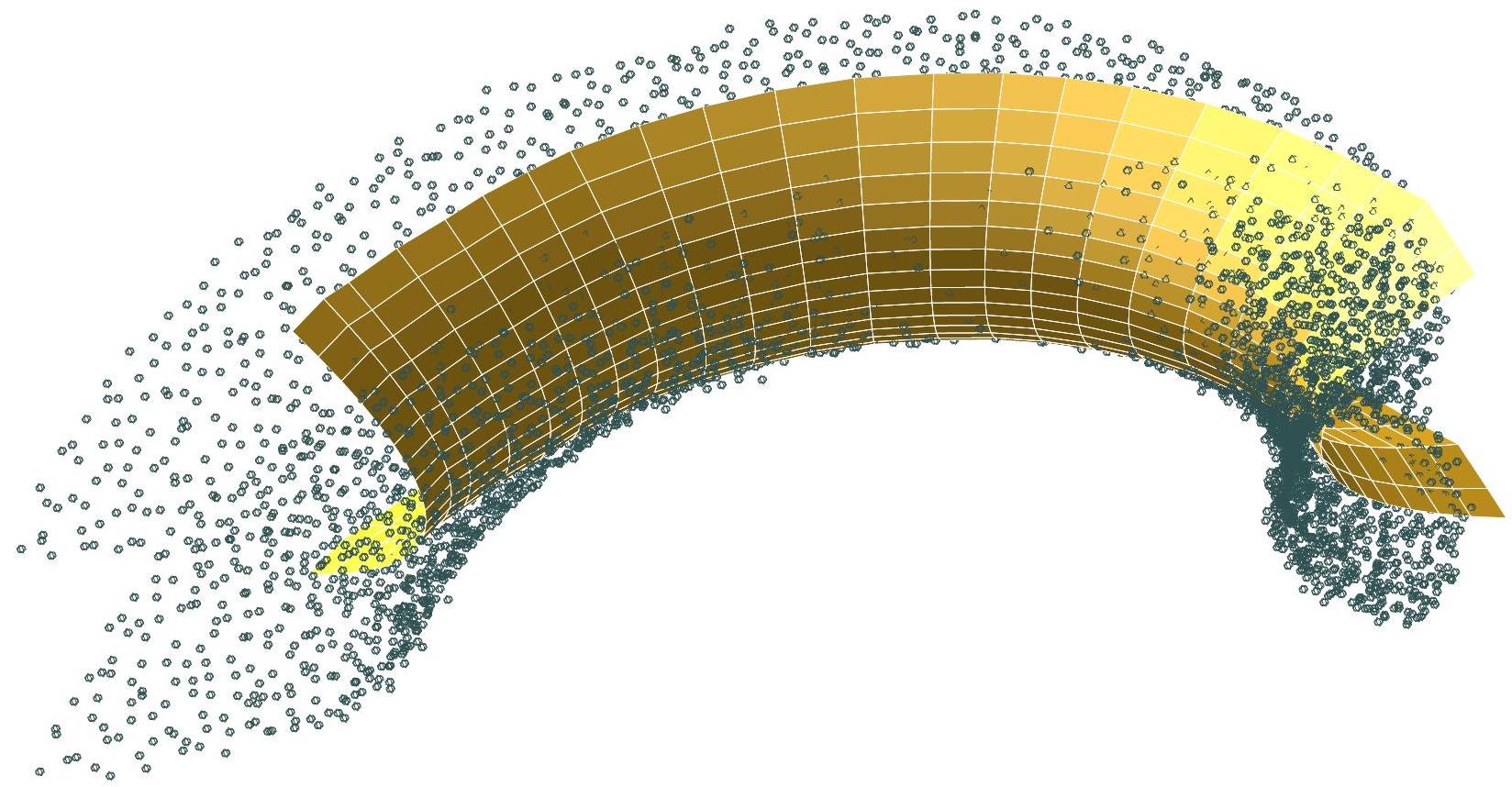}
        \caption{Final optimized mesh.}
    \end{subfigure}

    \begin{subfigure}[b]{.32\textwidth}
        \includegraphics[height = 27mm]{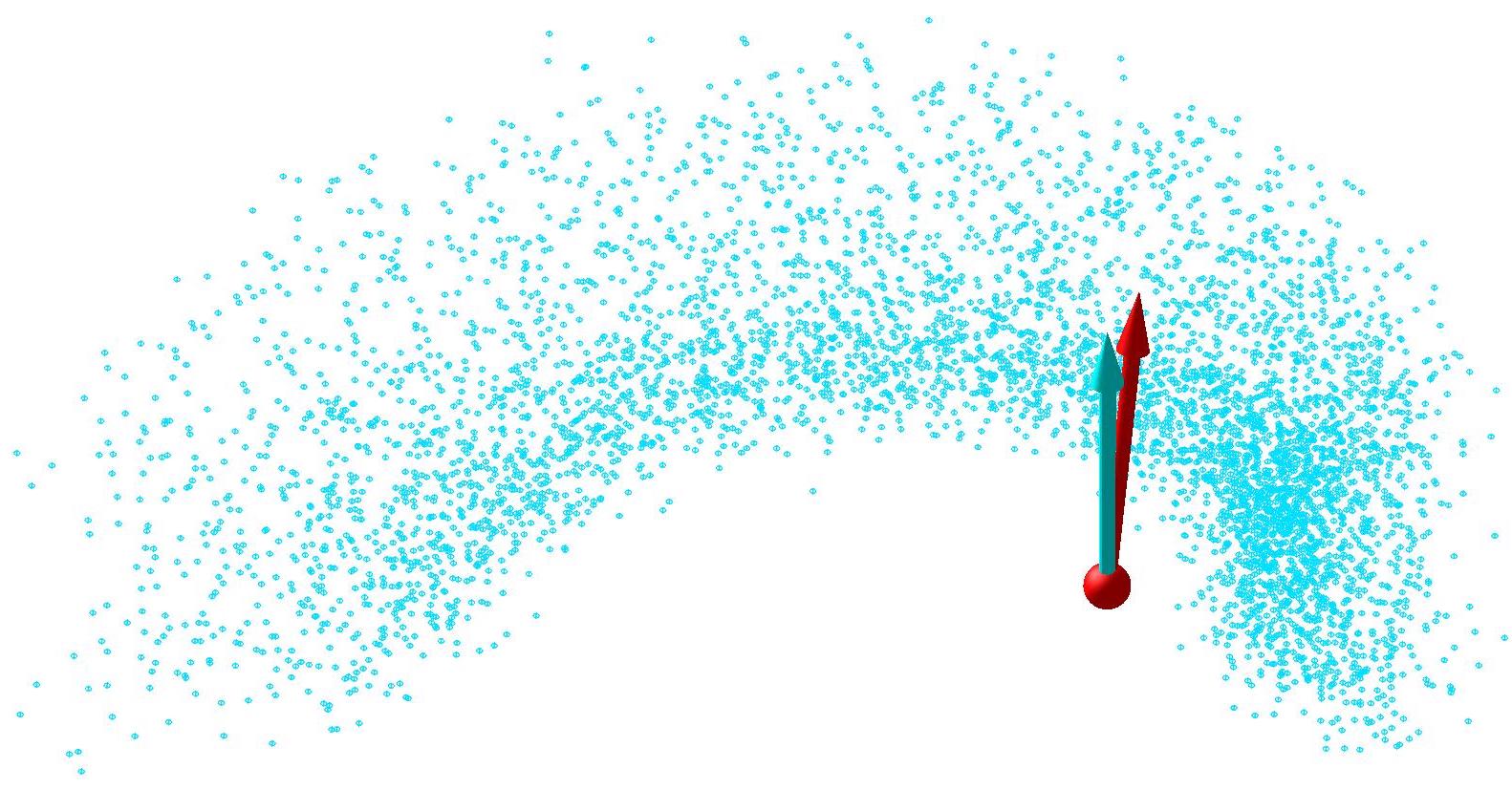}
        \caption{Axis $\mathbf{a}^\ast$ versus the original axis.}
    \end{subfigure}\hfill
    \begin{subfigure}[b]{.32\textwidth}
        \includegraphics[height = 28mm]{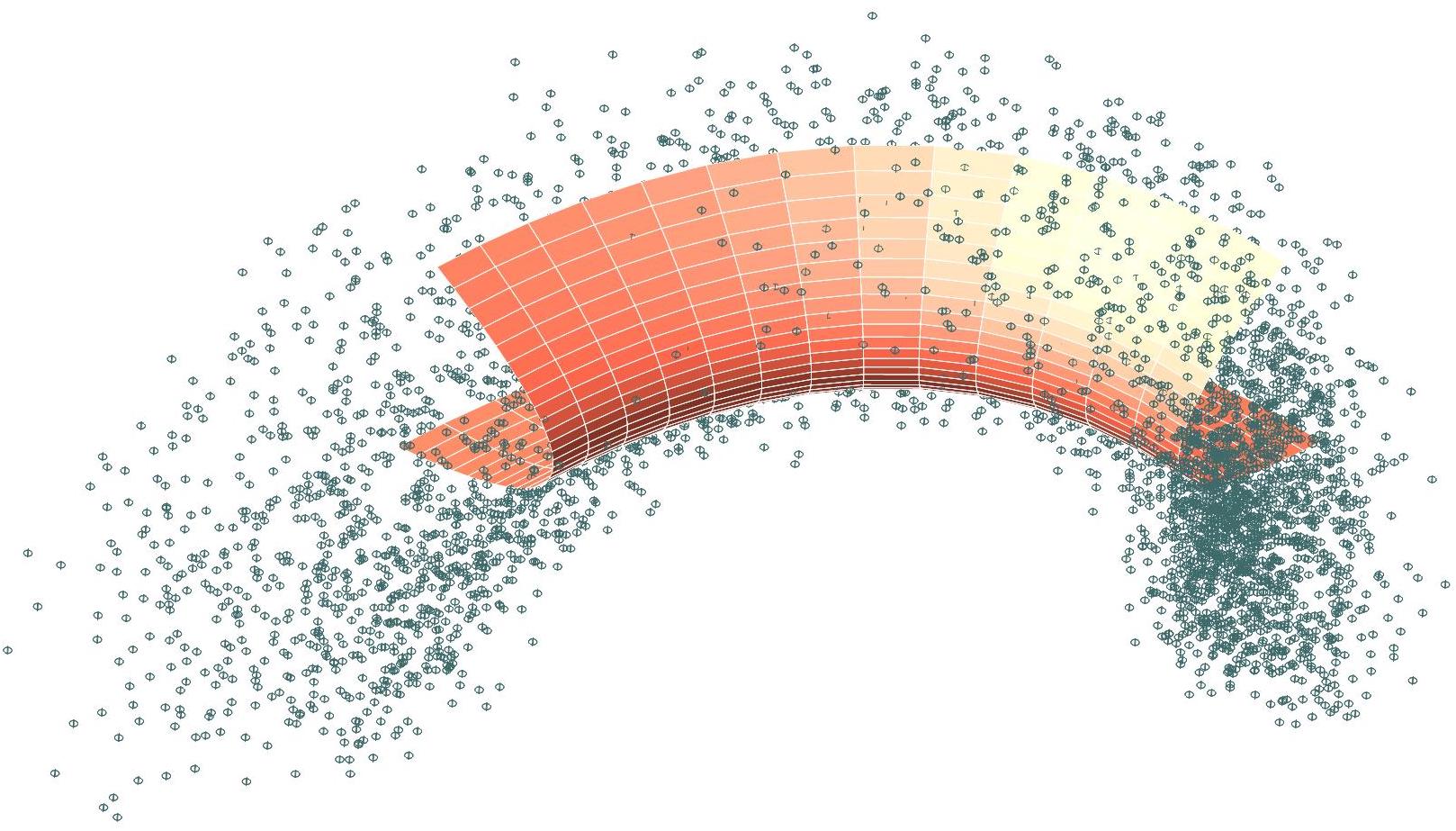}
        \caption{Initial guess.}
    \end{subfigure}\hfill
    \begin{subfigure}[b]{.32\textwidth}
        \includegraphics[height = 28mm]{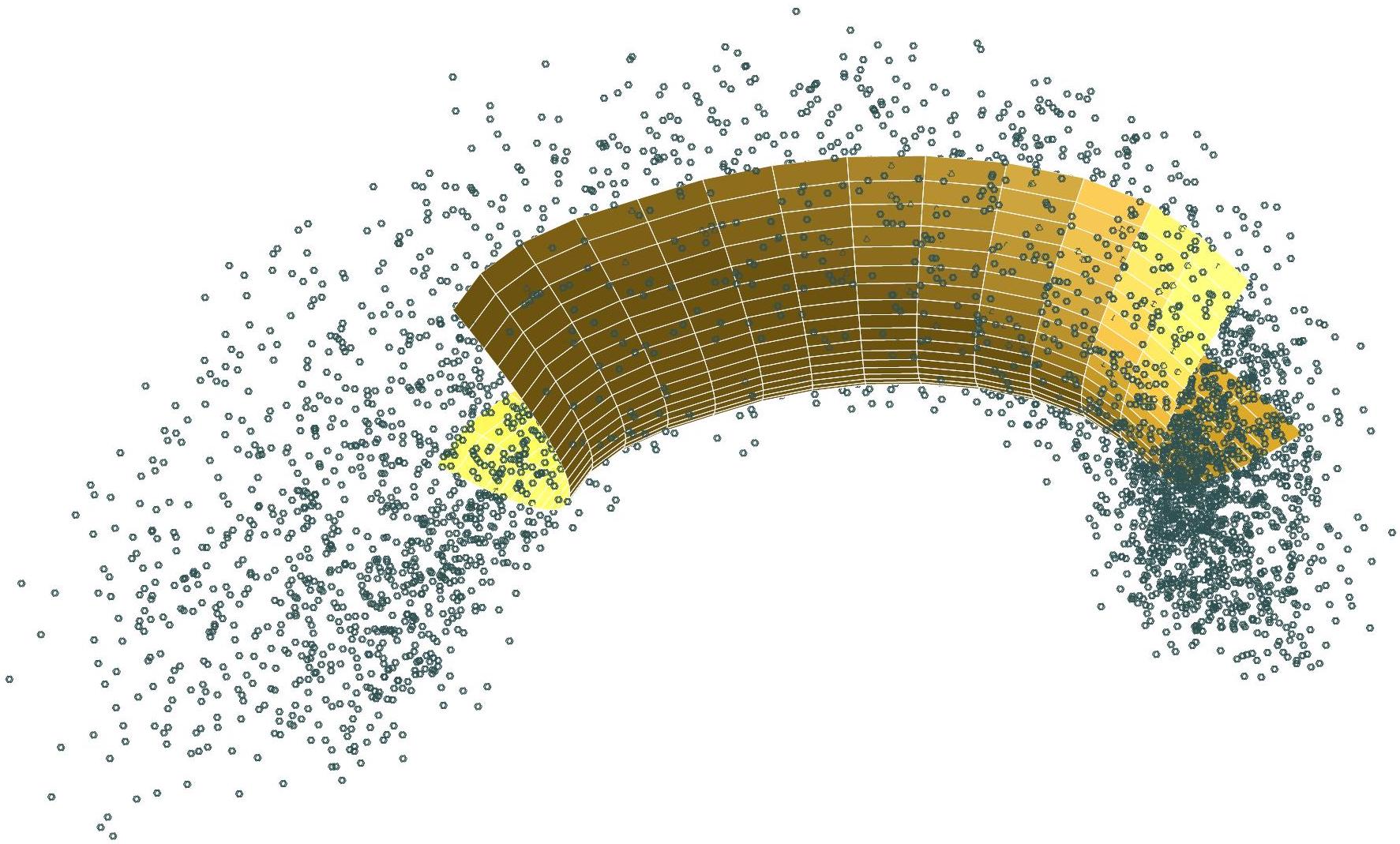}
        \caption{Final optimized mesh.}
    \end{subfigure}

    \caption{Visualization of T-surface point cloud (with everywhere negative Gaussian curvature) surface reconstruction with added noise. The noise is added using a normal distribution with $0$ mean and variance of $10^{-16}$ (top row), $10^{-3}$ (middle row) and $10^{-1}$ (bottom row).
    (a,d,g) Contrasts the determined axis $\mathbf{a}^\ast$ (shown in red) against the T-surface's original axis (depicted in cyan). (b,e,h) Depicts the construction of the initial guess. (c,f,i) Presents the final optimized T-hedron. }
	\label{fig:noisy}
\end{figure*}

In Section \ref{sec:initial_guess}, we detailed the construction of the initial guess for the mesh in Example \ref{example:axis}, outlining each step of the process. In Section \ref{sec:global_opt}, we discussed the optimization procedure in cases where the axis is inaccurately identified. Another challenge often encountered in generating point clouds, either through object scanning or 3D sketching, is the presence of noise. Such irregularities are a common feature in real-world point cloud data. In this section, we aim to address this issue by demonstrating our algorithm's robustness against noise.

To illustrate this, we base our example on a smooth T-surface, as enabled by Theorem \ref{theorem:parametrization:type:III}, which offers a framework for constructing a variety of examples. We choose a case where the scaling factor is $\xi(t) = 1 + t/2$, and the \emph{directrix curve} $\mathbf{d}$ is a unit circle in $xy$-plane. Additionally, we assume the first profile curve $\mathbf{p}_0$ is a half-circle chosen in such a way that the T-surface would have everywhere negative Gaussian curvature. We construct the surface following the parametrization in Eq.\ (\ref{eq:Tsurface:general:case}) with the real axis of the T-surface aligned with the $z$-axis. To simulate real-world conditions, we introduce noise into the point cloud by adding a random matrix with entries following a normal distribution, mean $0$, and variances of different values of $10^{-16}$ (Fig.\ \ref{fig:noisy} top row), $10^{-3}$ (Fig.\ \ref{fig:noisy} middle row) and $10^{-1}$ (Fig.\ \ref{fig:noisy} bottom row).

Upon running the algorithm, we obtain an axis that is slightly different from the original whenever the noise is substantial. It is noteworthy that in the absence of the noise, the obtained axis is very closely aligned with the T-surface axis. Table\ \ref{table:comparisson} gives a comparative analysis of the angular deviation of the found axes with the positive direction of the $z$-axis.
\begin{table}[b]
\centering
\begin{tabular}{ |p{0.35\columnwidth} p{0.15\columnwidth} p{0.15\textwidth} p{0.15\textwidth}|}
 \hline
 \multicolumn{4}{|c|}{Comparative Analysis of Angular Deviations from the Positive \( z \)-axis} \\
 \hline
 Variance                                            & $10^{-16}$                & $10^{-3}$          & $10^{-1}$\\
 \hline
 Angular Deviation Before Elapse                     & ${10^{-13}}^\circ$        & {$4.17^\circ$}      & $13.18^\circ$\\
 \hline
 Angular Deviation After 1st Elapse                  & -                         & {$4.05^\circ$}      & {$7.59^\circ$}\\
 \hline
 Angular Deviation After 2nd Elapse                  & -                         & {$3.75^\circ$}      & {$3.61^\circ$}\\
 \hline
\end{tabular}
\caption{A summary of the angular deviations of the identified axis (illustrated in cyan in Fig. \ref{fig:noisy}) relative to the positive direction of the $z$-axis.}
\label{table:comparisson}
\end{table}
Subsequently, by slicing the point cloud parallel to this axis we can obtain the $\beta$-path. Now, by determining the profile planes, we acquire the first profile polyline. This leads to the initial guess shown in Fig.\ \ref{fig:noisy}-(b,e,h), which appears rough and engulfed within the cloud. However, after optimization, we achieve a surface that closely approximates the point cloud as pictured in Fig.\ \ref{fig:noisy}-(c,f,i).

\begin{remark}
   In scenarios with moderate noise levels, the cuts with identical $\beta$-angles may not encompass the entire range of the cut. This can lead to the mesh covering only a portion of the point cloud, as shown in Fig.\ \ref{fig:noisy}-(e). However, this issue can be substantially mitigated through subsequent optimization and adjustment of the mesh, as demonstrated in Fig.\ \ref{fig:noisy}-(f). \hfill $\diamond$
\end{remark}

\begin{remark}
    When faced with significant noise, the parallel cuts are likely to become uneven, leading to considerable variation in their lengths. Consequently, even the most optimal paths may differ substantially in length. Additionally, upon the application of the VPO technique, these paths might exhibit only partial similarity in their $\beta$-angles. Such discrepancies can lead to a reconstruction of a part of the point cloud. Fig\ \ref{fig:noisy}-(g,h,i) illustrates an instance of this phenomenon, where the variation in distortion results in only a partial covering of the mesh. 
    \hfill $\diamond$
\end{remark}

After adjusting the surface, we can recompute the angular deviation from the positive direction of $z$-axis for the resulting surface. This serves as an evaluation technique to assess how closely the reconstructed surface resembles the original point cloud. This process can be iteratively performed multiple times. Table\ \ref{table:comparisson} presents data showing how the angular deviation decreases with each optimization cycle. These evaluations are labeled as \emph{Angular Deviation After 1st Elapse} and \emph{Angular Deviation After 2nd Elapse}. However, in cases with a very low variance, such as $10^{-16}$, further iterations are irrelevant, as the initial result already closely matches the original surface.


\section{Conclusion}\label{sec:conclusion}

In this paper we presented an algorithm for the reconstruction of T-surfaces from a point cloud using methods of kinematical geometry and theoretical results on generalized evolutes and involutes, respectively.  
The given algorithm consists of two major phases; namely the determination of a first approximation of the point cloud by a T-hedron (Section \ref{sec:initial_guess}), which is used as inital guess for a global optimization (Section \ref{sec:global_opt}).
The algorithm is demonstrated and validated by several examples.

As already pointed out in Section \ref{sec:Tsurface}  our algorithm does not cover a certain sub-class of T-surfaces; namely translational surfaces. But for this special case other reconstruction methods are already available in the literature (e.g.\ an approach using active contours \cite{peternell}). 

Note that the presented algorithm can also be adapted for fitting equiform moulding surfaces, which is an interesting sub-class of the equiform profile surfaces mentioned as future research in \cite{kovacs}.

 Recently, also a generalization of T-hedra to non-planar quads was achieved in \cite{aikyn}. If smooth analogs of them exist, which is an open question, it would also be interesting to come up with an algorithm for their reconstruction.


\section*{Acknowledgement}
This work was supported by Austrian Science Fund (FWF) project F~77 (SFB ``Advanced Computational Design'').
Further thanks go out to the authors of 
\cite{jiang} for providing us the data set used in Example \ref{example:axis}.

\appendix

\section{Singularities of the $\beta$-evolutes}\label{sec:singularities}

\begin{theorem}\label{theorem:beta:evolute:singularity}
Let $\mathbf{c}: I \rightarrow \mathbb{R}^2$ be a regular arc-length parameterized curve and $\mathbf{c}_{\ast}$ be its $\beta$-evolute. Then the singular points of $\mathbf{c}_{\ast}$ correspond to the those points of $\mathbf{c}$ where its curvature equals
\begin{equation}
    \kappa = \frac{\cos(\beta)}{\int_{\,0}^{\,t}\,\sin(\beta)\,\mathrm{d}t + C} - \dot{\beta},\quad\quad\text{with}\quad\quad C = \frac{\cos(\beta)}{\kappa + \dot{\beta}}\left|_{t = 0}\right.
\end{equation}
\end{theorem}
\begin{proof}
To have a compact notation, let us identify the plane $\mathbb{R}^2$ with the complex line $\mathbb{C}$. 
Let $\dot{\mathbf{c}} = \mathbf{f}_t = \mathrm{e}^{\,\mathrm{i}\,\phi}$ and consequently let $\mathbf{f}_n = \mathrm{i}\,\mathrm{e}^{\,\mathrm{i}\,\phi}$, $\kappa = \dot{\phi}$ (we assume that the curvature is positive) and $\mathbf{R} = \mathrm{e}^{\,\mathrm{i}\,\beta}$ then we have
\begin{equation*}
    \dot{\mathbf{c}}_\ast = \mathrm{e}^{\,\mathrm{i}\,\phi} + \frac{\mathrm{d}}{\mathrm{d}t}\left[\frac{\cos(\beta)}{\dot{\phi} + \dot{\beta}}\,\mathrm{i}\,\mathrm{e}^{\,\mathrm{i}\,(\phi + \beta)}\right].
\end{equation*}
By putting $\dot{\mathbf{c}}_{\ast} = 0$ we intend to compute the points at which $\mathbf{c}_\ast$ is singular. Therefore, expanding the above equation gives
\begin{equation*}
    \mathrm{e}^{\,\mathrm{i}\,\phi} = -\frac{\mathrm{d}}{\mathrm{d}t}\left[\frac{\cos(\beta)}{\dot{\phi} + \dot{\beta}}\right]\,\mathrm{i}\,\mathrm{e}^{\,\mathrm{i}\,(\phi + \beta)} + \cos(\beta)\,\mathrm{e}^{\,\mathrm{i}\,(\phi + \beta)} \implies
    \mathrm{e}^{\,-\mathrm{i}\,\beta} = \cos(\beta) -\frac{\mathrm{d}}{\mathrm{d}t}\left[\frac{\cos(\beta)}{\dot{\phi} + \dot{\beta}}\right]\,\mathrm{i}.
\end{equation*}
Finally, expanding the last equation gives
\begin{equation*}
    \sin(\beta) = \frac{\mathrm{d}}{\mathrm{d}t}\left(\frac{\cos(\beta)}{\dot{\phi} + \dot{\beta}}\right).
\end{equation*}
The rest is obtained from integrating the sides of the above equation.
\end{proof}
It is noteworthy to mention the special cases of $\beta = 0$ and $\beta = \mathrm{const.}\neq 0$ separately.
\begin{corollary}\label{coro:singularities}
Let $\mathbf{c}: I \rightarrow \mathbb{R}^2$ be an arc-length parametrized curve and $\mathbf{c}_\ast$  be its $\beta$-evolute. Then we have
\begin{itemize}
    \item if $\beta = 0$, the singular points of $\mathbf{c}_\ast$ correspond to critical points of the curvature of $\mathbf{c}$,
    \item if $\beta = \mathrm{conts.} \neq 0$, then the singular points of $\mathbf{c}_\ast$ correspond to the points on $\mathbf{c}$ where 
    \begin{equation*}
        \dot{\left(\frac{1}{\kappa}\right)} = \tan{(\beta)} \quad\quad\implies\quad\quad \kappa = \frac{\cos({\beta})}{t\,\sin(\beta) + C\,\cos(\beta)} \quad\text{with} \quad \kappa(0) = \frac{1}{C}.
    \end{equation*}
\end{itemize}
\end{corollary}

    From a practical perspective, Theorem \ref{theorem:beta:evolute:singularity} indirectly highlights the sensitivity involved in reconstructing a $\beta$-evolute. For the sake of simplicity we explain it on the simple case of $\beta = 0$. In this case the evolute is the envelope of lines spanned by the normals of the curve $\mathbf{c}$. However, any noise in $\mathbf{c}$ may result in the curvature of the curve to increase and decrease locally resulting in $\dot{\kappa} = 0$ at a point in between. Therefore, in this case some singularities in the $\beta$-evolute appear.   

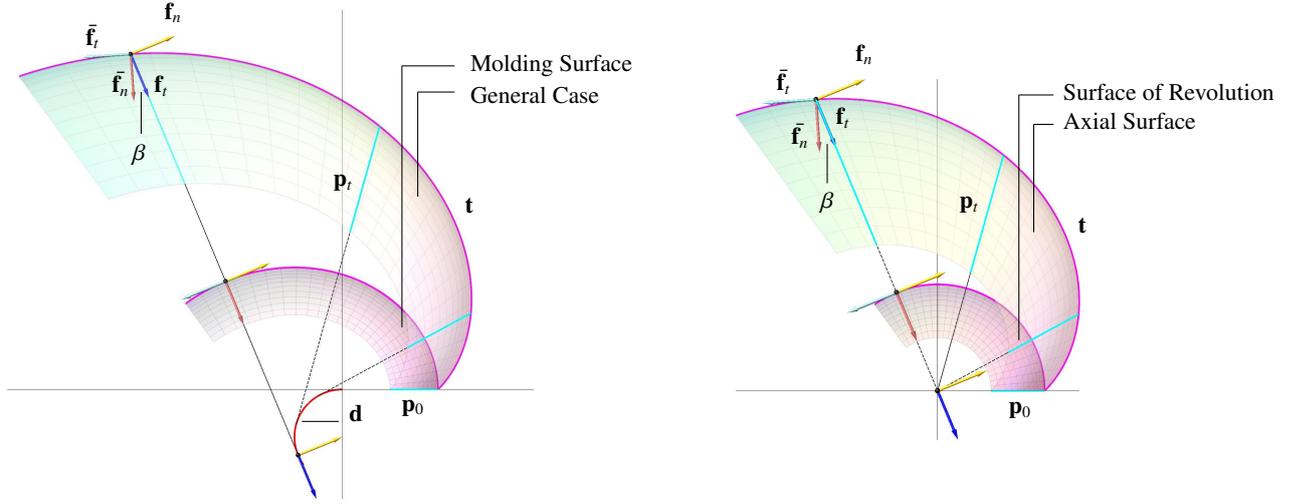
\begin{figure}[t!]
\begin{center}
 \begin{minipage}{.49\textwidth}
  \centering
  \begin{overpic}[height = 65 mm]{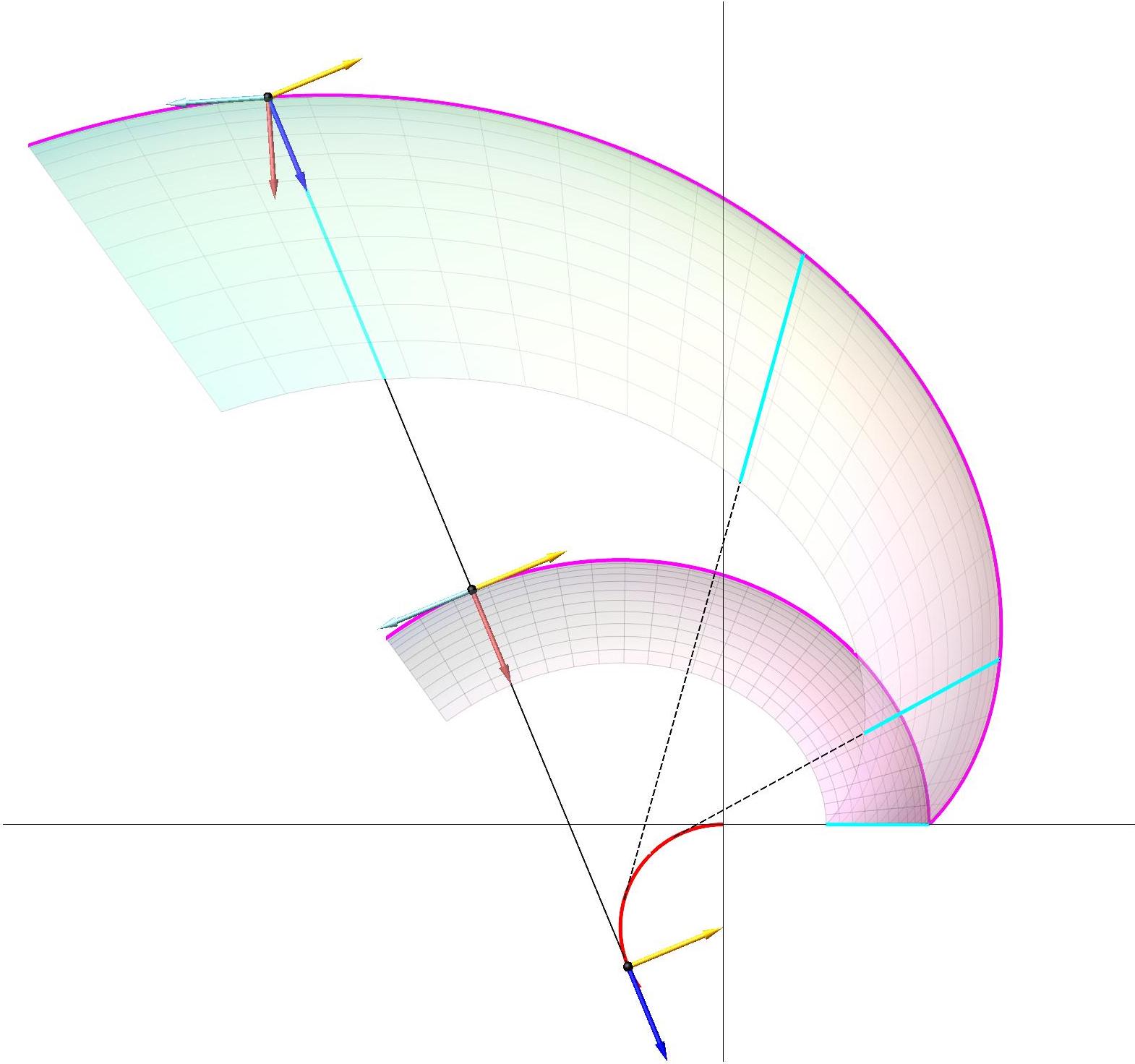}
	\begin{small}
	\put(28,78){$\mathbf{f}_t$}
	\put(30,92){$\mathbf{f}_n$}
	\put(15,87){$\Bar{\mathbf{f}_t}$}
	\put(20,78){$\Bar{\mathbf{f}_n}$}
	\put(24,65){$\beta$}
	\put(25.5,70){\color{black}\line(0,1){8}}
	\put(87,55){$\mathbf{t}$}
	\put(65,15){$\mathbf{d}$}
	\put(62,60){$\mathbf{p}_t$}
	\put(75,17){$\mathbf{p}_0$}
	\put(75,33){\color{black}\line(0,1){50}}
	\put(78,58){\color{black}\line(0,1){20}}
	\put(75,83){\color{black}\line(1,0){10}}
	\put(78,78){\color{black}\line(1,0){7}}
	\put(88,82){Molding Surface}
	\put(88,76){General Case}
	\put(63,15){\color{black}\line(-1,0){7}}
	\end{small}     
  \end{overpic}
  \end{minipage}
  \hfill
  \begin{minipage}{.49\textwidth}
  \centering
  \begin{overpic}[height = 65 mm,trim={15cm 10.5cm 15cm 10cm},clip]{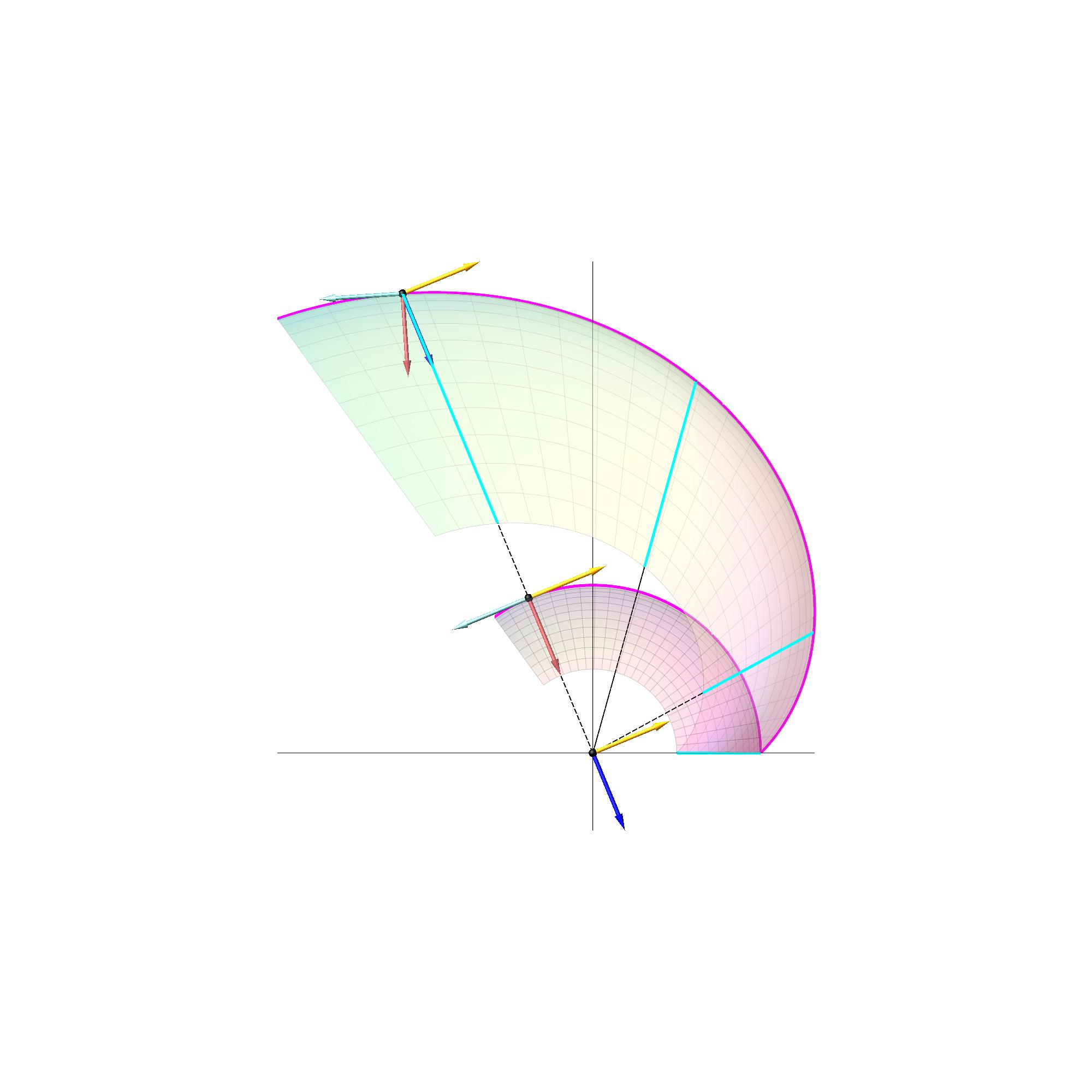}
	\begin{small}
	\put(26,77){$\mathbf{f}_t$}
	\put(30,90){$\mathbf{f}_n$}
	\put(14,84){$\Bar{\mathbf{f}_t}$}
	\put(17,73){$\Bar{\mathbf{f}_n}$}
	\put(23,60){$\beta$}
	\put(24.2,65){\color{black}\line(0,1){8}}
	\put(75,55){$\mathbf{t}$}
	\put(51,60){$\mathbf{p}_t$}
	\put(61,18){$\mathbf{p}_0$}
	\put(63,33){\color{black}\line(0,1){50}}
	\put(66,55){\color{black}\line(0,1){22}}
	\put(63,83){\color{black}\line(1,0){7}}
	\put(66,77){\color{black}\line(1,0){4}}
	\put(72,82){Surface of Revolution}
	\put(72,76){Axial Surface}
	\end{small}     
  \end{overpic}
  \end{minipage}
\end{center}
	\caption{Illustration of the trajectory curve (magenta) and directrix (red) for T-surfaces of different types with the same profile curve (cyan). The dashed lines show the top view of the profile planes. Furthermore, the symbols $\Bar{\mathbf{f}_t}$ and $\Bar{\mathbf{f}_n}$ denote the Frenet frame of the trajectory curve $\mathbf{t}$ while ${\mathbf{f}_t}$ and ${\mathbf{f}_n}$ stand for the Frenet frame of the directrix curve. The Frenet frame of the directric curve's origin is moved in order to give a visual comparison to the Frenet frame of $\mathbf{t}$.}
	\label{fig:frames}
\end{figure}

\section{Sub-classes of T-surfaces}\label{appendix:subclasses}


The formulation of Eq.\ (\ref{eq:Tsurface:general:case}) allows one to obtain different classes of the T-surfaces
(with the exception of \emph{translational} type)
in the following way {(cf.\ Fig.\ \ref{fig:frames})}:
\begin{itemize}
    \item Substituting $\beta = 0$ results in $\xi = 1$ and therefore Eq.\ (\ref{eq:Tsurface:general:case}) gives us a \emph{molding surface}. {Note that in this case the $\beta$-involutes get simplified to classical involutes.}
    \item The T-surface of \emph{axial surface} type can be seen as the limit of a T-surface of \emph{general case} type in the following way:
          Let the directrix curve be parametrized in such a way that $\|\,\dot{\mathbf{d}}\,\| = R$, where $R$ is a real constant. Then we have
          \begin{equation*}
                {d}_x(t) = R\int_{\,0}^{\,t}\cos\,(\theta\,(w))\,dw,\quad\quad\quad\quad {d}_y(t) = R\int_{\,0}^{\,t}\sin\,(\theta\,(w))\,dw. 
          \end{equation*}
          Now, write Eq.\ (\ref{eq:Tsurface:general:case}) as 
          \begin{equation*}
              \mathbf{x}(s,t) = \left(\begin{array}{>{\displaystyle}c}
              - \,\xi(t)\,p_x(s)\,\frac{\dot{\mathbf{d}}(t)}{\|\,\dot{\mathbf{d}}(t)\,\|}\\[.5cm]
              p_z(s)\end{array}\right) + 
              \underbrace{\left(\begin{array}{>{\displaystyle}c}
              \mathbf{d}(t) - \,\xi(t)\,\left(\int_{\,0}^{\,t}\frac{\|\,\dot{\mathbf{d}}(w)\,\|}{\xi(w)}\,dw\right)\,\frac{\dot{\mathbf{d}}(t)}{\|\,\dot{\mathbf{d}}(t)\,\|}\\[0.5cm]
              0\end{array}\right)}_{=:\,\star}.
          \end{equation*}
          Taking limit over $\star$ with respect to $R$ results in 
            \begin{equation*}
                \lim_{R\,\rightarrow\,0} \star = \lim_{R\,\rightarrow\,0}\, 
                \left(\begin{array}{c}
                R\left(\int_{\,0}^{\,t}\,\cos\,(\theta(w))\,\mathrm{d}w -\xi(t)\,\cos\,(\theta(t))\,\int_{\,0}^{\,t}\frac{1}{\xi(w)}\,dw\right)\\[0.2 cm]
                R\left(\int_{\,0}^{\,t}\,\sin\,(\theta(w))\,\mathrm{d}w -\xi(t)\,\sin\,(\theta(t))\,\int_{\,0}^{\,t}\frac{1}{\xi(w)}\,dw\right)\\[0.2 cm]
                0\end{array}\right) = \mathbf{0}.
            \end{equation*}
            Consequently, after simplification $\mathbf{x}$ can be written as 
             \begin{equation}\label{eq:axial:surface}
              \mathbf{x}(s,t) = \left(\begin{array}{>{\displaystyle}c}
              - \,\xi(t)\,p_x(s)\,\cos{(\theta(t))}\\[.2cm]
              - \,\xi(t)\,p_x(s)\,\cos{(\theta(t))}\\[.2cm]
              p_z(s)\end{array}\right),
          \end{equation}
          which is a stretch-rotation surface and hence an \emph{axial surface}. {Following two interesting special cases exist:}
    \begin{enumerate}[$\star$]
           \item {For constant $\beta \neq 0$ the trajectories are \emph{logarithmic spirals}.}
        \item {$\beta = 0$ results in $\xi = 1$ and therefore Eq.\ (\ref{eq:axial:surface}) gives us a \emph{surface of revolution}.}
    \end{enumerate}
\end{itemize}
\bibliography{main}

\end{document}